\documentstyle[preprint,aps,axodraw]{revtex} 
\begin{document}
\draft 
\preprint{${\scriptstyle{\rm ZU-TH\ 19/97} \atop
\scriptstyle{\rm UMHEP-445}}$}
\tightenlines
\title{Two-loop Analysis of Axialvector Current Propagators
in Chiral Perturbation Theory}
\author{Eugene Golowich} 
\address{Department of Physics and Astronomy,
University of Massachusetts \\
Amherst MA 01003 USA}
\author{Joachim Kambor}
\address{Institut f\"ur Theoretische Physik,
Universit\"at Z\"urich \\
CH-8057 Z\"urich, Switzerland}
\maketitle
\def\Tr{\,{\rm Tr}\,}
\def\beq{\begin{equation}}
\def\eeq{\end{equation}}
\def\beqa{\begin{eqnarray}}
\def\eeqa{\end{eqnarray}}
\begin{abstract}
\noindent
We perform a calculation of the isospin and hypercharge
axialvector current propagators ($\Delta_{{\rm A}3}^{\mu\nu}(q)$
and $\Delta_{{\rm A}8}^{\mu\nu}(q)$) to two loops in 
$SU(3)\times SU(3)$ chiral perturbation theory.  A large number 
of ${\cal O}(p^6)$ divergent counterterms are fixed, and 
complete two-loop renormalized expressions for the pion and 
eta masses and decay constants are given.  The calculated 
isospin and hypercharge axialvector polarization functions 
are used as input in new chiral sum rules, valid to second order 
in the light quark masses.  Some phenomenological implications of 
these sum rules are considered.
\end{abstract}
\section{\bf Introduction}
Although low energy quantum chromodynamics remains analytically 
intractable, the calculational scheme of chiral perturbation 
theory~\cite{sw} (ChPT) has led to many valuable contributions.  
Following the seminal papers of Gasser and Leutwyler~\cite{gl1,gl2}, 
numerous studies conducted in the following decade convincingly 
demonstrated the power of ChPT.  The state of the art up to 1994 
is summarized in several reviews {\it e.g.}~\cite{eck95,bkm} 
(see also \cite{daph2,bh95}).  The exploration of ChPT continues 
to this day, and two-loop studies represent an active frontier area 
of research.  These include processes which have leading 
contributions in the chiral expansion at order 
$p^4$~\cite{bgs94,gk1,bur96,jet96,bt97,ps97} or even $p^6$~\cite{gk2},  
as well as systems for which precision tests will soon be 
available, {\it e.g.} the low-energy behaviour of 
$\pi\pi$ scattering~\cite{kmsf96,betal96,betal97}. 
While the case of $SU(2)\times SU(2)$ ChPT to two-loop order 
has been relatively well explored (in particular see \cite{betal97}), 
works in $SU(3)\times SU(3)$ ChPT are still few in 
number~\cite{gk1,jet96,ps97,can1,can2}.  

Recently, we performed a calculation of the isospin and 
hypercharge vector current propagators ($\Delta_{{\rm
V}3}^{\mu\nu}(q)$ and $\Delta_{{\rm V}8}^{\mu\nu}(q)$) 
to two-loop order in $SU(3)\times SU(3)$ chiral perturbation 
theory~\cite{gk1}.  A partial motivation for working 
in the three-flavour sector stems from its inherently richer 
phenomenology.  In particular, it becomes 
possible to derive new chiral sum rules which explicitly 
probe the $SU(3)$-breaking sector.  With the aid of improved 
experimental information on spectral 
functions with strangeness content, it should become possible 
to evaluate and test these sum rules.

We have completed this program of calculation by determining the
corresponding isospin and hypercharge axialvector current propagators
($\Delta_{{\rm A}3}^{\mu\nu}(q)$ and $\Delta_{{\rm A}8}^{\mu\nu}(q)$).
Determination of axialvector propagators is much more 
technically demanding than for the vector propagators, 
but at the same time yields an extended set of results, 
among which are: 
\begin{enumerate}
\item a large number of constraints on the set of ${\cal O}(p^6)$ 
counterterms, 
\item predictions for threshold behaviour of the $3\pi$, ${\bar K}K\pi$, 
${\bar K}K\pi$, $\eta \pi \pi$, {\it etc} axialvector spectral 
functions, 
\item an extensive analysis of the so-called `sunset' diagrams, 
\item new axialvector spectral function sum rules, 
\item a new contribution to the Das-Mathur-Okubo sum 
rule~{\cite{dmo}}, and 
\item a complete two-loop renormalization of the masses 
and decay constants of the pion and eta mesons.  
This final item places the axialvector problem 
at the heart of two-loop studies in $SU(3)\times SU(3)$ ChPT. 
\end{enumerate}

We begin the presentation in Sect.~II by presenting basic 
definitions and describing the calculational approach.  
To illustrate the procedure, we summarize the results for 
the tree-level and one-loop sectors.  Our calculation of the 
two-loop amplitudes and the corresponding enumeration of the 
${\cal O}(p^6)$ counterterms is given in Sect.~III.  
The construction of a proper renormalization procedure forms 
the subject of Sect.~IV.  It provides the framework for the 
removal of divergences, described in Sect.~V, and leads to 
finite renormalized expressions for the meson masses, decay 
constants, and polarization functions.  We give explicit expressions 
for the isospin polarization functions in Sect.~VI and continue 
the presentation of finite results in Appendix B.  Sect.~VII deals 
with the determination of spectral functions. The subject of chiral 
sum rules is discussed in Sect.~VIII and Sect.~IX contains 
our conclusions.  Various technical details regarding sunset 
integrals are presented in Appendix A and as mentioned, 
our final expressions for meson masses, decay constants 
and the hypercharge polarization amplitudes are collected in Appendix B. 
In addition, at several points in the paper we compare results 
as expressed in the `$\overline\lambda$-subtraction' renormalization
used here with a variant of the $\overline{MS}$  scheme.  
\section{Tree-level and One-loop Analyses}
The one-loop chiral analysis of the isospin axialvector-current
propagator was first carried out by Gasser
and Leutwyler who used the background-field formalism and
worked in an $SU(2)$ basis of fields~\cite{gl1}.
We shall devote this section to a re-calculation of the isospin
axialvector propagator through one-loop order, but now done within
the context of a Feynman diagram calculation and
using an $SU(3)$ basis of fields.
\subsection{Basic Definitions and Calculational Procedure}
Our normalization for the $SU(3)$ octet of axialvector currents 
is standard,
\begin{equation}
A^\mu_k = {\bar q}{\lambda_k \over 2}\gamma^\mu \gamma_5 q
\qquad (k = 1,\ldots,8) \ \ .
\label{curr}
\end{equation}
In this paper, we shall deal with the axialvector current propagators 
\begin{equation}
\Delta_{{\rm A}a}^{\mu\nu}(q) \equiv i \int d^4x~ e^{iq\cdot x}~
\langle 0|T\left( A^\mu_a (x) A^\nu_a (0)\right)|0\rangle 
\qquad (a = 3,8 \ {\rm not \ summed}) \ \ ,
\label{aa1}
\end{equation}
having the spectral content 
\begin{equation}
{1\over \pi} {\cal I}m~\Delta_{{\rm A}a}^{\mu\nu}(q) = 
(q^\mu q^\nu - q^2 g^{\mu\nu}) \rho_{{\rm A}a}^{(1)}
(q^2 ) + q^\mu q^\nu \rho_{{\rm A}a}^{(0)} (q^2) \ , 
\label{aa2}
\end{equation}
where $\rho_{{\rm A}a}^{(1)}$ and $\rho_{{\rm A}a}^{(0)}$ are 
the spin-one and spin-zero spectral functions.  This motivates 
the following tensorial decomposition usually adopted in the 
literature, 
\begin{equation}
\Delta_{{\rm A}a}^{\mu\nu}(q) = 
(q^\mu q^\nu - q^2 g^{\mu\nu}) \Pi_{{\rm A}a}^{(1)}
(q^2 ) + q^\mu q^\nu \Pi_{{\rm A}a}^{(0)} (q^2) \ \ ,
\label{aa2a}
\end{equation}
where $\Pi_{{\rm A}a}^{(1)}$ and $\Pi_{{\rm A}a}^{(0)}$ are 
the spin-one and spin-zero polarization functions.  
The low-energy behaviour of the spin-zero spectral function 
$\rho_{{\rm A}a}^{(0)}$ is dominated by the pole contribution 
associated with propagation of a Goldstone mode,
\beq
\rho_{{\rm A}a}^{(0)}(s) \equiv ~F_a^2 ~ \delta ( s - M_a^2) + 
{\bar \rho}_{{\rm A}a}^{(0)}(s) \ \ .
\label{pole}
\eeq
In the chiral limit, one has $M_a\to 0$ and 
${\bar \rho}_{{\rm A}a}^{(0)}(s) \to 0 $ as well.  

The lowest-order chiral lagrangian ${\cal L}^{(2)}$ is given by
\begin{equation}
{\cal L}^{(2)} = {F^2_0\over 4} \Tr \left( D_\mu U D^\mu U^\dagger\right) +
{F^2_0\over 4} \Tr \left(\chi U^\dagger + U\chi ^\dagger\right) \ ,
\label{P5}
\end{equation}
where $F_0$ is the pseudoscalar meson decay constant
to lowest order and $\chi = 2B_0 {\bf m}$ is proportional to
the quark mass matrix ${\bf m} = {\rm diagonal}\left(
{\hat m}, {\hat m}, m_s\right)$ with
\beq
B_0 = - {1\over F_0^2} \langle {\bar q}q \rangle \ \ .
\label{P5a}
\eeq
Note that we work in the isospin symmetric limit of $m_u = m_d \equiv
{\hat m}$ and that to lowest order we may use the Gell~Mann-Okubo
relation, 
\beq
m_\eta^2 = {1\over 3} \left( 4 m_K^2 - m_\pi^2 \right) \ \ .
\label{gmo}
\eeq
The field variable $U$ is defined in terms of the octet of
pseudoscalar meson fields $\{ \phi_k \}$,
\begin{equation}
U \equiv \exp (i\lambda_k\cdot\phi_k/F_0 ) \ ,\qquad
\label{P6}
\end{equation}
and we construct the covariant derivative $D_\mu U$ via 
external axialvector sources $a_\mu$,\footnote{Throughout this 
paper we adopt the phase employed in Ref.~\cite{dgh}, which is 
opposite to that used in Ref.~\cite{gl2}.}
\begin{equation}
D_\mu U \equiv \partial_\mu U + ia_\mu U + iU a_\mu \ \ .
\label{DP}
\end{equation}
The axialvector source $a_\mu$ has a component $a_\mu^k$ for each of the
$SU(3)$ flavours,
\beq
a_\mu \equiv {1 \over 2} \lambda_k ~a_\mu^k \ \ .
\label{P8}
\eeq
 
Adopting the approach carried out in Ref.~\cite{gk1}, we make 
use of external axialvector sources to determine the axialvector 
propagators.  The procedure is simply to compute the ${\cal S}$-matrix
element connecting initial and final states of an axialvector 
source.  Analogous to the analysis of Ref.~\cite{gk1}, the invariant 
amplitude is then guaranteed to be the axialvector-current 
propagator, say of flavour `$a$',  
\beq
\langle a_a (q',\lambda')|{\cal S} -1 |a_a (q,\lambda)\rangle 
= i (2\pi)^4 \delta^{(4)} (q' - q) ~\epsilon_\mu^\dagger (q',\lambda') 
\Delta^{\mu\nu}_{{\rm A}a} (q) \epsilon_\nu (q,\lambda) \ .
\label{P11}
\eeq
We shall typically use the invariant amplitude symbol 
${\cal M}_{\mu\nu a}$ to denote various individual contributions 
(tree, tadpole, counterterm, 1PI, {\it etc}) to the full propagator.  
\subsection{Tree-level Analysis}
For definiteness, the analysis in the remainder of this section 
will refer to the isospin flavour.  Given the lagrangian of 
Eq.~(\ref{P5}), it is straightforward to determine
the lowest-order propagator contributions, 
\beq
{\cal M}_{\mu\nu 3}^{\rm (tree)} = F^2_0 g_{\mu\nu} - 
{F^2_0 \over q^2 - m_\pi^2 + i \epsilon}
q_\mu q_\nu \ \ .
\label{lo}
\eeq
The two terms represent respectively contributions from a contact 
interaction (Fig.~1(a)) and a pion-pole term (Fig.~1(b)).  Although 
${\cal M}_{\mu\nu 3}^{\rm (tree)}$ has an exceedingly simple form, it is 
nonetheless worthwhile to briefly point out two of its features.  
First, as follows from unitarity there is an imaginary part
corresponding to the pion single-particle intermediate state, 
\beq
{\cal I}m~{\cal M}_{\mu\nu 3}^{\rm (tree)} = \pi F_0^2 
\delta (q^2 - m^2_\pi) q_\mu q_\nu \ \ .
\label{loa}
\eeq
However, there is also a non-pole contribution to 
${\cal M}_{\mu\nu 3}^{\rm (tree)}$.  Its presence is needed to 
ensure the proper behaviour in the chiral limit
($\partial_\mu A^\mu_3 = 0$), where 
${\cal R}e~{\cal M}_{\mu\nu 3}^{\rm (tree)}$ 
is required to obtain a purely spin-one (or `transverse') form.  
This is indeed the case, as we find by taking $m_\pi^2 \to 0$ 
in Eq.~(\ref{lo}), 
\beq
{\cal M}_{\mu\nu 3}^{\rm (tree)}\bigg|_{m_\pi = 0} 
= -{F^2_0 \over q^2}~(q_\mu q_\nu - q^2 g_{\mu\nu}) \ \ .
\label{form}
\eeq
\subsection{One-loop Analysis}
At one-loop level, the axialvector-current propagators are
determined from the lagrangian of Eq.~(\ref{P5}) and correspond
to the Feynman diagrams appearing in Fig.~2.  The loop correction
to the lowest-order contact amplitude appears in Fig.~2(a) and
Figs.~2(b),(c) depict corrections (whose significance we shall
consider shortly) to the pion-pole amplitude.  We find for the 
one-loop (`tadpole') contributions to the isospin propagator, 
\beqa
\lefteqn{{\cal M}_{\mu\nu 3}^{\rm (tadpole)} = -i\left( 2 A(m_\pi^2)
+ A(m_K^2) \right)g_{\mu\nu}  + {4i\over 3} { 2 A(m_\pi^2)
+ A(m_K^2) \over q^2 - m_\pi^2 } q_\mu q_\nu}  \nonumber \\
& & + {i\over 6} {  A(m_\pi^2) ( m_\pi^2 - 4q^2 )
+ 2A(m_K^2) ( m_K^2 - q^2 ) + A(m_\eta^2) m_\pi^2 \over
(q^2 - m_\pi^2 )^2} q_\mu q_\nu \ , 
\label{lp1}
\eeqa
where all masses occurring in pole denominators are understood 
to have infinitesimal negative imaginary parts.  
In the above, $A$ is the scalar integral
\beq
A ( m^2 ) \equiv \int d{\tilde k} 
 ~ { 1 \over k^2 - m^2 } \ \ ,
\label{a1}
\eeq
where $d{\tilde k} \equiv d^d k/(2\pi)^d$.  Hereafter, 
any integration measure accompanied by a super-tilde 
will have a similar meaning. The evaluation of $A$ is standard, 
and we have
\beqa
A ( m^2 ) &=& {-i \over (4 \pi)^{d/2} }
{\mu^{4-d} \over \mu^{4-d}} 
\Gamma\left( 1 - {d\over 2} \right)
(m^2 )^{(d - 2 )/2} \label{a2} \\
&=& \mu^{d-4} \left[ -2i m^2 {\overline \lambda}
- { im^2 \over 16 \pi^2} \log \left( {m^2 \over 
\mu^2} \right) + \dots \right] \ \ ,
\label{a3}
\eeqa
with 
\beq
\lambda = \mu^{d - 4} {\overline \lambda} =
{ \mu^{d - 4} \over 16 \pi^2} \left[ {1 \over d - 4} - {1\over 2}
\left( \log {4\pi} - \gamma + 1 \right) \right] \ \ .
\label{a4}
\eeq
The quantity $\mu$ introduced in Eq.~(\ref{a2}) is the mass scale
which enters the calculation via the use of dimensional regularization.
The $\mu^{d - 4}$ prefactor in Eq.~(\ref{a3}) ensures that 
$A(m^2)$ has the proper units in $d$-dimensions.
 
To deal with divergences arising from the loop corrections, 
one must include counterterm amplitudes.  It suffices to employ the 
well-known list of counterterms $\{ L_i \}$ ($i = 1,\dots, 10$) 
and $\{ H_j \}$ ($j = 1,2$) appearing in the ${\cal O}(p^4 )$ 
chiral lagrangian of Gasser and 
Leutwyler~\cite{gl2}.\footnote{See also the discussion surrounding 
Eq.~(26) of Ref.~\cite{gk1}.}  
Analogous to Eq.~(\ref{a3}) for the $A(m^2)$ integral, each 
${\cal O}(p^4 )$ counterterm is expressible as an expansion 
in ${\overline \lambda}$,  
\begin{equation}
L_\ell = \mu^{(d - 4)} \sum_{n=1}^{-\infty} \ L_\ell^{(n)}(\mu) 
~{\overline\lambda}^n 
= \mu^{(d - 4)} \left[~ L_\ell^{(1)}(\mu) ~{\overline\lambda}
\ + \ L_\ell^{(0)}(\mu) \ + L_\ell^{(-1)}(\mu) ~{\overline\lambda}^{-1}
+ \dots ~\right] \ \ ,
\label{aa14} \\
\end{equation}
where the leading degree of singularity is seen to be 
linear.  The counterterm diagrams involve a contact term 
(Fig.~2(d)) as well as contributions to the pion pole 
term (Figs.~2(e),(f)) and 
yield the following isospin counterterm amplitude, 
\beqa
\lefteqn{{\cal M}_{\mu\nu 3}^{({\rm CT})} =
2(L_{10} - 2H_1) (q_\mu q_\nu - q^2 g_{\mu\nu})
+ 8\left[ (m_\pi^2 + 2m_K^2 )L_4 + m_\pi^2 L_5 \right] g_{\mu\nu} }
\nonumber \\
& & - 16 {  (m_\pi^2 + 2m_K^2 )L_4 + m_\pi^2 L_5
\over  q^2 - m_\pi^2 } q_\mu q_\nu \label{ct4} \\
& & + 8 {  q^2 \left( (m_\pi^2 + 2m_K^2 )L_4 + m_\pi^2 L_5 \right)
-2 m_\pi^2 \left( (m_\pi^2 + 2m_K^2 )L_6 + m_\pi^2 L_8 \right)
\over  (q^2 - m_\pi^2 )^2} q_\mu q_\nu \ .\nonumber
\eeqa
 
\begin{center}
{\bf Results through One-loop Order}
\end{center}
 
Our complete expression for the isospin axialvector current propagator 
through one-loop order is given by the sum of
Eqs.~(\ref{lo}),(\ref{lp1}),(\ref{ct4}),
\beq
\Delta_{{\rm A}\mu\nu3} = {\cal M}_{\mu\nu 3}^{\rm (tree)} +
{\cal M}_{\mu\nu 3}^{\rm (tadpole)} +
{\cal M}_{\mu\nu 3}^{({\rm CT})} \ \ .
\label{res}
\eeq
The resulting expression is complicated and seems to lack immediate 
physical interpretation because we have not yet 
accounted for renormalizations of the pion's mass and decay constant.  
A detailed account of the renormalization procedure is 
deferred to Sect.~IV.  However, we note here that the 
renormalized masses and decay constants have the expansions 
\begin{eqnarray}
F^2 &=& F^{2(0)} + F^{2(2)} + F^{2(4)} + \dots  \nonumber \\
M^2 &=& M^{2(2)} + M^{2(4)} + M^{2(6)} + \dots  
\label{fm1}
\end{eqnarray}
where we have temporarily suppressed flavour notation and 
the superscript indices $\{ (i) \}$ denote quantities 
evaluated at chiral order $\{ p^i \}$.  To one loop~\cite{gl2}, 
the explicit expressions are given by\footnote{At 
one-loop order, our counterterms $\{ L_i^{(0)} \}$ and 
$\{ H_i^{(0)} \}$ are equivalent to the $\{ L_i^r \}$ and 
$\{ H_i^r \}$ of Ref.~\cite{gl2}.}
\beqa
F_\pi^2 &=& F^2_0 +
8\left[ (m_\pi^2 + 2m_K^2 )L_4^{(0)} + m_\pi^2
L_5^{(0)} \right]
- {2m_\pi^2 \over 16\pi^2} \ln{m_\pi^2 \over \mu^2}
- {m_K^2 \over 16\pi^2} \ln{m_K^2 \over \mu^2} \nonumber \\
M_\pi^2 &=& m_\pi^2 + {1\over F_0^2} \left[
{m_\pi^4 \over 32\pi^2} \ln{m_\pi^2 \over \mu^2}
- {m_\pi^2 m_\eta^2 \over 96\pi^2} \ln{m_\eta^2 \over \mu^2}
\right. \nonumber \\
& & \left. -  8m_\pi^2 \left( (m_\pi^2 + 2m_K^2)(L^{(0)}_4 -
2L^{(0)}_6) + m_\pi^2 (L^{(0)}_5 - 2 L^{(0)}_8)\right)\right]\ .
\label{fm2}
\eeqa
Upon combining the information gathered in 
Eqs.~(\ref{res}),(\ref{fm2}) as well as the divergent part 
of the one-loop functional given in Ref.~\cite{gl2}, one finds 
through one-loop for the renormalized isospin propagator, 
\beq
\Delta^{\mu\nu}_{{\rm A}3} = F_\pi^2 g^{\mu\nu} + 
2(L_{10}^{(0)} - 2H_1^{(0)}) (q^\mu q^\nu - q^2 g^{\mu\nu})
 - {F_\pi^2 \over q^2 - M_\pi^2} q^\mu q^\nu \ .
\label{tot}
\eeq

The hypercharge propagator $\Delta^{\mu\nu}_{{\rm A}8}$ can be 
obtained analogously.  We comment that both 
the isospin and hypercharge amplitudes contain the 
regularization dependent constant $H_1^{(0)}$, and are therefore
unphysical.  A physically observable quantity is obtained from 
the difference,  
\beq
\Delta^{\mu\nu}_{{\rm A}3} - \Delta^{\mu\nu}_{{\rm A}8} = 
\left( F_\pi^2 - F_\eta^2 \right) g^{\mu\nu} 
- q^\mu q^\nu \left[ {F_\pi^2 \over q^2 
- M_\pi^2} - {F_\eta^2 \over q^2 - M_\eta^2} \right] \ \ .
\label{diff}
\eeq
\section{\bf Two-loop Analysis} 
The general structure of propagator corrections is 
displayed in Fig.~3, which includes the one-particle irreducible 
(1PI) diagrams of Fig.~3(a) and the one-particle reducible (1PR) 
diagrams of Fig.~3(b).  The latter consists of both 
vertex and self-energy corrections.  Throughout this section, 
we continue to focus on the isospin sector when giving explicit 
expressions for two-loop amplitudes.  
\subsection{Two-loop Analysis: 1PI Graphs}
First, we consider the 1PI diagrams of Fig.~4 and Fig.~5.  
The graphs of Figs.~4(a)--(c) are simple 
in the sense that they are either equal to or
the negative of identical graphs in which external vector
sources occur.  As such, they can be read off from the work
of Ref.~{\cite {gk1}}.  This is only partly true of Fig.~4(d), 
and the graph of Fig.~5 (the so-called `sunset graph') has no 
counterpart in the vector system.  For convenience, we 
compile definitions and explicit representations for the 
sunset-related functions in App.~A, leaving detailed derivation 
of these results for another setting~\cite{gk3}.  

The amplitudes of Figs.~4(a)--(b) are of the general form 
\beq
{\cal M}_{\mu\nu 3}[4a,4b] = {g_{\mu\nu} \over F_0^2} ~
\sum_{k,\ell}\ a_{k\ell}^{(a,b)} A(k)~A(\ell)   \ \ ,
\label{4a} 
\eeq
where the $\{ a_{k\ell}^{(a,b)} \}$ are numerical coefficients
(in some cases dependent on dimension $d$) and the 
$\{ A(k) \}$ are the integrals of Eq.~(\ref{a1}).  
The sum over the indices $k,\ell$ simply reflects the 
need to perform flavour sums independently for each of the 
two loops.  The amplitude of Fig.~4(c) can be expressed 
similarly, 
\beq
{\cal M}_{\mu\nu 3}[4c] = {g_{\mu\nu} \over F_0^2} ~
\sum_{k,\ell}\ a_{k\ell}^{(c)} A(k)~L_\ell   \ \ ,
\label{4b} 
\eeq 
except now the coefficients$\{ a_{k\ell}^{(c)} \}$ are proportional 
to squared meson masses and there are $\{ L_{\ell} \}$ factors 
arising from the ${\cal O}(p^4)$ lagrangian.  

Each of the above amplitudes will diverge for $d\to 4$, and we write 
\beq
{\cal M}_{\mu\nu 3}[4a,4b,4c] = {g_{\mu\nu} \over F_0^2} ~
\left[ {\overline a}_2^{(a,b,c)} {\bar\lambda}^2 + 
{\overline a}_1^{(a,b,c)} {\bar\lambda}^1 + \dots \right]\ \ ,
\label{4c} 
\eeq
where the singular quantity ${\overline \lambda}$ has been 
previously defined in Eq.~(\ref{a4}), ${\overline a}_2^{(a,b,c)}$ 
and ${\overline a}_1^{(a,b,c)}$ are numerical quantities,  
and the ellipses refer to terms which are nonsingular 
at $d=4$.  As expected for two-loop amplitudes, the leading 
singularity goes as ${\overline \lambda}^2$. 

For Fig.~4(d), the contribution 
from $L_{10}$ is the negative of the vector case, 
but there are also many new terms, 
\beqa
\lefteqn{{\cal M}_{\mu\nu 3}[4d] = -(q_\mu q_\nu - g_{\mu\nu}q^2) 
{4 i \over F_0^2} ( 2A(\pi) + A(K)) L_{10} } \nonumber \\
& & + {i g_{\mu\nu} \over F_0^2} \Bigg[
A(\pi)\bigg( m_\pi^2 [ (48 + {32\over d})L_1
+ (16 + {64\over d})L_2  + (24 + {16\over d}) L_3 ] \nonumber \\
& & \phantom{xxxxxxxxx} - 8 (5 m_\pi^2 + 4 m_K^2 ) L_4 - 28 m_\pi^2 L_5
\bigg) \nonumber \\
& & + A(K)\bigg( m_K^2 [ 64 L_1
+ {64\over d} L_2 + (16 + {16\over d} )L_3]  \label{4d} \\
& & \phantom{xxxxxxxxx} - 8 ( m_\pi^2 + 6 m_K^2 ) L_4 - 8 (m_\pi^2 +
m_K^2) L_5 \bigg) \nonumber \\
& & + A(\eta)\bigg( m_\eta^2 [ 16 L_1
+ {16\over d} L_2 + ({8\over 3} + {16\over 3d}) L_3  ]
- 8 m_\eta^2 L_4 - {4\over 3} m_\pi^2 L_5 \bigg) \Bigg] \ .\nonumber 
\eeqa
Aside from the presence of a transverse component 
proportional to $(q_\mu q_\nu - g_{\mu\nu}q^2)$,  it resembles 
the form of Fig.~4(c) and shares the degree of singularity 
shown in Eq.~(\ref{4c}).  

Finally, there is the sunset contribution of Fig.~5, which 
we write as 
\beqa
\lefteqn{{\cal M}_{\mu\nu 3}[5] = {4\over 9} {\cal H}_{\mu\nu} 
(q^2, m_\pi^2 , m_\pi^2 ) + {1\over 6} {\cal H}_{\mu\nu} (q^2,
m_\eta^2 , m_K^2)} \nonumber \\
& & \phantom{xxxxxxx} 
+ {1\over 18} {\cal H}_{\mu\nu} (q^2, m_\pi^2 , m_K^2 ) + 
{\cal L}_{\mu\nu} (q^2, m_\pi^2 , m_K^2 ) \ .
\label{5}
\eeqa
The quantities ${\cal H}_{\mu\nu}$ and ${\cal L}_{\mu\nu}$ 
are defined in App.~A by Eqs.~(\ref{ss3}),(\ref{ss4}).  
The flavour dependence of the individual sunset contributions 
can be read off from the arguments of the above functions.  
The defining representations for ${\cal H}_{\mu\nu}$ and 
${\cal L}_{\mu\nu}$ ({\it cf} Eqs.~(\ref{ss3}),(\ref{ss4})) 
are given as multidimensional Feynman integrals.  
It takes considerable analysis to reduce these integrals 
to forms which can be compared directly with the other 
two-loop amplitudes, and one finds~\cite{gk3} that the sunset 
amplitudes share the singular behaviour of Eq.~(\ref{4c}).  
\subsection{\bf Vertex Corrections}
The 1PR graphs are of two types, the vertex modifications 
of Figs.~6(a)--(e) and the self-energy effects of Figs.~7(a)--(e). 

The two-loop vertex amplitude $\Gamma^{(4)}_{\mu a}$ is defined by 
\beq
\langle P_a (q') | {\cal S} - 1 | a_a (q, \lambda) 
\rangle_{\rm Fig.~6} = 
i (2\pi)^4 \delta^{(4)}(q' - q)~\epsilon^\mu (q,\lambda) 
\Gamma^{(4)}_{\mu a}  \qquad (a = 3,8) \ \ .
\label{v0}
\eeq
A typical example is the vertex amplitude of Fig.~6(b), 
\beqa
& & \Gamma^{(4)}_{\mu3}[b] = {iq_\mu\over 18F_0^3}  
\bigg[ A(\pi) \bigg( 16 A (\pi)  + 
12 m_\pi^2 B(0,\pi) \bigg) + 11 A(K) A(\pi)  \label{v3}\\ 
& & +  6 A^2 (K) + A(\eta) \bigg( 3 A(K) - 4 m_\pi^2 B(0,\pi) 
+ 4 m_K^2 B(0,K) \bigg) \bigg] \ \ , 
\nonumber 
\eeqa 
where $B(0,m^2)$ is defined and evaluated as 
\begin{equation}
B(0,m^2) \equiv \int d {\tilde k} ~{ 1 \over (k^2 - m^2)^2} 
= {1\over m^2} \left[ A(m^2) - {4 - d\over 2}A(m^2) \right] \ \ .
\label{v3a}
\end{equation}

The amplitudes of Figs.~6(a),(b) have the generic form 
\beq
\Gamma^{(4)}_{\mu 3}[a,b] = {q_\mu \over F_0^2} ~
\sum_{k,\ell}\ v_{k\ell}^{(a,b)} A(k)~A(\ell)   \ \ ,
\label{v3b} 
\eeq
with numerical coefficients $v_{k\ell}^{(a,b)}$ 
whereas those of Figs.~6(c),6(d) are 
\beq
\Gamma^{(4)}_{\mu 3}[c,d] = {q_\mu \over F_0^2} ~
\sum_{k,\ell}\ v_{k\ell}^{(c,d)} A(k)~L_\ell   \ \ ,
\label{v3c} 
\eeq
with coefficients $v_{k\ell}^{(c,d)}$ proportional to the 
squared meson masses.  As in Eq.~(\ref{v3})), all the invariant 
parts of non-sunset vertex functions are found to be independent 
of the propagator momentum $q^2$. 

For the sunset contribution of Fig.~6(e), we write 
\beqa
\lefteqn{\Gamma^{(4)}_{\mu 3}[e] = 
{i \over F_0^3}\bigg[ {2\over 9}I_{1\mu}
\left(q^2;m_\pi^2;m_\pi^2;m_\pi^2\right) 
+ {1\over 36}I_{1\mu} \left(q^2;m_\pi^2;m_K^2;2(m_\pi^2 + 
m_K^2)\right)} \nonumber \\
& & + {1\over 12}I_{1\mu} \left(q^2;m_\eta^2;m_K^2;{2\over 3}(m_\pi^2 -
m_K^2)\right) 
+ {1\over 2}I_{2\mu} \left(q^2;m_\pi^2;m_K^2\right) 
\bigg]\ \ . \label{6e} 
\eeqa 
The vector-valued quantities $I_{1\mu}$ and $I_{2\mu}$ 
are defined in Eqs.~(\ref{int1}),(\ref{int2}) of App.~A, and 
are found to share the singular behaviour of Eq.~(\ref{4c}).  

\subsection{\bf Self-energy Corrections}
The two-loop self-energy $\Sigma_a^{(6)}$ 
arises from the meson-to-meson matrix element, 
\beq
\langle P_a (q') | {\cal S}- 1 | P_a (q) \rangle_{\rm Fig.~7} 
= i (2\pi)^4 \delta^{(4)}(q' - q)~\Sigma_a^{(6)} \qquad (a = 3,8) 
\ \ .
\label{se1}
\eeq

The non-sunset contributions of Figs.~7(a)--7(b) are 
proportional to $\{ A(k) A(\ell) \}$ and those of 
Figs.~7(c)--7(d) are proportional to $\{ A(k) L_\ell \}$.  
As an example, the isospin self-energy contribution of Fig.~7(a) is 
\beqa
F_0^4 \Sigma^{(6)}_3[a] &=& \bigg[ - {4q^2 \over 9} + \left( {7\over 24} 
-{8\over 9} \right)m_\pi^2  \bigg] A^2 (\pi) 
+ {5m_\pi^2 \over 36} A(\pi) A(\eta) \nonumber \\
&+& \bigg[ - {7q^2 \over 18}  - {7 m_\pi^2 + 2 m_K^2 \over 18} + 
{2 m_\pi^2 +  m_K^2 \over 9}\bigg]A(\pi) A(K)  \nonumber \\
&+& \bigg[ - {2q^2 \over 15} - {17 m_K^2  \over 45}
+ { 7 (m_\pi^2 +  2 m_K^2) \over 90}\bigg] A^2 (K) \label{se2} \\
&+& \bigg[ - {q^2 \over 30}  - {6 m_K^2 +  m_\eta^2 \over 30} + 
{m_\pi^2 +  m_K^2 \over 45}\bigg]A(K) A(\eta)  + 
{m_\pi^2 \over 72} A^2 (\eta) \ \ . \nonumber 
\eeqa
As in Eq.~(\ref{se2}), the remaining non-sunset self-energies are at 
most linear in the propagator momentum $q^2$.  

Finally, the sunset contribution of Fig.~7(e) is given by 
\beqa
& & F_0^4 \Sigma^{(6)}_3[e] = {m_\pi^4 \over 6} 
S(q^2; m_\pi^2; m_\pi^2) + {m_\pi^4 \over 18} S(q^2; m_\pi^2; m_\eta^2) 
\nonumber \\
& & \phantom{xxxxx} + {1 \over 9} R(q^2; m_\pi^2; m_\pi^2; m_\pi^2) + 
{1 \over 72} R(q^2; m_\pi^2; m_K^2; 2 (m_\pi^2 + m_K^2)) 
\nonumber \\
& & \phantom{xxxxx} + {1 \over 24} R(q^2; m_\eta^2; m_K^2; 
{2\over 3}(m_\pi^2 - m_K^2)) + {1\over 4} U(q^2; m_\pi^2; m_K^2) \ \ ,
\label{se6}
\eeqa
where the quantities $S$, $R$ and $U$ 
are yet new sunset functions defined by 
Eqs.~(\ref{se6aa}),(\ref{se7}),(\ref{se7a}) of App.~A.  
Both they and the non-sunset 
self-energy contributions share the degree of singularity 
of the 1PI and vertex amplitudes ({\it cf} Eq.~(\ref{4c})).  

\subsection{\bf ${\cal O}(p^6)$ Counterterms}
To construct the counterterm amplitudes needed to subtract off 
divergences and scale dependence contained in the two-loop graphs, 
we refer to the lagrangian of ${\cal O}(p^6)$ counterterms 
of Fearing and Scherer~{\cite{fs}}.  From their compilation, 
we extract $23$ counterterm operators which contribute to the 
axialvector propagators.  

The ${\cal O}(p^6)$ counterterm amplitudes are 
computed in like manner to the two-loop contributions 
considered thus far in this section.  For example, 
the isospin 1PI two-loop amplitudes 
will require a corresponding counterterm contribution 
${\cal M}_{\mu\nu 3}^{({\rm ct})}$ as computed from 
\beq
\langle a_3 (q',\lambda')|{\cal S}^{({\rm ct})} -1 
|a_3 (q,\lambda)\rangle = i (2\pi)^4 \delta^{(4)} (q' - q) 
~\epsilon^{\mu^\dagger} (q',\lambda') ~
{\cal M}_{\mu\nu 3}^{({\rm ct})}
~\epsilon^\nu (q,\lambda) \ , 
\label{aa}
\eeq
and one obtains 
\beqa
& & {\cal M}_{\mu\nu 3}^{({\rm ct})} = 
(q_\mu q_\nu - q^2 g_{\mu\nu}) \bigg[ 4 (m_\pi^2 + 2m_K^2 ) 
(B_{29} - 2B_{49}) \nonumber \\
& & \phantom{xxxxx} - 4 q^2 (B_{33} - 2B_{32}) + 
8 m_\pi^2 (B_{28} - B_{46}) \bigg] 
\label{ctaa3} \\
& & \phantom{xxxxx} + g_{\mu\nu} \bigg[ -4 (m_\pi^2 + 2m_K^2 )^2 B_{21} 
-4(3 m_\pi^4 - 4 m_\pi^2 m_K^2 + 4 m_K^4) B_{19} \nonumber \\
& & \phantom{xxxxx} - 4 m_\pi^2 (m_\pi^2 + 2m_K^2 )( B_{16} + B_{18}) 
+ 4m_\pi^4 (2B_{14} - B_{11} - B_{17} ) \bigg] \ . \nonumber
\eeqa
This has three independent ${\cal O}(p^6)$ counterterms 
for the `transverse' amplitude proportional to $q_\mu q_\nu - q^2
g_{\mu\nu}$ and five for the `longitudinal' amplitude proportional 
to $g_{\mu\nu}$.  

Employing a relation analogous to Eq.~(\ref{v0}) for 
the vertex counterterm amplitude, one finds for the isospin case, 
\beqa
& & \Gamma_{\mu 3}^{({\rm ct})} = 
i~{q_\mu\over F_0} \bigg[ 4 m_\pi^4 (B_{14} - B_{17}) 
-4 \left( 4m_K^4 - 4m_K^2 m_\pi^2 + 3 m_\pi^4 \right) B_{19} 
\nonumber \\
& & \phantom{xxxxx} - (m_\pi^2 + 2 m_K^2) 
\left( 2 m_\pi^2 (B_{16} + 2B_{18}) + 4(m_\pi^2 + 2 m_K^2) 
B_{21}) \right) \bigg] \ \ ,
\label{ctam3} 
\eeqa
and a relation analogous to Eq.~(\ref{se1}) leads to 
the isospin counterterm self-energy 
\beqa
& & \Sigma^{({\rm ct})}_3 = 
-{2\over F_0^2} \bigg[ m_\pi^6 \left( 3B_{1} + 2B_{2} \right) 
+ m_\pi^2 (5 m_\pi^4 + 4 m_K^4) B_3 
\nonumber \\
& & \phantom{xxxxx} + (m_\pi^2 + 2 m_K^2) \left( 2 m_\pi^4 B_4 
+ 3 m_\pi^2 (m_\pi^2 + 2 m_K^2) B_6 \right) \bigg] 
\nonumber \\
& & \phantom{xxxxx} - 4 {q^2\over F_0^2} \bigg[ m_\pi^4 B_{17} 
+ (4m_K^4 - 4m_K^2 m_\pi^2 + 3 m_\pi^4 ) B_{19} 
\nonumber \\ 
& & \phantom{xxxxx} + (m_\pi^2 + 2 m_K^2) \left(  m_\pi^2 B_{18} 
+ (m_\pi^2 + 2 m_K^2) B_{21}\right) \bigg] \ .
\label{ctmm3} 
\eeqa

Finally, we express the $\{B_\ell\}$ in dimensional regularization 
as\footnote{The dependence of 
$L_\ell^{(n)}(\mu)$ and $B_\ell^{(n)}(\mu)$  upon the scale 
$\mu$ is determined from the renormalization group equations 
and has been explicitly given in Ref.~\cite{gk1}.}     
\begin{equation}
B_\ell = \mu^{2(d - 4)} \sum_{n=2}^{-\infty} \ B_\ell^{(n)}(\mu) 
~{\overline\lambda}^n 
= \mu^{2(d - 4)} \left[~ B_\ell^{(2)}(\mu) ~{\overline\lambda}^2 
\ + \ B_\ell^{(1)}(\mu) ~{\overline\lambda} \ + \ B_\ell^{(0)}(\mu) \ 
+ \dots ~\right] \ .
\label{aa14a}
\end{equation}
This representation, together with Eq.~(\ref{a3}) for the $A$-integral 
and Eq.~(\ref{aa14}) for the ${\cal O}(p^4)$ counterterms 
$\{ L_\ell \}$, expresses the axialvector propagator 
as an expansion in the singular quantity ${\overline\lambda}$.  
\section{\bf Renormalization Procedure}
The axialvector propagator $\Delta^{\mu\nu}_{{\rm A}a}$ 
will have contributions from both the $1PI$ part 
${\cal M}^{\mu\nu}_a$ and the 1PR pole term, 
\begin{equation}
\Delta^{\mu\nu}_{{\rm A}a} (q) \ = \ {\cal M}^{\mu\nu}_a (q) - 
q^\mu q^\nu {\Gamma^2_a (q^2) \over q^2 - m_a^2 + 
\Sigma_a (q^2)} \qquad (a = 3,8) \ \ .
\label{r1}
\end{equation}
In the above, the ${\cal O}(p^6)$ counterterms are understood to 
be already included in ${\cal M}^{\mu\nu}_a$, in the 
self energy $\Sigma_a$ and also in $\Gamma_a (q^2)$.  The latter 
is defined in terms of the vertex amplitude $\Gamma^\mu_a (q)$ as 
\begin{equation}
\Gamma^\mu_a (q) \equiv i q^\mu \Gamma_a (q^2) \ \ .
\label{r2}
\end{equation}
The renormalized mass $M_a$ and decay constant $F_a$ are defined 
as parameters occurring in the meson pole term 
\begin{equation}
q^\mu q^\nu {\Gamma^2_a (q^2) \over q^2 - m_a^2 +
\Sigma_a (q^2)}  \equiv  q^\mu q^\nu \left(
{F_a^2 \over q^2 - M_a^2} + R_a (q^2) \right) \ \ ,
\label{r3}
\end{equation}
where $R_a (q^2)$ is a remainder term having no poles.  

\subsection{\bf Identification of the Meson Mass}
From Eqs.~(\ref{r1}),(\ref{r3}), it follows that 
$M_a^2$ is a solution of the implicit relation 
\begin{equation}
M_a^2 = m_a^2 - \Sigma_a (M_a^2) \ \ .
\label{rm0}
\end{equation}
Since we have already calculated $\Sigma (m^2)$ (in the following, 
we temporarily omit flavour indices), 
it makes sense to expand the self-energy $\Sigma (M^2)$ as
\begin{equation}
\Sigma (M^2) = \Sigma (m^2) + \Sigma' (m^2) (M^2 - m^2) 
+ \dots \ \ .
\label{rm1}
\end{equation}
Then, expressing the squared-mass perturbatively, 
\begin{equation}
M^2 = M^{(2)2} + M^{(4)2} + M^{(6)2} + \dots \ \ ,
\label{rm2}
\end{equation}
and similarly for the self-energy, we obtain the 
perturbative chain 
\begin{eqnarray}
M^{(2)2} &=& m^2 \ , \label{rm3a} \\ 
M^{(4)2} &=& - \Sigma^{(4)} (m^2)  \ , \label{rm3b} \\ 
M^{(6)2} &=& - \Sigma^{(6)} (m^2)  
+ \Sigma^{(4)'}(m^2)~\Sigma^{(4)}(m^2) \ , \label{rm3c}
\end{eqnarray}
where we have noted that 
\begin{equation}
M^2 - m^2 = {\cal O}(q^4) = - \Sigma^{(4)}(m^2)  + \dots \ \ .
\label{rm4}
\end{equation}

We briefly exhibit this procedure at one-loop level.  
Using the fourth-order isospin self-energy 
\begin{eqnarray}
& & \Sigma_3^{(4)} (q^2) = {i\over F_0^2} \left[ 
 {m_\pi^2 - 4 q^2 \over 6}A(\pi) + {m_\pi^2 - q^2 \over 3}A(K) 
+ {m_\pi^2 \over 6}A(\eta) \right]
\nonumber \\
& & + {8 \over F_0^2} \left[ (m_\pi^2 + 2 m_K^2)( q^2 L_4 
- 2 m_\pi^2 L_6 ) + m_\pi^2 (q^2 L_5 - 2 m_\pi^2 L_8) \right] \ ,
\label{rm5}
\end{eqnarray}
we obtain from Eq.~(\ref{rm3b}), 
\begin{equation}
M_\pi^{(4)2} = {i\over F_0^2} \left[ 
 {3 A(\pi) - A(\eta) \over 6} \right]
- { 8 m_\pi^2 \over F_0^2} \left[ (m_\pi^2 + 2 m_K^2)( L_4 
- 2 L_6 ) + m_\pi^2 (L_5 - 2 L_8) \right] \ .
\label{rm6}
\end{equation}
Expanding the $A$-integrals in a 
Laurent series around $d=4$, we readily obtain the result 
cited earlier in Eq.~(\ref{fm2}).  
In like manner, the one-loop hypercharge self-energy
\begin{eqnarray}
& & \Sigma_8^{(4)} (q^2) = {i\over F_0^2} \left[ 
 {m_\pi^2 \over 2}A(\pi) + {16 m_K^2 - 7 m_\pi^2 \over 18}A(\eta) 
- (q^2 + {m_\pi^2 \over 3}) A(K) \right]
\nonumber \\
& & + {8 \over F_0^2} \bigg[ (m_\pi^2 + 2 m_K^2)( q^2 L_4 
- 2 m_\eta^2 L_6 ) + m_\eta^2 q^2 L_5 
\nonumber \\
& & - {16\over 3} (m_K^2 - m_\pi^2)^2 L_7 - {2\over 3} 
(8 m_K^4 - 8 m_K^2 m_\pi^2 + 3 m_\pi^4 ) L_8 \bigg] \ ,
\label{rm7}
\end{eqnarray}
yields 
\begin{eqnarray}
& & M_\eta^{(4)2} = {i\over F_0^2} \left[ 
- {m_\pi^2 \over 2} A(\pi) - {16 m_K^2 - 7 m_\pi^2 \over 18}A(\eta) 
+ {4 m_K^2\over 3}A(K) \right]
\nonumber \\
& & - {8 \over F_0^2} \bigg[ m_\eta^2 (m_\pi^2 + 2 m_K^2)( L_4 
- 2 L_6 ) + m_\eta^4 L_5 
\nonumber \\
& & - {16\over 3} (m_K^2 - m_\pi^2)^2 L_7 - {2\over 3} 
(8 m_K^4 - 8 m_K^2 m_\pi^2 + 3 m_\pi^4 ) L_8 \bigg] \ .
\label{rm8}
\end{eqnarray}
These agree, of course, in the $SU(3)$ limit of equal quark mass.  
The analysis at two-loop level proceeds analogously. 
\subsection{\bf Identification of the Meson Decay Constant}
With the aid of Eq.~(\ref{rm0}), it is straightforward to 
write the meson pole term in the form (again temporarily 
suspending flavour indices) 
\begin{eqnarray}
{\Gamma^2 (q^2) \over q^2 - m^2 - \Sigma (q^2)} &=& 
{\Gamma^2 (q^2) \over q^2 - M^2}\cdot 
{1 \over 1 + {\tilde \Sigma}(q^2)}  \nonumber \\
&=& {\Gamma^2 (q^2)\left( 1 - {\tilde \Sigma}(q^2) + 
{\tilde \Sigma}^2 (q^2) + \dots \right)
 \over q^2 - M^2} \ \ , \label{rv1}
\end{eqnarray}
where ${\tilde \Sigma}(q^2)$ is the divided difference 
\begin{equation}
{\tilde \Sigma}(q^2) \equiv {\Sigma(q^2) - \Sigma(M^2) 
\over q^2 - M^2} \ \ .
\label{rv2}
\end{equation}
From the definition in Eq.~(\ref{r3}) of the squared decay
constant $F^2$, we have 
\begin{equation}
F^2 = \lim\limits_{q^2 = M^2} \Bigg[ \Gamma^2 (q^2) 
\left( 1 - {\tilde \Sigma}(q^2) + 
({\tilde \Sigma}(q^2))^2  + \dots \right) \bigg] \ .
\label{rv4}
\end{equation}
Let us analyze this relation perturbatively.

We begin by writng the vertex quantity $\Gamma (q^2)$ 
(evaluated at $q^2 = M^2$) as 
\begin{equation}
\Gamma (M^2) = \Gamma^{(0)} + \Gamma^{(2)} + \Gamma^{(4)}(M^2) 
+ \dots \ \ , 
\label{rv5}
\end{equation}
where both $\Gamma^{(0)}$ and $\Gamma^{(2)}$
are independent of $q^2$.  Expanding $\Gamma^{(4)}(M^2)$ as  
\begin{equation}
\Gamma^{(4)}(M^2) = \Gamma^{(4)}(m^2) + \Gamma^{(4)'}(m^2) (M^2 - m^2)
+ \dots \ \ ,
\label{rv6}
\end{equation}
we see from chiral counting that to the order at which we are 
working, one is justified in replacing 
$\Gamma^{(4)}(M^2)$ by $\Gamma^{(4)}(m^2)$.   As for the 
self energy dependence in Eq.~(\ref{rv4}), we first observe that 
\begin{equation}
{\tilde \Sigma}(M^2) = \lim\limits_{q^2 = M^2} {\tilde \Sigma}(q^2) 
= \Sigma '(M^2) \ \ .
\label{rv3}
\end{equation}
Recalling from Eqs.~(\ref{rm5}),(\ref{rm7}) that $\Sigma^{(4)}(q^2)$ 
is linear in $q^2$, we have the perturbative expression 
\begin{equation}
\Sigma '(M^2) = \Sigma^{(4)'} + \Sigma^{(6)'}(M^2) + \dots 
= \Sigma^{(4)'} + \Sigma^{(6)'}(m^2) + \dots \ \ ,
\label{rv7}
\end{equation}
where the error in replacing $\Sigma^{(6)'}(M^2)$ by 
$\Sigma^{(6)'}(m^2)$ appears in higher order.  Thus, the perturbative 
content of Eq.~(\ref{rv4}) reduces to 
\begin{eqnarray}
F^2 &=& \left( \Gamma^{(0)} + \Gamma^{(2)} + \Gamma^{(4)}(m^2)\right)^2 
\nonumber \\
& & \times \left( 1 - \Sigma^{(4)'} - \Sigma^{(6)'}(m^2) 
+ (\Sigma^{(4)'})^2  \right) + \dots \ .
\label{rv8}
\end{eqnarray}
Upon organizing terms in ascending chiral powers, 
we obtain the following expression, valid through two-loop order, 
for the decay constant
\begin{eqnarray}
F &=& \Gamma^{(0)} \ + \ \left[ \Gamma^{(2)} 
- {1\over 2}\Gamma^{(0)}\Sigma^{(4)'} \right] \ 
+ \ \bigg[ \Gamma^{(4)}(m^2) -  {1\over 2}\Gamma^{(2)}\Sigma^{(4)'} 
\nonumber \\
&+& \Gamma^{(0)} \left( -{1\over 2} \Sigma^{(6)'}(m^2) + 
{3\over 8} (\Sigma^{(4)'})^2 \right) \bigg] + \dots 
\ \ , \label{rv9}
\end{eqnarray}
where we have collected together terms of a given order.  

As an example, Eq.~(\ref{rv9}) readily provides a 
determination of the isospin and hypercharge decay constants 
through one-loop order.  The corresponding vertex quantities are 
\begin{eqnarray}
\Gamma_\pi &=& F_0 - {2i\over 3F_0}\left[ 2 A(\pi) + A(K) \right] 
+ {8\over F_0}\left[ (m_\pi^2 + 2 m_K^2)L_4 + m_\pi^2 L_5 \right] +
\dots \ ,
\nonumber \\
\Gamma_\eta &=& F_0 - {2i\over F_0} A(K) 
+ {8\over F_0}\left[ (m_\pi^2 + 2 m_K^2)L_4 + m_\eta^2 L_5 \right] 
+ \dots \ .
\label{rv10}
\end{eqnarray}
Upon inferring $\Sigma_k^{(4)'}\ (k = 3,8)$ from 
Eqs.~(\ref{rm5}),(\ref{rm7}), we obtain  
\begin{eqnarray}
& & F_\pi^{(0)} + F_\pi^{(2)} = F_0 - {i\over 2F_0}\left[ 
2 A(\pi) + A(K) \right] 
\nonumber \\
& & \phantom{xxxxxxx} + {4\over F_0}\left[ 
(m_\pi^2 + 2 m_K^2)L_4 + m_\pi^2 L_5 \right] \ , \label{rv11} \\
& & F_\eta^{(0)} + F_\eta^{(2)} = F_0 - {3i\over 2F_0} A(K) 
+ {4\over F_0}\left[ (m_\pi^2 + 2 m_K^2)L_4 + m_\eta^2 L_5 \right] 
\ . \nonumber 
\end{eqnarray}
The one-loop expressions for the pion decay constant (appearing 
in Eq.~(\ref{fm2})) and the eta decay constant (not shown)
follow from the above relation.  The two-loop corrections 
are found analogously.  
\subsection{\bf The Remainder Term}
The preceding work allows extraction of the remainder term $R(q^2)$ 
defined earlier in Eq.~(\ref{r3}).  Making use of 
Eqs.~(\ref{rm2})--(\ref{rm3c}) as well as Eq.~(\ref{rv9}), 
a straightforward calculation yields the expression 
\begin{eqnarray}
\lefteqn{R(q^2) = 2 F_0 {\Gamma^{(4)}(q^2) - \Gamma^{(4)}(m^2) 
\over q^2 - m^2}} \nonumber \\
& & - F_0^2~ {\Sigma^{(6)}(q^2) - \Sigma^{(6)}(m^2) - 
(q^2 - m^2) \Sigma^{(6)'}(m^2) \over (q^2 - m^2)^2} \ .
\label{rmdr1}
\end{eqnarray}
It is manifest that $R(q^2)$ has no pole at $q^2 = m^2$.  Moreover, 
since the non-sunset vertex functions $\Gamma^{(4)}[a]$--$[d]$ of 
Sect.~3 are constant in $q^2$, they do not contribute to $R(q^2)$. 
Nor do the non-sunset self-energies $\Sigma^{(6)}[a]$--$[d]$ 
since they are at most linear in $q^2$.  Thus, the only contributors 
to $R(q^2)$ are sunset amplitudes, and it 
is straightforward to obtain the isospin and hypercharge 
remainder functions directly from Eq.~(\ref{rmdr1}).  
\subsection{\bf The Polarization Amplitudes 
${\hat \Pi}^{(1)}$ and ${\hat \Pi}^{(0)}$}
From the tree-level and one-loop results, we anticipate that 
the two-loop $1PI$ amplitude ${\cal M}^{(4)}_{\mu\nu}$ will contain an 
additive term involving the meson decay constant, and can thus 
be written 
\begin{equation}
{\cal M}^{(4)}_{\mu\nu} \equiv g_{\mu\nu} (F^2)^{(4)} + 
{\hat {\cal M}}_{\mu\nu} \ \ ,
\label{1PI3}
\end{equation} 
where ${\hat {\cal M}}_{\mu\nu}$ denotes the residual part 
of the $1PI$ amplitude.  It turns out that much of the 
rather complicated content in the $1PI$ amplitudes of Sect.~3 
is attributable to the two-loop squared decay constant 
$(F^2)^{(4)}$, as can be verified from decay 
constant results derived earlier in this section.  First, we 
re-express the expansion of Eq.~(\ref{rv8}) as 
\begin{equation}
F^2 = (F^2)^{(0)} + (F^2)^{(2)} + (F^2)^{(4)} + \dots \ \ ,
\label{1PI1}
\end{equation}
with $(F^2)^{(0)} = F_0^2$, $(F^2)^{(2)} = 2F_0 F^{(2)}$ and  
\begin{equation}
(F^2)^{(4)} = (F^{(2)})^2 + 2 F_0 F^{(4)} \ \ .
\label{1PI2}
\end{equation}
One then compares the $g_{\mu\nu}$ part of 
${\cal M}^{(4)}_{\mu\nu}$ with $(F^2)^{(4)}$.  
The residual amplitude ${\hat {\cal M}}_{\mu\nu}$ is 
simply the difference of these.  

It is convenient to replace the 
residual amplitude ${\hat {\cal M}}^{\mu\nu}$ and 
the remainder term $- q^\mu q^\nu 
R (q^2)$ (which arises from the meson pole term but itself 
contains no poles) by equivalent quantities ${\hat \Pi}^{(1)}$ 
and ${\hat \Pi}^{(0)}$, 
\begin{equation}
{\hat {\cal M}}^{\mu\nu}(q)  - q^\mu q^\nu R (q^2) \equiv 
 (q^\mu q^\nu - q^2 g^{\mu\nu}) {\hat \Pi}^{(1)} (q^2) 
+ g^{\mu \nu} {\hat \Pi}^{(0)}(q^2) \ \ .
\label{sb5}
\end{equation}
We shall employ ${\hat \Pi}^{(1)}$ and ${\hat \Pi}^{(0)}$
throughout the rest of the paper.  
\section{\bf Removal of Divergences}
Thus far, we derived lengthy expressions 
for the various two-loop components of the axialvector propagators 
and then determined the renormalization structure of the 
masses and decay constants.  At this stage of the calculation, 
there are many terms which diverge as $d \to 4$ and which 
must therefore be removed from the description.  Below, we carry 
out the subtraction procedure by expanding the relevant quantities 
in powers of the parameter ${\bar \lambda}$ and then using 
${\cal O}(p^6)$ and ${\cal O}(p^4)$ counterterms to cancel 
the singular contributions.  In particular, this process will 
determine a subset of the so-called $\beta$-functions of the complete 
${\cal O}(p^6)$ lagrangian.  This is of special interest because 
the divergence structure of the generating functional to two-loop 
level can be obtained in closed form~\cite{ecker}, and our results 
derived in the following can be used as checks of such future calculations.  
\subsection{\bf Removal of ${\bar \lambda}^2$ Dependence}
It will simplify the following discussion 
to define the counterterm combinations
\begin{eqnarray}
A &\equiv& 2 B_{14} - B_{17}\ , \qquad B \ \equiv \ B_{16} + B_{18} \ ,
\qquad C \ \equiv \ B_{15} - B_{20} \ ,
\nonumber \\
D &=& B_{19} + B_{21} \ ,\qquad \ E \ \equiv \  B_{19} - B_{21} \ ,
\qquad \ F \ \equiv \ 3 B_{1} + 2 B_{2} \ \ .
\label{sb8} 
\end{eqnarray} 

First we consider the decay constant sector.  
Upon demanding that all ${\bar \lambda}^2$ dependence 
vanish, we obtain six equations containing five variables.  
There are {\it six} equations because the pion and eta constants 
each have explicit dependence on $m_\pi^4$, $m_\pi^2 m_K^2$ and 
$m_K^4$ factors and the singular behaviour must be subtracted 
for each of these.  

We find the equation set to be degenerate, and 
one obtains just the following four conditions, 
\begin{eqnarray}
A^{(2)} - 3 E^{(2)} &=& - {31\over 24 F_0^2} \ ,\qquad 
B^{(2)} - 2 E^{(2)} \ = \  - {53\over 72 F_0^2} \ ,
\nonumber \\
C^{(2)} + E^{(2)} &=& {13 \over 18 F_0^2} \ ,\qquad \qquad
D^{(2)} \ = \  {73\over 144 F_0^2} \ \ . 
\label{sb9} 
\end{eqnarray}
The information contained in the 
above set is unique and any other way of expressing the solution 
must be equivalent.  

In the mass sector, we find 
subtraction at the  ${\bar \lambda}^2$ level to yield {\it seven}
equations in eleven variables.  The number of equations follows 
from the dependence of each mass on $m_\pi^6$, $m_\pi^4 m_K^2$ 
and $m_\pi^2 m_K^4$ factors (this implies six equations), along 
with the fact that $m_K^6$ dependence is absent from 
$M_\pi^{(6)2}$.  This latter fact arises becauses there is 
no $m_K^6$ counterterm contribution in $M_\pi^{(6)2}$, and 
thus there must be a cancellation between sunset and nonsunset 
numerical terms.  Such a cancellation occurs and constitutes 
an important check on our determination of the sunset contribution.  
The equation set for the $M^{(6)2}$ masses is found to be degenerate 
and just five constraints can be obtained, {\it e.g.} 
\begin{eqnarray}
B_3^{(2)} &=& {4\over 27 F_0^2} - {1\over 6} F^{(2)} - B_4^{(2)} 
- B_5^{(2)} - 3 B_7^{(2)} \ ,
\nonumber \\
B_6^{(2)} &=& - {16\over 81 F_0^2} + {1\over 18} F^{(2)} 
+ {1\over 3} B_4^{(2)} + {1\over 3} B_5^{(2)} + B_7^{(2)} \ ,
\nonumber \\
B_{14}^{(2)} &=& {1\over 48} \left[ - {37 \over F_0^2} 
+ 72 B_4^{(2)} + 72 B_5^{(2)} + 216 B_7^{(2)} \right] \ ,
\nonumber \\
B_{15}^{(2)} &=& {307\over 216 F_0^2} - {1\over 3} F^{(2)} - 
B_4^{(2)} - 2 B_5^{(2)} - 3 B_7^{(2)} \ ,
\nonumber \\
B_{16}^{(2)} &=& - {91\over 72 F_0^2} + {1\over 6} F^{(2)} + 
2 B_4^{(2)} + B_5^{(2)} + 3 B_7^{(2)} \ \ .
\label{sb10} 
\end{eqnarray}

Finally, removal of ${\bar \lambda}^2$-dependence 
for the polarization functions ${\hat \Pi}^{(0)}$ and ${\hat \Pi}^{(1)}$ 
yields a final set of constraints at this order,
\begin{eqnarray}
B_{11}^{(2)} = B_{13}^{(2)} &=& 0 \ ,\phantom{xxxxxxxxxxxxx}~
B_{33}^{(2)} \ = \ \ 2 B_{32}^{(2)} \ ,
\nonumber \\
B_{28}^{(2)} - B_{46}^{(2)} &=& - {3 \over 16 F_0^2} \ ,\qquad 
B_{29}^{(2)} - 2 B_{49}^{(2)} = - {1 \over 8 F_0^2} \ \ .
\label{sb11} 
\end{eqnarray}
\subsection{\bf Removal of ${\bar \lambda}$ Dependence}
In a similar manner, we obtain constraints for the $\{ B^{(1)}\}$ 
counterterms.  We list these below without further comment, 
beginning with those following from decay constants, 
\begin{eqnarray}
A^{(1)} - 3 E^{(1)} &=& {1\over F_0^2}\left[ - {175\over 9216\pi^2} 
- {28 \over 3}L_1^{(0)} - {34 \over 3}L_2^{(0)} 
- {25 \over 3}L_3^{(0)} \right.
\nonumber \\ 
& & \left. \phantom{x} + {26 \over 3} L_4^{(0)} - 
{8 \over 3} L_5^{(0)} - 12 L_6^{(0)} + 12 L_8^{(0)} \right] \ ,
\nonumber \\ 
B^{(1)} - 2 E^{(1)} &=& {1\over F_0^2} \left[ 
{19\over 1536 \pi^2} - {32 \over 9}L_1^{(0)} - {8 \over 9}L_2^{(0)} 
- {8 \over 9}L_3^{(0)} \right.
\nonumber \\ 
& & \left. \phantom{x} + {106 \over 9} L_4^{(0)} - 
{22 \over 9} L_5^{(0)} - 20 L_6^{(0)} \right] \ ,
\nonumber \\ 
C^{(1)} + E^{(1)} &=& {1\over F_0^2} \left[ {691\over 82944 \pi^2} 
+ {28 \over 9}L_1^{(0)} + {34 \over 9}L_2^{(0)} 
+ {59 \over 18}L_3^{(0)} \right.
\nonumber \\ 
& & \left. \phantom{x} - {26 \over 9} L_4^{(0)} + 3 
L_5^{(0)} + 4 L_6^{(0)} - 6 L_8^{(0)} \right] \ ,
\nonumber \\ 
D^{(1)} &=& {1\over F_0^2} \left[ {43 \over 3072 \pi^2 } 
+ {104 \over 9}L_1^{(0)} + {26 \over 9}L_2^{(0)} 
+ {61 \over 18}L_3^{(0)} \right. 
\nonumber \\ 
& & \left. \phantom{x} - {34 \over 9} L_4^{(0)} + 
L_5^{(0)} - 4 L_6^{(0)} - 2 L_8^{(0)} \right] \ \ , 
\label{sb12} 
\end{eqnarray}
then masses, 
\begin{eqnarray}
B_3^{(1)} &=& {1\over F_0^2} \left[ {5\over 216 \pi^2} 
- {20 \over 3}L_4^{(0)} - {23 \over 3}L_5^{(0)} 
+ {40 \over 3}L_6^{(0)} \right. 
\nonumber \\ 
& & \left. \phantom{x} + 40 L_7^{(0)} + {86 \over 3} L_8^{(0)}\right] 
- {1\over 6} F^{(1)} - B_4^{(1)} - B_5^{(1)} - 3 B_7^{(1)} \ ,
\nonumber \\
B_6^{(1)} &=& {1\over 648} \left[ {1\over F_0^2} \bigg( - {5\over \pi^2} 
- 3168 L_4^{(0)} - 24 L_5^{(0)} + 6336 L_6^{(0)} - 9792 L_7^{(0)} \right.
\nonumber \\ 
& & \left. \phantom{x} - 3216 L_8^{(0)} \bigg)
+ 36 F^{(1)} + 216 B_4^{(1)} + 216 B_5^{(1)} + 648 B_7^{(1)} \right] \ ,
\nonumber \\
B_{14}^{(1)} &=& {1\over F_0^2} \left[ - {167\over 4608 \pi^2 } 
+ 8 L_6^{(0)} - 64 L_7^{(0)} - {62 \over 3} L_8^{(0)} \right]
\nonumber \\
& & \phantom{x} + {3\over 2} B_4^{(1)} + {3\over 2} B_5^{(1)} 
+ {9\over 2} B_7^{(1)} \ ,
\nonumber \\
B_{15}^{(1)} &=& {1\over F_0^2} \left[ {371\over 20736 \pi^2} 
+ 2 L_5^{(0)} - {16 \over 3}L_6^{(0)} 
+ 72 L_7^{(0)} + 24 L_8^{(0)} \right]
\nonumber \\
& & \phantom{x} - {1\over 3} F^{(1)} - B_4^{(1)} 
- 2 B_5^{(1)} - 3 B_7^{(1)} \ ,
\nonumber \\
B_{16}^{(1)} &=& {1\over F_0^2} \left[ - {9\over 512 \pi^2} 
- L_5^{(0)} - {152 \over 3} L_7^{(0)} - {62 \over 3} L_8^{(0)} \right]
\nonumber \\
& & \phantom{x} + {1\over 6} F^{(1)} + 2 B_4^{(1)} + B_5^{(1)} 
+ 3 B_7^{(1)} \ \ , 
\label{sb13} 
\end{eqnarray}
and finally from the polarizations ${\hat \Pi}^{(0)}$ and ${\hat \Pi}^{(1)}$,  
\begin{eqnarray}
& & B_{11}^{(1)} = - {49 \over 576} {1\over 16 \pi^2 F_0^2} \ ,
\qquad B_{13}^{(1)} = {173  \over 5184} {1\over 16 \pi^2 F_0^2} \ ,
\qquad 2 B_{32}^{(1)} - B_{33}^{(1)} = 
{3 \over 64} {1\over 16 \pi^2 F_0^2} \ ,
\nonumber \\
& & B_{28}^{(1)} - B_{46}^{(1)} = {1\over F_0^2} \left[ - 
{5\over 64} {1\over 16 \pi^2} + {3 \over 2} L_{10}^{(0)} \right] \ ,
\qquad B_{29}^{(1)} - 2 B_{49}^{(1)} = {1\over F_0^2} \left[ 
- {17\over 96} {1\over 16 \pi^2}+ L_{10}^{(0)} \right] \ \ .
\label{sb14} 
\end{eqnarray}
This completes the subtraction part of the calculation.
\subsection{\bf ${\overline\lambda}$-Subtraction and 
${\overline {MS}}$ Renormalization Schemes}
The renormalization procedure employed originally in Ref.~\cite{gk1} 
and adopted in this paper amounts to ${\overline
\lambda}$-subtraction, {\it cf} Eqs.~(\ref{aa14}),(\ref{aa14a}).  
Alternatively, one could employ minimal subtraction (MS),
\beqa
L_\ell (d) &=& {\mu^{2 \omega} \over (4 \pi )^2} \left[ 
{\Gamma_\ell \over 2 \omega} + L_{\ell,r}^{\rm {MS}} (\mu ,\omega) 
+ \dots \right] \nonumber \\
B_\ell (d) &=& {\mu^{4 \omega} \over (4 \pi )^4} \left[ 
{B_\ell^{(2){\rm MS}} \over (2 \omega)^2} + 
{B_\ell^{(1){\rm MS}} \over 2 \omega} + 
B_{\ell,r}^{(0){\rm MS}} (\mu ,\omega) + \dots \right] \ ,
\label{ms1}
\eeqa 
where $\omega \equiv d/2 - 2$, or modified 
minimal subtraction (${\overline {\rm MS}}$), 
\beqa
L_\ell (d) &=& {(\mu c)^{2 \omega} \over (4 \pi )^2} \left[ 
{\Gamma_\ell \over 2 \omega} + L_{\ell,r}^{\overline{\rm {MS}}} (\mu ,\omega) 
+ \dots \right] \nonumber \\
B_\ell (d) &=& {(\mu c)^{4 \omega} \over (4 \pi )^4} \left[ 
{B_\ell^{(2){\rm MS}} \over (2 \omega)^2} + 
{B_\ell^{(1){\rm MS}} \over 2 \omega} + (4\pi)^4 
B_{\ell,r}^{(0){\overline{\rm MS}}} (\mu) + \dots \right] \ ,
\label{ms2}
\eeqa 
where we make the standard ChPT choice
\beq
\ln c = - {1 \over 2} \left[ 1 - \gamma_{\rm E} + \ln(4\pi) \right] 
\equiv - C \ \ .
\label{ms3}
\eeq
Of course, there must be only finite differences between these 
three procedures, amounting to additional finite renormalizations. 

As an illustration, we relate the $\{B_\ell^{(n)} \}$ and 
$\{L_\ell^{(n)} \}$ renormalization constants employed here 
to those defined in the ${\overline {\rm MS}}$ 
approach of Ref.~\cite{betal97}.  In the latter scheme, 
one writes further that 
\beq
L_{\ell,r}^{\overline{\rm {MS}}} (\mu ,\omega) = 
L_{\ell,r}^{\overline{\rm {MS}}} (\mu , 0) + 
L_{\ell,r}^{\overline{\rm {MS}}'} (\mu , 0) ~\omega  + \dots 
\label{ms4}
\eeq
and sets 
\beq
L_{\ell,r}^{\overline{\rm {MS}}'} (\mu , 0) = 0 \ \ .
\label{ms4a}
\eeq
The `convention' established by Eq.~(\ref{ms4a}) is 
allowed because it can be shown~\cite{betal97} 
that the effect of the quantity 
$L_{\ell,r}^{\overline{\rm {MS}}'}(\mu ,0)$ 
is to add a local contribution at order $p^6$, which can always be
abosrbed into the couplings of the ${\cal O}(p^6)$ lagrangian.   
Comparison of the two approaches yields 
\beqa
L_\ell^{(1)} &=& \Gamma_\ell \ , \nonumber \\
L_\ell^{(0)} &=& 
{1 \over (4 \pi)^2} L_{\ell,r}^{\overline{\rm {MS}}} (\mu , 0) 
\equiv L_\ell^r (\mu) \ ,
\label{ms5} \\
L_\ell^{(-1)} &=& {C \over (4 \pi)^4} \left[  
- L_{\ell,r}^{\overline{\rm {MS}}} (\mu , 0) + 
{ C \Gamma_\ell \over 2} \right] \ ,
\nonumber 
\eeqa
and 
\beqa
B_\ell^{(2)} &=& B_\ell^{(2){\rm MS}} \ ,\nonumber \\ 
B_\ell^{(1)} &=& {1 \over (4 \pi )^2} B_\ell^{(1){\rm MS}} \ , 
\label{ms6} \\ 
B_\ell^{(0)}(\mu) &=& 
B_{\ell,r}^{{\overline{\rm MS}}} (\mu) - 
{C  \over (4 \pi )^4 } B_\ell^{(1){MS}} + 
{C^2  \over (4 \pi )^4 } B_\ell^{(2){MS}} \ .
\nonumber 
\eeqa 
We stress that the content of Eqs.~(\ref{ms5}),(\ref{ms6}) 
is partly a reflection of the convention of Eq.~(\ref{ms4a}).  
Finally, one can combine the 
relations of Eq.~(\ref{ms6}) to write 
\beq
B_{\ell,r}^{{\overline{\rm MS}}} (\mu) = B_\ell^{(0)}(\mu) 
+ {C \over (4 \pi)^2} B_\ell^{(1)} - 
{C^2 \over (4 \pi)^4} B_\ell^{(2)}  \ \ .
\label{ms7}
\eeq

We shall return to the comparison between the ${\overline 
\lambda}$-subtraction and ${\overline {\rm MS}}$ 
renormalizations at the ends of Sect.~VI and of App.~B.
\section{\bf The Renormalized Isospin Polarization Functions}
Having performed the removal of ${\bar \lambda}^2$ and 
${\bar \lambda}^1$ singular dependence from the theory, we are 
left with the ${\bar \lambda}^0$ sector in which all quantities 
are finite.  We shall express such contributions entirely 
in terms of physical quantities by replacing the tree-level parameters 
$m_\pi^2, m_K^2, m_\eta^2, F_0$ with $M_\pi^2, M_K^2, M_\eta^2, F_\pi$. 
Any error thereby induced would appear in still higher orders.  

For any observable ${\cal O}$ ({\it e.g.} ${\cal O} = 
{\hat \Pi}_3^{(1)}$, $M_\pi$, $F_\pi$, {\it etc}) evaluated 
at two-loop level, there will be generally three kinds of 
finite contributions,  
\begin{equation}
{\cal O} = {\cal O}_{\rm rem} + {\cal O}_{\rm CT} 
+ {\cal O}_{\rm YZ} \ \ .  
\label{fin1}
\end{equation}
${\cal O}_{\rm rem}$ refers to the finite 
${\bar \lambda}^0$ `remnants' in $\{ A(k)A(\ell)\}$ or 
$\{ A(k)L_\ell\}$ contributions and 
arise from either the product of two ${\bar \lambda}^0$ factors or 
the product of ${\bar \lambda}^1$ and ${\bar \lambda}^{-1}$ 
factors.  ${\cal O}_{\rm CT}$ denotes any term containing 
the $\{ B_\ell^{(0)}\}$ $p^6$-counterterms, whereas 
${\cal O}_{\rm YZ}$ represents contributions from the 
finite $Y$, $Z$ integrals ({\it cf} Eqs.~(\ref{fn1}),(\ref{fn2}) 
of App.~A) which occur solely in sunset amplitudes.   

Before proceeding, we address a technical issue related 
to the presence of $L_\ell^{(-1)}$ terms appearing in the 
`remnant' amplitudes.  Such terms are always multiplied by 
polynomials in the quark masses and hence can be absorbed 
by the ${\cal O}(p^6)$ counterterms, as expected from general 
renormalization theorems~\cite{betal97}.   In the vector 
current analysis of Ref.~\cite{gk1}, we defined 
the following dimensionless quantities (now expressed 
in terms of the $\{ B_\ell \}$),  
\beqa
P &\equiv& 4 F_\pi^2 \left( 
-2 B_{30}^{(0)} + B_{31}^{(0)} \right) + 4 L_9^{(-1)} \ \ ,
\nonumber \\
Q &\equiv& 2 F_\pi^2 B_{47}^{(0)} - 3 \left( L_9^{(-1)} + 
L_{10}^{(-1)} \right) \ \ ,
\nonumber \\
R &\equiv& 2 F_\pi^2 B_{50}^{(0)} - \left( L_9^{(-1)} + 
L_{10}^{(-1)} \right) \ \ .
\label{ct}
\eeqa
and thus removed all explicit $L_\ell^{(-1)}$ dependence.  
We repeat that procedure here by defining axialvector quantities 
\beqa
P_{\rm A} &\equiv& 4 F_\pi^2 \left( 
-2 B_{32}^{(0)} + B_{33}^{(0)} \right) \ \ ,\nonumber \\
Q_{\rm A} &\equiv& 2 F_\pi^2 \left( 
- B_{28}^{(0)} + B_{46}^{(0)} \right) + 3 L_{10}^{(-1)} \ \ ,
\nonumber \\
R_{\rm A} &\equiv& F_\pi^2 \left( 
- B_{29}^{(0)} + 2 B_{49}^{(0)} \right) + L_{10}^{(-1)} \ \ ,
\label{ctA}
\eeqa
such that the counterterm dependence of $\Pi_{{\rm A}3,8}^{(1)}$ 
({\it cf} Eqs.~(\ref{fin2a1c}),(\ref{fin2b1c})) 
is identical to that established originally 
for $\Pi_{{\rm V}3,8}$~\cite{gk1}. 

Since our finite results are quite lengthy, 
we restrict the discussion in this section to just the {\it isospin} 
polarization functions.  Expressions for all the other 
observables (masses, decay constants and hypercharge polarization 
functions) are compiled in Appendix B.
\subsection{\bf The Spin-one Isospin Polarization Amplitude}
Turning now to the isospin transverse polarization amplitude, 
we have for the remnant piece, 
\begin{eqnarray}
& & F_\pi^2 {\hat \Pi}_{3,{\rm rem}}^{(1)} (q^2) = 
{M_\pi^2 \over \pi^4} \left[ {49 \over 13824} - {C \over 192} \right.
\nonumber \\ 
& & \left. + \log{M_\pi^2 \over \mu^2} \left( {1 \over 288} - 
{\pi^2 \over 2} L_{10}^{(0)} - {1 \over 768} \log{M_\pi^2 \over \mu^2}
- {1 \over 768} \log{M_K^2 \over \mu^2} \right) \right.
\nonumber \\ 
& & \left. + \log{M_K^2 \over \mu^2} \left( {1 \over 576} +  
{1 \over 1536} \log{M_K^2 \over \mu^2} \right) \right]
\nonumber \\ 
& & + {M_K^2 \over \pi^4} \left[ {5 \over 36864} - {17~C \over 3072} 
+ \log{M_K^2 \over \mu^2} \left( {17 \over 3072} - 
{\pi^2 \over 4} L_{10}^{(0)} - {1 \over 1024} \log{M_K^2 \over \mu^2} 
\right) \right] \nonumber \\ 
& & + {q^2 \over \pi^4} \left[ - {283 \over 294912} + {3~C \over 4096} 
- {1 \over 3072} \log{M_\pi^2 \over \mu^2} 
- {5 \over 12288} \log{M_K^2 \over \mu^2} \right] \ \ ,
\label{fin2a1a} 
\end{eqnarray}
where the constant $C$ has the same meaning as in Ref.~{\cite{gk1}}, 
\begin{equation}
C \equiv {1 \over 2} \left[ 1 - \gamma_{\rm E} + \ln(4\pi) \right] \ \ .
\end{equation}
Next is the counterterm contribution, 
\begin{equation}
{\hat \Pi}_{3,{\rm CT}}^{(1)} (q^2) = 
- {q^2 \over F_\pi^2} P_{\rm A} - {8 M_K^2 \over F_\pi^2} R_{\rm A} 
- {4 M_\pi^2 \over F_\pi^2} \left( Q_{\rm A} + R_{\rm A} \right) \ , 
\label{fin2a1c} 
\end{equation}
and finally upon defining 
\begin{equation}
{\cal H}^{qq} \equiv S - 6 {\overline S} + 9 S_1 \ \ ,
\label{fin2a9}
\end{equation}
we have the so-called YZ piece, 
\begin{eqnarray}
\lefteqn{F_\pi^2 {\hat \Pi}_{3,{\rm YZ}}^{(1)} (q^2) =
{4\over 9} {\cal H}^{qq}_{\rm YZ} (q^2, M_\pi^2, M_\pi^2) + 
{1\over 6} {\cal H}^{qq}_{\rm YZ} (q^2, M_\eta^2, M_\pi^2)} 
\nonumber \\
& & + {1\over 18} {\cal H}^{qq}_{\rm YZ} (q^2, M_\pi^2, M_K^2) + 
{1\over 3} {\cal K}_{1,{\rm YZ}} (q^2, M_\pi^2, M_K^2)  
- R_{3,{\rm YZ}} (q^2) \ \ .
\label{fin2a1b1}
\end{eqnarray}
The quantity $R_{3,{\rm YZ}}$ is rather complicated, so 
before writing it down explicitly we first develop some 
useful notation.  For a quantity $f(q^2,\dots)$, 
we define the auxiliary functions 
\begin{eqnarray}
{\overline f}(q^2,\dots) &\equiv& f(q^2,\dots) - f(M^2,\dots) \nonumber \\
{\breve f}(q^2,\dots) &\equiv& {\overline f}(q^2,\dots) - 
(q^2 - M^2) f'(M^2,\dots) \ \ , 
\label{aux}
\end{eqnarray}
where the `$M^2$' quantities become $M_\pi^2$ for the case 
of isospin flavour and $M_\eta^2$ for hypercharge flavour.
Then we have for $R_{3,{\rm YZ}}$ the expression 
\begin{eqnarray}
& & R_{3,{\rm YZ}} (q^2) \equiv 
{2 \over q^2 - M_\pi^2} 
\bigg[ {2\over 9} 
{\overline I}_{1,{\rm YZ}} (q^2 ; M_\pi^2; M_\pi^2 ; M_\pi^2) 
\nonumber \\ 
& & + {1\over 36} {\overline I}_{1,{\rm YZ}} (q^2 ; M_\pi^2; M_K^2 
;2(M_\pi^2 + M_K^2)) 
\nonumber \\
& & + {1\over 12} {\overline I}_{1,{\rm YZ}} (q^2 ; M_\eta^2; M_K^2 ;
{2(M_\pi^2 - M_K^2)\over 3}) 
+ {1\over 2} {\overline I}_{2,{\rm YZ}} (q^2 ; M_\pi^2; M_K^2) 
\bigg] 
\nonumber \\
& & - {1 \over (q^2 - M_\pi^2)^2}\bigg[ 
{M_\pi^4 \over 6} {\breve S}_{\rm YZ} (q^2 ; M_\pi^2 ; M_\pi^2) 
+ {M_\pi^4 \over 18} {\breve S}_{\rm YZ} (q^2 ; M_\pi^2 ; M_\eta^2) 
\nonumber \\
& & + {1\over 4} {\breve U}_{\rm YZ} (q^2 ; M_\pi^2 ; M_K^2) 
+ {1\over 9} {\breve R}_{\rm YZ} (q^2 ; M_\pi^2 ; M_\pi^2;
M_\pi^2)) \label{fin2a1b8} \\ 
& & + {1\over 72} {\breve R}_{\rm YZ} (q^2 ; M_\pi^2 ; M_K^2; 
2(M_\pi^2 + M_K^2)) 
\nonumber \\
& & + {1\over 24} {\breve R}_{\rm YZ} (q^2 ; M_\eta^2 ; M_K^2; {2(M_\pi^2 -
M_K^2)\over 3}) \bigg] \ .
\nonumber 
\end{eqnarray}
For example, we obtain at $q^2 = 0$ the numerical value 
\begin{equation}
F_\pi^2 {\hat \Pi}_{3,{\rm YZ}}^{(1)} (0) = 1.927 \times 
10^{-6}~{\rm GeV}^2 \ \ .
\label{val133}
\end{equation}
\subsection{\bf The Spin-zero Isospin Polarization Amplitude}
For the isospin polarization function ${\hat \Pi}_3^{(0)}$, we find 
for the remnant contribution, 
\begin{eqnarray}
& & F_\pi^2 {\hat \Pi}_{3,{\rm rem}}^{(0)} (q^2) = 
{M_\pi^4 \over \pi^4} \left[  - {361 \over 294912} + 
{49~C \over 36864} - {1\over 3072} \log{M_\pi^2 \over \mu^2} \right.
\nonumber \\ 
& & \left. - {11\over 12288} \log{M_K^2 \over \mu^2} 
- {1\over 9216} \log{M_\eta^2 \over \mu^2} \right] \ ,
\label{fin2a2a}
\end{eqnarray}
whereas the counterterm piece is given by 
\begin{equation}
{\hat \Pi}_{3,{\rm CT}}^{(0)} = - 4 M_\pi^4 B_{11}^{(0)} \ \ .
\label{fin2a2c}
\end{equation}
For the piece coming from the finite functions, we have 
\begin{eqnarray}
\lefteqn{F_\pi^2 {\hat \Pi}_{3,{\rm YZ}}^{(0)} (q^2) = 
4S_{2,{\rm YZ}} (q^2, M_\pi^2, M_\pi^2) + 
{3\over 2} S_{2,{\rm YZ}} (q^2, M_\eta^2, M_\pi^2)} 
\nonumber \\
& & + {1\over 2} S_{2,{\rm YZ}}(q^2, M_\pi^2, M_K^2) + 
{1\over 3} {\cal K}_{2,{\rm YZ}} (q^2, M_\pi^2, M_K^2)  
\nonumber \\
& & + q^2 \bigg( {4\over 9} {\cal H}^{qq}_{\rm YZ} (q^2, M_\pi^2, M_\pi^2) + 
{1\over 6} {\cal H}^{qq}_{\rm YZ} (q^2, M_\eta^2, M_\pi^2)
\nonumber \\
& & + {1\over 18} {\cal H}^{qq}_{\rm YZ} (q^2, M_\pi^2, M_K^2) + 
{1\over 3} {\cal K}_{1,{\rm YZ}} (q^2, M_\pi^2, M_K^2)  
- R_{3,{\rm YZ}} (q^2)\bigg) 
\nonumber \\
& & - 2 \bigg[ {2\over 9}I_{1,{\rm YZ}} 
\left(M_\pi^2;M_\pi^2;M_\pi^2;M_\pi^2\right)
+ {1\over 2}I_{2,{\rm YZ}} \left(M_\pi^2;M_\pi^2;M_K^2\right) 
\nonumber \\
& & + {1\over 36}I_{1,{\rm YZ}} \left(M_\pi^2;M_\pi^2;M_K^2;2(M_\pi^2 + 
M_K^2)\right) 
\label{1PI4a} \\
& & + {1\over 12}I_{1,{\rm YZ}} 
\left(M_\pi^2;M_\eta^2;M_K^2;{2\over 3}(M_\pi^2 - M_K^2)\right) 
\nonumber \\
& & - {M_\pi^4 \over 12} S'_{\rm YZ} (M_\pi^2 ; M_\pi^2 ; M_K^2 ) -
{M_\pi^4 \over 36} S'_{\rm YZ} (M_\pi^2 ; M_\pi^2 ; M_\eta^2 ) 
\nonumber \\
& & - {1\over 18} R'_{\rm YZ} (M_\pi^2 ; M_\pi^2 ; M_\pi^2 ; M_\pi^2 ) 
- {1\over 144} R'_{\rm YZ} (M_\pi^2 ; M_\pi^2 ; M_K^2 ; 2(M_\pi^2 + M_K^2)) 
\nonumber \\
& & - {1\over 48} R'_{\rm YZ} (M_\pi^2 ; M_\eta^2 ; M_K^2 ; 
{2\over 3}(M_\pi^2 - M_K^2)) - {1\over 8} 
U'_{\rm YZ} (M_\pi^2 ; M_\pi^2 ; M_K^2) \bigg] \ \ .
\label{fin2a2b} 
\end{eqnarray}
The numerical value of $q^2 = 0$ is found to be 
\begin{equation}
F_\pi^2 {\hat \Pi}_{3,{\rm YZ}}^{(0)} (0) = - 2.954 \times 
10^{-9}~{\rm GeV}^4 \ \ .
\label{val033}
\end{equation}

As mentioned earlier, the remaining finite results 
(hypercharge polarization amplitudes, {\it etc}) appear 
in Appendix B. 
\subsection{\bf Isospin Polarization Functions and 
${\overline {MS}}$ Renormalization} 
The discussion in Sect.~VI-C allows one to 
re-express the isospin polarization functions 
${\hat \Pi}_3^{(1,0)}$ in the 
${\overline {MS}}$ renormalization.  The result of this is, 
in essence, to replace the finite renormalization constants 
obtained in $\overline\lambda$-subtraction by the 
corresponding $\overline{\rm MS}$ quantities and at the same 
time to omit all terms from the remnant contributions 
containing the constant $C$. 

As an example, let us determine the relation 
between the renormalization constants $P_{\rm A}$ 
and $P_{\rm A}^{{\overline{\rm MS}}}$.  
Starting from Eq.~(\ref{ctA}) and applying the 
relations in Eq.~(\ref{ms6}), we find 
\beq
P_{\rm A}^{{\overline{\rm MS}}} = P_{\rm A} + 
{F_\pi^2 C \over \pi^2} \left( - 2 B_{32}^{(1)} + B_{33}^{(1)} 
\right) - {F_\pi^2 C^2 \over (2 \pi)^4} 
\left( - 2 B_{32}^{(2)} + B_{33}^{(2)} \right) \ \ .
\label{msb0}
\eeq 
Then from Eqs.~(\ref{sb11}),(\ref{sb14}), we obtain 
\beq
P_{\rm A}^{{\overline{\rm MS}}} 
= P_{\rm A} - {3 \over 16} {C \over (4 \pi)^4} \ \ .
\label{msb1}
\eeq
This result is entirely consistent with the 
form obtained earlier in this section for 
${\hat \Pi}_3^{(1)}$.  That is, from 
Eqs.~(\ref{fin2a1a}),(\ref{fin2a1c}) we have 
\beq
{\hat \Pi}_3^{(1)} (q^2) = 
{q^2 \over F_\pi^2} \left[ - P_{\rm A} + 
{3 C \over 4096 \pi^4} + \dots \right] \ \ .
\label{msb2}
\eeq
But this is just the combination of factors appearing in 
Eq.~(\ref{msb1}) and we conclude 
\beq
{\hat \Pi}_3^{(1){\overline{\rm MS}}} (q^2) = 
{q^2 \over F_\pi^2} \left[ - P_{\rm A}^{{\overline{\rm MS}}} + 
\dots \right] \ \ .
\label{msb3}
\eeq

Analogous steps lead to the further relations 
\beqa
Q_{\rm A}^{{\overline{\rm MS}}} 
&=& Q_{\rm A} + {5 \over 32} {C \over (4 \pi)^4}
\nonumber \\
R_{\rm A}^{{\overline{\rm MS}}} 
&=& R_{\rm A} + {17 \over 96} {C \over (4 \pi)^4}
\label{msb11} \\
B_{11,r}^{{\overline{\rm MS}}} 
&=& B_{11}^{(0)} - {49 \over 576} {C \over (4 \pi)^4 F_\pi^2}
\nonumber \\
B_{13,r}^{{\overline{\rm MS}}} 
&=& B_{13}^{(0)} + {173 \over 5184} {C \over (4 \pi)^4 F_\pi^2} \ \ .
\nonumber 
\eeqa
Again, the forms for ${\hat \Pi}_3^{(1,0)}$ obtained by us 
in $\overline\lambda$-subtraction are found to convert to 
${\overline {MS}}$ renormalization by simply removing all 
$C$ dependence from the polarization functions and employing
the ${\overline {MS}}$ finite counterterms.

Finally, we point out that for the renormalization constants 
$P$, $Q$, and $R$ ({\it cf} Eq.~(\ref{ct})) which appeared 
in our two-loop analysis of vector-current propagators~\cite{gk1}, 
there is no difference between the $\overline\lambda$-subtraction 
and ${\overline {MS}}$ schemes. This can be traced to the fact 
that the quantities $P^{(1)}$, $Q^{(1)}$, and $R^{(1)}$ have 
only contributions from $L_{9}^{(0)}$ and $L_{10}^{(0)}$. 
\section{\bf Spectral Functions}
As noted in Sect.~II, there will generally 
exist {\it two} spectral functions, $\rho_{{\rm A}a}^{(1)}(q^2)$ 
and $\rho_{{\rm A}a}^{(0)}(q^2)$ for the system of axialvector 
propagators.  In the following, we 
determine the $3\pi$ contribution to the spectral functions for the 
isospin case $a = 3$.  The ${K{\bar K}\pi}$ and 
$K{\bar K}\eta$ components will have thresholds 
at higher energies.  
\subsection{\bf Three-pion Contribution to Isospin Spectral Functions}

Spectral functions can be determined from the imaginary parts 
of polarization functions ({\it cf} Eqs.~(\ref{aa2}),(\ref{aa2a})), 
\beq
\rho_{{\rm A}a}^{(1)}(q^2) = {1\over \pi} ~{\cal I}m~\Pi^{(1)}_{a}(q^2) 
\ \qquad {\rm and} \ \qquad 
\rho_{{\rm A}a}^{(0)}(q^2) = {1\over \pi} ~{\cal I}m~\Pi^{(0)}_{a}(q^2) 
\ \ .
\label{im14}
\eeq
In two-loop ChPT, the imaginary parts arise solely from 
sunset graphs, and in terms of the notation established in App.~A 
we find the $3\pi$ component of $\rho_{{\rm A}3}^{(1)}$ to be 
\beqa
& & \rho_{{\rm A}3}^{(1)}[3\pi] = 
- {2 \over \pi F_\pi^2} {1\over (16 \pi^2)^2} 
{\cal I}m~\bigg[ q^2 \left( 2 {\bar Y}^{(3)}_0 - 
3 {\bar Y}^{(2)}_0 + {\bar Y}^{(1)}_0 \right) \nonumber \\ 
& & + M_\pi^2 \left( 4 {\bar Y}^{(1)}_0 - 
3 {\bar Y}^{(2)}_0 - {\bar Y}^{(0)}_0 \right) 
+ 4 M_\pi^2 \left( 2{\bar Z}^{(2)}_0 - {\bar Z}^{(1)}_0 \right) 
\nonumber \\ 
& & + {M_\pi^4 \over q^2} \left( {\bar Y}^{(1)}_0 - {\bar Y}^{(0)}_0 
- 4 {\bar Z}^{(1)}_0 \right) \bigg] \ \ , 
\label{im15}
\eeqa
where the above finite functions are evaluated with $3\pi$ mass 
values.  There is a comparable, but rather more complicated, expression 
for $\rho_{{\rm A}3}^{(0)}[3\pi]$ which we do not display here.  

\subsection{\bf Unitarity Determination of $\rho_{{\rm A}3}^{(0,1)}[3\pi]$}

Unitarity provides an alternative determination of 
the three-pion component of the isospin spectral functions 
$\rho_{{\rm A}3}^{(1)}[3\pi]$ and 
$\rho_{{\rm A}3}^{(0)}[3\pi]$.   The first step is to relate 
the spectral functions to the fourier transform of a 
non-time-ordered product, 
\begin{equation}
\rho_{{\rm A}3}^{(1)} (q^2)~(q^\mu q^\nu - q^2 g^{\mu\nu}) 
+ \rho_{{\rm A}3}^{(0)} (q^2)~q^\mu q^\nu 
= {1\over 2\pi} \int d^4 x ~e^{iq\cdot x}
\langle 0|A^\mu_3 (x) A^\nu_3 (0) |0\rangle \ .
\label{u1}
\end{equation}
One obtains the three-pion contribution by simply inserting the 
$3 \pi^0$ and $\pi^+\pi^0 \pi^-$ intermediate states in the above 
integral.  To determine the relevant ${\cal S}$-matrix element, 
we employ the lowest order chiral langrian of Eq.~(\ref{P5}) to find 
\begin{eqnarray}
& & \langle a_3 (q,\lambda)|{\cal S}|\pi^0 (p_1) \pi^0 (p_2) 
\pi^0 (p_3)\rangle =  \nonumber \\
& & \phantom{xxxxxxxx} i (2\pi)^4 \delta^{(4)} 
(q - p_1 - p_2 - p_3) \epsilon_\mu^* (q,\lambda)
{\cal M}^\mu_{000} \ , \nonumber \\
& & \langle a_3 (q,\lambda)|{\cal S}|\pi^+ (p_1) \pi^- (p_2) 
\pi^0 (p_3)\rangle =  \nonumber \\
& & \phantom{xxxxxxxx} i (2\pi)^4 \delta^{(4)} 
(q - p_1 - p_2 - p_3) \epsilon_\mu^* (q,\lambda){\cal M}^\mu_{+-0} \ ,
\label{u2}
\end{eqnarray}
where the invariant amplitudes are given by 
\begin{eqnarray}
{\cal M}^\mu_{000} &=& {i \over \sqrt{6} F_\pi} ~q^\mu 
{ M_\pi^2 \over q^2 - M_\pi^2} \ , 
\nonumber \\
{\cal M}^\mu_{+-0} &=& {i \over F_\pi} ~
\left[ 2~p_0^\mu + q^\mu~{M_\pi^2 - 
2q \cdot p_0 \over q^2 - M_\pi^2} \right]\ .
\label{u3}
\end{eqnarray}
Observe that ${\cal M}^\mu_{+-0}$ has two distinct contributions, a
direct coupling and a pion pole term, whereas ${\cal M}^\mu_{000}$ 
has only a pion pole term.  In the chiral limit of 
massless pions, the above amplitudes are conserved 
($q_\mu {\cal M}^\mu_{3\pi} = 0$) as required by chiral symmetry. 

To determine the spectral functions from Eq.~(\ref{u1}), 
we take $\mu = \nu = 3$ for $\rho_{{\rm A}3}^{(1)}[3\pi]$ 
and $\mu = \nu = 0$ for $\rho_{{\rm A}3}^{(0)}[3\pi]$.   
Throughout, we work in the Lorentz frame where $q_\mu = 
(q^0, {\bf 0})$.  With the tacit understanding that we 
consider only the three-pion component and take 
$q^2 > 9M_\pi^2$ in the following, this leads to 
\begin{eqnarray}
& & q^2 \rho_{{\rm A}3}^{(1)}(q^2) = {1 \over 128 \pi^6 F_\pi^2} 
I^{(1)} \ \ ,\nonumber \\
& & I^{(1)} \equiv \int {d^3 p_1 \over E_1} 
{d^3 p_2 \over E_2} {d^3 p_3 \over E_3} 
(p_3^3)^2 \delta^{(4)}(q - p_1 - p_2 - p_3) \ \ , 
\label{u4a}
\end{eqnarray}
and to 
\begin{eqnarray}
& & q^2 \rho_{{\rm A}3}^{(0)}(q^2) = {1 \over 512 \pi^6 F_\pi^2} 
\cdot {M_\pi^4 \over (q^2 - M_\pi^2 )^2} I^{(0)}  \ \ , 
\label{u4b} \\
& & I^{(0)} \equiv \int 
{d^3 p_1 \over E_1} {d^3 p_2 \over E_2} {d^3 p_3 \over E_3} 
\left( {7 \over 6} q^2 - 4 p_3^0 \sqrt{q^2} + 4 (p_3^0)^2 \right) 
\delta^{(4)}(q - p_1 - p_2 - p_3)\ . \nonumber 
\end{eqnarray}
The former experiences only the $\pi^+\pi^-\pi^0$ 
contribution whereas the latter has contributions from 
both the $\pi^+\pi^-\pi^0$ and $3\pi^0$ intermediate states.

Integration of the above integrals is staightforward and one obtains 
a form involving integration over the invariant squared-mass of the 
$\pi_1 \pi_2$ subsystem.  For example, for the spectral function 
$\rho_{{\rm A}3}^{(1)}(q^2)$ of Eq.~(\ref{u4a}) we obtain 
\begin{equation}
q^2 \rho_{{\rm A}3}^{(1)}(q^2) = {1 \over 768 \pi^4 F_\pi^2}~{1\over (q^2)^3}
\int_{4 M_\pi^2}^{(\sqrt{q^2} - M_\pi)^2} ~da ~
\sqrt{ a - 4 M_\pi^2 \over a} \lambda^{3/2}(q^2, a , M_\pi^2) \ ,
\label{u5}
\end{equation}
where 
\begin{equation}
\lambda(x,y,z) \equiv x^2 + y^2 + z^2 - 2 xy - 2 xz - 2 yz \ \ .
\label{u6}
\end{equation}
With the change of variable 
\begin{equation}
a = {1\over 2} \left( \sqrt{q^2} - M_\pi\right)^2 + 2 M_\pi^2  + 
\bigg( {1\over 2} \left( \sqrt{q^2} - M_\pi\right)^2 - 2 M_\pi^2 
\bigg) \cos\phi \ ,
\label{u7}
\end{equation}
we obtain the final form, 
\begin{equation}
\rho_{{\rm A}3}^{(1)}(q^2) = {1 \over 768 \pi^4}\cdot 
{q^2 \over F_\pi^2} \theta(q^2 - 9 M_\pi^2)~{\bar I}^{(1)}(x) \ ,
\label{u8}
\end{equation}
where $x \equiv M_\pi / \sqrt{q^2}$ and 
${\bar I}^{(1)}(x)$ is the dimensionless integral  
\begin{eqnarray}
& & {\bar I}^{(1)}(x) = \left[ {1\over 2} \left( 1 - x \right)^2 - 2 x^2 
\right]^{3/2} \int_0^\pi d\phi \ \sin\phi ~(1 + \cos\phi)^{1/2} 
\nonumber \\
& & \times {\left[ - 4 x^2 + \left( {1\over 2} (1 - x)^2 + x^2 - 1 + 
\left[ {1\over 2} (1 - x)^2 - 2  x^2 \right] \cos\phi \right)^2 
\right]^{3/2} \over \left[ {1\over 2} (1 - x)^2 + 2 x^2 + 
\left[ {1\over 2} (1 - x)^2 - 2  x^2 \right] \cos\phi \right]^{1/2} }
\ . \nonumber \\
\label{u9}
\end{eqnarray}
In like manner, we find 
\begin{eqnarray}
& & \rho_{{\rm A}3}^{(0)}(q^2) = {1 \over 512 
\pi^4} \cdot {q^2 \over F_\pi^2} \cdot {M_\pi^4 \over (q^2 - M_\pi^2)^2}
\nonumber \\
& & \times \left[ {7\over 3} ~{\bar I}^{(0)}_0 (x) - 4 ~{\bar I}^{(0)}_1 (x) 
+ 2 ~{\bar I}^{(0)}_2(x) \right] 
\theta(q^2 - 9 M_\pi^2) \ ,
\label{u10}
\end{eqnarray}
where 
\begin{eqnarray}
& & {\bar I}^{(0)}_n (x) = \left[ {1\over 2} \left( 1 - x \right)^2 - 2 x^2 
\right]^{3/2} \int_0^\pi d\phi \ \sin\phi ~(1 + \cos\phi)^{1/2} 
\nonumber \\
& & \times \left[ - 4 x^2 + \left( {1\over 2} (1 - x)^2 + x^2 - 1 + 
\left[ {1\over 2} (1 - x)^2 - 2  x^2 \right] \cos\phi \right)^2 
\right]^{1/2} 
\nonumber \\
& & \times {\left(1 - x^2 - {1\over 2} (1 - x)^2 - 
\left[ {1\over 2} (1 - x)^2 - 2  x^2 \right] \cos\phi \right)^n
\over \left[ {1\over 2} (1 - x)^2 + 2 x^2 + 
\left[ {1\over 2} (1 - x)^2 - 2  x^2 \right] \cos\phi \right]^{1/2} }
\ . \nonumber \\
\label{u11}
\end{eqnarray}
The spectral functions obtained in the unitarity approach 
described here agree precisely with those obtained from 
the imaginary parts of the sunset amplitudes.
\section{Chiral Sum Rules}
In previous sections, we have 
determined the isospin and hypercharge axialvector propagators 
to two-loop order in ChPT.  Essential to the success of this 
program is the renormalization procedure by which the results 
are rendered finite.  As a consequence of the renormalization
paradigm, however, the physical results contain a number of 
undetermined finite counterterms.  In particular, the 
real parts of the polarization functions $\Pi_{{\rm A}a}^{(1)}$ and
$\Pi_{{\rm A}a}^{(0)}$\  ($a=3,8$) contain 
two such constants from one-loop order ($L_{10}^{(0)}$ and
$H_1^{(0)}$) and five independent combinations of the 
$\{B_\ell^{(0)}\}$ from two-loop order.  Of these, 
the constants $H_1^{(0)}$ and $B_{11}^{(0)}$ are 
related to contact terms which are regularization dependent and 
thus physically unobservable.  However, the remaining 
counterterms (called `low energy constants' or LEC) must be 
extracted from data.  In this section, we describe how 
some of the ${\cal O}(p^6)$ counterterm coupling constants 
are obtainable from chiral sum rules, and as an example 
we study a specific case involving broken $SU(3)$. 

A derivation of the chiral sum rules together with an 
application of one of them appears in 
Ref.~\cite{gkPRL}.  We refer the reader to that article for 
a general orientation.  For the purpose of writing dispersion 
relations it suffices to note that at low energies the polarization 
functions are already determined from the results obtained in 
previous sections, although some care must be taken with the 
kinematic poles at $q^2=0$ in the individual functions 
$\Pi_{{\rm A}a}^{(1)}$ and $\Pi_{{\rm A}a}^{(0)}$.  
As regards high energy behaviour, the large-$s$ limit of 
spectral functions relevant to our analysis can be read off 
from the work in Refs.~\cite{fnr,bnp}.  However, 
since we are calculating up to order $p^6$, terms up to and including 
quadratic dependence in the light quark masses must be included.  
Thus, for example, the asymptotic expansion for the isospin axialvector 
spectral function summed over spin-one and spin-zero reads
\begin{equation}
\rho_{{\rm A}3}^{(1+0)}(s)={1\over 8 \pi^2} \left( 
1 + {\alpha_s \over \pi} 
\left[ 1 + 12 { {\hat m}^2 \over q^2} \right] 
+{\cal O}(\alpha_s^2,1/s^2)  \right) \ .
\label{asymexamp}
\end{equation} 
Similar expressions hold for the other components, and we 
summarize their collective leading asymptotic behaviour by
\begin{equation}
\rho_{{\rm A}a}^{(1)}(s) \sim {\cal O}(1) \ , \quad 
\rho_{{\rm A}a}^{(0)}(s) \sim {\cal O}(s^{-1}) \ ,
\quad (\rho_{{\rm A}3}^{(1)} - \rho_{{\rm A}8}^{(1)})(s) \sim 
{\cal O}(s^{-1}) \ . 
\label{asym}
\end{equation}
The real parts of the polarization functions 
show exactly the same asymptotic behaviour as the imaginary parts,
{\it i.e.} expressions analogous to Eq.~(\ref{asym}) hold. 
The K\"all\'en-Lehmann spectral representation of two-point functions 
then implies the following dispersion relations for the 
axialvector polarization functions of a given flavour $a = 3,8$, 
\begin{equation}
q^2 \Pi_{{\rm A}a}^{(0)}(q^2) - \lim_{q^2 = 0}\left( q^2 
\Pi_{{\rm A}a}^{(0)}(q^2)\right) = q^2 \int_0^\infty ds ~
{\rho_{{\rm A}a}^{(0)}(s) \over s - q^2 - i \epsilon} 
\ \ , \label{aa5} 
\end{equation}
\begin{equation}
q^2 \Pi_{{\rm A}a}^{(1)}(q^2) - \lim_{q^2 = 0}\left( q^2 
\Pi_{{\rm A}a}^{(1)}(q^2)\right) - \lim_{q^2 = 0} {d\over dq^2}
\left( q^2 \Pi_{{\rm A}a}^{(1)}(q^2)\right) = q^4 \int_0^\infty ds~ 
{\rho_{{\rm A}a}^{(1)}(s) \over s \left( s - q^2 - i \epsilon \right)}
\ \ . \label{aa5'}
\end{equation} 
We work with $q^2 \Pi_{{\rm A}a}^{(0),(1)}(q^2)$ due to 
the presence of $q^2 = 0$ kinematic poles. Moreover, 
the subtraction constants have been placed on the left hand side in
Eqs.~(\ref{aa5}),(\ref{aa5'}) in order to equate only physically 
observable quantities. 
Dispersion relations involving $SU(3)$-breaking combinations 
have an improved asymptotic behaviour, such as 
\begin{equation}
(\Pi_{{\rm A}3}^{(1)} + \Pi_{{\rm A}3}^{(0)}
- \Pi_{{\rm A}8}^{(1)} - \Pi_{{\rm A}8}^{(0)})(q^2)
= \int_0^\infty ds~
{(\rho_{{\rm A}3}^{(1)}+\rho_{{\rm A}3}^{(0)}-
  \rho_{{\rm A}8}^{(1)}-\rho_{{\rm A}8}^{(0)})(s) 
\over s - q^2 - i \epsilon}
\label{aa5''}
\end{equation}
and 
\begin{equation}
q^2 (\Pi_{{\rm A}3}^{(1)} - \Pi_{{\rm A}8}^{(1)})(q^2)-
\lim_{q^2 = 0}\left( q^2 
\left( \Pi_{{\rm A}3}^{(1)}-\Pi_{{\rm A}8}^{(1)}\right) (q^2) \right)
= q^2 \int_0^\infty ds ~
{\left( \rho_{{\rm A}3}^{(1)}-\rho_{{\rm A}8}^{(1)} \right) (s)
\over s - q^2 - i \epsilon}   
\ . \label{aa5'''}
\end{equation}

Sum rules are obtained by evaluating arbitrary derivatives of such 
relations at $q^2 = 0$. For sum rules inferred from
Eqs.~(\ref{aa5}),(\ref{aa5'}) it is preferable to express the left hand 
side in terms of ${\hat \Pi}_{{\rm A}a}^{(0,1)}$, 
\begin{eqnarray}
& & {1\over n!}\left[{d \over dq^2}\right]^n 
{\hat \Pi}_{{\rm A}a}^{(0)}(0) = \int_0^\infty ds ~ 
{{\bar \rho}_{{\rm A}a}^{(0)}(s) \over s^n} \qquad (n \ge 1) \ ,
\label{aa11a} \\
& & {1\over (n -1)!}\left[{d \over dq^2}\right]^{n-1} 
{\hat \Pi}_{{\rm A}a}^{(1)}(0) - {1\over n!}\left[{d \over dq^2}\right]^n 
{\hat \Pi}_{{\rm A}a}^{(0)}(0) = \int_0^\infty ds ~ 
{\rho_{{\rm A}a}^{(1)}(s) \over s^n} \qquad (n \ge 2) \ \ ,
\label{aa11b} 
\end{eqnarray}
where ${\bar \rho}_{{\rm A}a}^{(0)}(s)$ is defined in 
Eq.~(\ref{pole}).  Finally, Eq.~(\ref{aa5'''}) leads directly to the 
following sequence of sum rules explicitly involving broken $SU(3)$, 
\begin{equation}
{1\over n!}\left[{d \over dq^2}\right]^n\left( 
\Pi_{{\rm A}3}^{(1)} + \Pi_{{\rm A}3}^{(0)}
- \Pi_{{\rm A}8}^{(1)} - \Pi_{{\rm A}8}^{(0)}\right)(0) 
= \int_0^\infty ds ~ {(\rho_{{\rm A}3}^{(1)} + \rho_{{\rm A}3}^{(0)} 
- \rho_{{\rm A}8}^{(1)} - \rho_{{\rm A}8}^{(0)})(s)  \over s^{n+1}} \ ,
\label{aa12} 
\end{equation}
where $n \ge 0$. 

For this last sum rule, let us consider in some detail the 
case $n=0$. An equivalent form, better suited for phenomenological
analysis, is given by 
\begin{equation}
\left( \hat\Pi_{{\rm A}3}^{(1)}-\hat\Pi_{{\rm A}8}^{(1)} \right) (0)-
{d\over dq^2} 
\left( \hat\Pi_{{\rm A}3}^{(0)}-\hat\Pi_{{\rm A}8}^{(0)} \right) (0)
=\int_0^\infty ds \ {\left( \rho_{{\rm A}3}^{(1)} - 
\rho_{{\rm A}8}^{(1)} \right) (s) \over s} \ \ .
\label{su3break}
\end{equation}
Evaluation of the left hand side (LHS) of this sum rule yields 
\begin{eqnarray}
{\rm LHS}=& & {16 \over 3} {M_K^2 - M_\pi^2 \over F_\pi^2} 
Q_{\rm A}(\mu) \ + \  0.001053 \nonumber \\
& &+{1\over F_\pi^2 (16 \pi^2)^2} \bigg[ 
M_\pi^2 \log {M_\pi^2 \over \mu^2} \left( {8\over 9}
-{1\over 3} \log {M_\pi^2 \over \mu^2}-{1\over 3} \log {M_K^2 \over
\mu^2} \right) \nonumber \\
& & \qquad\qquad\qquad +M_\pi^2 \log {M_K^2 \over \mu^2}
\left(-{1\over 18}+{1\over 6} \log {M_K^2 \over \mu^2} \right) \nonumber\\
& & \qquad\qquad\qquad +M_K^2 \log {M_K^2 \over \mu^2}
\left( -{5\over 6}+{1\over 2} \log {M_K^2 \over \mu^2} \right) \nonumber\\
& & \qquad\qquad\qquad +128 \pi^2 L_{10}^{(0)} 
\left(M_K^2 \log {M_K^2\over \mu^2} - M_\pi^2 \log {M_\pi^2 \over
\mu^2}\right) \bigg]  \ .
\label{LHS}
\end{eqnarray}
Recall from the discussion at the beginning of Sect.~VI that 
there are three distinct sources for the finite low energy 
terms: (i) ${\cal O}(p^6)$ CTs $\{ B_\ell^{(0)}\}$, 
(ii) the `remnant' contributions, and (iii) the finite $Y$, $Z$ 
integrals ({\it cf} Eq.~(\ref{fin1})).  It is a combination 
of the latter two which give rise to the numerical term 
(which is scale-independent and vanishes in the $SU(3)$ 
limit of equal masses) in the first line.  We have displayed 
all chiral logs explicitly, and $Q_{\rm A} (\mu)$ is the 
${\cal O}(p^6)$ counterterm defined earlier in Eq.~(\ref{ctA}).  
Since the full expression is scale-independent, this
allows one to directly read off the variation of the contributing 
counterterm combination at renormalization scale $\mu$.

We can use the sum rule of Eq.~(\ref{LHS}) to numerically estimate 
$Q_{\rm A} (\mu)$.  One needs to evaluate the 
spectral integral on the right hand side (RHS) of
Eq.~(\ref{su3break}).  For our purposes, it is sufficient to 
approximate the contribution of the isospin spectral function 
in terms of the $a_1$ resonance taken in narrow width approximation, 
$\rho_{{\rm A}3}^{(1){\rm res}}(s) \simeq g_{a_1} \delta (s - M^2_{a_1})$.  
Employing resonance parameters as obtained from the fit in
Ref.~\cite{dg}, we obtain 
\begin{equation}
\int_0^\infty ds~ {\rho_{{\rm A}3}^{(1){\rm res}} \over s} \simeq
0.0189 \ .
\label{isovalue}
\end{equation}
Although consistency with QCD dictates that we also include 
the large-$s$ continuum~\cite{blr}, the leading-order contributions 
would cancel in Eq.~(\ref{su3break}) and the remaining mass
corrections are small.  As regards the hypercharge spectral 
function $\rho_{{\rm A}8}^{(1)}$, little is presently known. 
The lowest lying resonances which contribute are $f_1~$(1285) 
and $f_1$~(1510) but the couplings of these resonances to the 
axialvector current have not been determined.  Since the corresponding
sum rule for vector current spectral functions \cite{gk2} 
exhibits large cancellations between the contributions from 
$\rho$~(770), $\omega$~(782) and $\Phi$~(1020), 
we expect a similar cancellation to be at work in the axialvector 
sector.  To obtain a rough estimate, we assume 
the two resonances $f_1$~(1285) and $f_1$~(1510) can be approximated
by a single effective resonance with spectral function 
$\rho_{{\rm A}8}^{(1){\rm eff}}(s) \simeq g_{a_8} \delta (s-M^2_{a_8})$.
Assuming further $g_{a_8} \simeq g_{a_1}$ and $M_{a_8} \simeq 1.4$~GeV
we estimate the hypercharge contribution to the RHS of
Eq. (\ref{su3break}) as $0.012$. Allowing for a $50~\%$ error in
this estimate places the RHS in the range $0.001 \le RHS \le 0.013$ 
and leads finally to 
\begin{equation}
0.000043 \le Q_{\rm A}(M_{a_1}) \le 0.000130 \ \ ,
\label{QAguess}
\end{equation}
where the renormalization scale $\mu=M_{a_1}$ has been adopted.  
This is clearly to be taken as just a rough estimate. Only 
experimental determination of the missing coupling constants can 
provide a more reliable estimate. In addition, a more thorough 
phenomenological analysis will involve use of the entire spectrum. 
However, this example serves to illustrate the general procedure.

We have not touched on chiral sum rules 
involving {\it both} vector and axialvector spectral functions. The most
prominent example of this type is the Das-Mathur-Okubo (DMO) sum 
rule~\cite{dmo} which, in modern terminology, has been employed to
determine the LEC $L_{10}^{(0)}$~\cite{gl2,Eckeretal}.  In 
Ref.~\cite{gkPRL} we
have shown how the DMO sum rule must be modified to be
valid to second order in the light quark masses. Recently, $\tau$-decay 
data has renewed interest in this sum rule from the experimental
side~\cite{hamburg97}. We have begun a phenomenological study of 
the DMO sum rule using the two-loop results of polarization functions 
obtained both here and in Ref.~\cite{gk1}.  Results will be 
reported elsewhere.  

Finally, there are also those sum rules involving {\it no} 
${\cal O}(p^6)$ counterterm coupling constants, {\it i.e.} 
those obtained by taking appropriately many derivatives of the 
dispersion relations Eqs.~(\ref{aa5})-(\ref{aa5'''}). From 
experience with the corresponding inverse moment sum rules of 
vector current spectral functions~\cite{gk2}, 
we expect these sum rules in general not to be
verified.  This is because the relevant physics 
(which involves the low-lying resonances) enters the relations 
only in higher order of the chiral
expansion. A quantitative study of these sum rules is deferred to a
forthcoming publication.   
\section{\bf Conclusions}
Our analysis of axialvector current propagators in 
two-loop ChPT has led to a complete two-loop renormalization 
of the pion and eta masses and decay constants as well as 
the real-valued parts of the isospin and hypercharge 
polarization functions.  It has yielded predictions for 
axialvector spectral functions and has allowed the derivation 
of spectral function sum rules.  

Despite the complexity of many of the individual steps and results, 
the sum of tree, one-loop and two-loop contributions to the 
axialvector propagator yields a simple overall structure, 
\begin{equation}
\Delta_{{\rm A},\mu\nu}(q^2) =
(F^2 + {\hat \Pi}_{\rm A}^{(0)}(q^2)) 
g_{\mu\nu}  - {F^2 \over q^2 - M^2} q_\mu q_\nu 
+ ( 2L_{10}^{(0)} - 4H_1^{(0)} + {\hat \Pi}_{\rm A}^{(1)}(q^2))
 (q_\mu q_\nu - q^2 g_{\mu\nu}) \ , 
\label{two-loop}
\end{equation}
where flavor labelling is suppressed.  Comparing this to the general 
decomposition of Eq.~(\ref{aa2a}) yields 
\begin{eqnarray}
\Pi_{{\rm A}}^{(1)}(q^2) &=& 
2L_{10}^{(0)} - 4H_1^{(0)} + {\hat \Pi}_{\rm A}^{(1)}(q^2)
- {F^2 + {\hat \Pi}_{{\rm A}}^{(0)}(q^2) \over q^2} \ , \nonumber \\
\Pi_{{\rm A}}^{(0)}(q^2) &=& {{\hat \Pi}_{{\rm A}}^{(0)}(q^2) \over q^2} 
- {F^2 M^2 \over q^2 (q^2 - M^2)} \ \ .
\label{relate}
\end{eqnarray}
As noted earlier, there are kinematic poles at $q^2 = 0$ in both 
the spin-one and spin-zero polarization functions, but the sum 
$\Pi_{\rm A}^{(1)}$ + $\Pi_{\rm A}^{(0)}$ is free of 
such singularities.  

A large number of ${\cal O}(p^6)$ counterterms entered the 
axialvector calculation, and many constraints among them were obtained 
from the subtraction procedure.  Thus given the total of 
$23$ ${\cal O}(p^6)$ counterterms which appeared by employing 
the basis of Ref.~\cite{fs}, 
each of the ${\bar \lambda}^2$ and ${\bar \lambda}$ subtractions 
were found to yield $14$ constraints.  The analysis of vector 
propagators in Ref.~\cite{gk1} yielded another $3$ conditions for 
each of the ${\bar \lambda}^2$ and ${\bar \lambda}$ subtractions.   
This total of $17$ conditions constraining the $\{ B_\ell^{(2)}\}$ 
and $\{ B_\ell^{(1)}\}$ counterterms is of course universal 
and can be used together with results of other two-loop studies.  
We can summarize the remaining nine finite ${\cal O}(p^6)$ 
counterterms as 
\begin{eqnarray}
{\rm Polarization \ Amplitudes}:& & \ P_{\rm A},Q_{\rm A},R_{\rm A},
B_{11}^{(0)},B_{13}^{(0)} 
\nonumber \\
{\rm Decay \ Constants}:& & \ \{ {\tilde B}_\ell \} \qquad (\ell = 1,\dots 4)
\label{summct} \\
{\rm Masses}:& & \ \{ {\tilde B}_\ell \} \qquad (\ell = 1,\dots 9) \ \ ,
\nonumber 
\end{eqnarray}
where $P_{\rm A},Q_{\rm A},R_{\rm A}$ are defined in Eq.~(\ref{ctA}) 
and the $\{ {\tilde B}_\ell \}$ in Eqs.~(\ref{effdc}),(\ref{effmm}).  
We have made preliminary numerical estimates for $Q_{\rm A}$ in 
this paper ({\it cf} Eq.~(\ref{QAguess})) and for $P_{\rm A}$ 
in Ref.~\cite{gkPRL}.  The counterterm $B_{11}^{(0)}$ is related 
to a contact term and is regularization dependent, much the same 
as the constant $H_1^{(0)}$ appearing in Eq.~(\ref{two-loop}).  However,
these terms always drop out when physical observable quantities are 
considered.  The constant $R_{\rm A}$ is seen to contribute equally 
to all flavour components of the axialvector polarization functions.  
It cannot therefore be accessed by the chiral sum rules involving 
broken $SU(3)$ considered in the previous section.  However, by 
combining the results obtained here with the two-loop analysis of 
the vector current two-point functions, the combination 
$R - R_{\rm A}$ is seen to constitute a mass correction to the 
Das-Mathur-Okubo sum rule.  The details of this analysis will be 
presented elsewhere.  Finally, little is known about the nine 
constants $\{ {\tilde B}_\ell \}$ ($\ell = 1, \dots, 9$) which 
determine (together with the calculated loop contributions) the 
$p^6$ corrections to masses and decay constants.  We have not 
attempted to estimate these constants ({\it e.g.} by the 
resonance saturation hypothesis) since as far as the axialvector 
two-point function is concerned, their contribution is implicit.  
However, the explicit expressions given here can be used in further 
ChPT studies when expressing bare masses and decay constants in terms
of fully renormalized physical quantities.  

In view of the length and difficulty of the 
calculation, it is reassuring that 
a broad range of independent checks 
was available to gauge the correctness of our results.  
We list the most important of them here:
\begin{enumerate}
 \item Because the calculation involved {\it independent} 
determinations of isospin and hypercharge channels 
at each stage, the $SU(3)$ limit of equal masses provided 
numerous tests among the set of isospin and hypercharge 
decay constants, masses and polarization functions.  
As a by-product, it also revealed the presence of previously 
unnoticed identities among the sunset amplitudes.  
 \item As shown in Sect.~VII, it was possible to 
determine spectral functions directly from the two-loop 
analysis or equivalently from a unitarity approach which employed 
one-loop amplitudes as input.  In this way, both the 
specific sunset integrals as well as the structural relations 
of Eq.~(\ref{relate}) were able to be tested successfully.  
 \item It turned out that although most of the divergent terms 
could be subtracted away with counterterms, there occurs 
no $m_K^6$ counterterm contribution in $M_\pi^{(6)2}$. 
To avoid disaster, there must thus be a cancellation between 
sunset and nonsunset numerical terms.  Such a cancellation 
indeed occurs and constitutes a nontrivial check 
on our determination of the sunset contribution.  
 \item Given that ${\hat \Pi}_{\rm A}^{(0)}(q^2)$ can be 
shown to vanish in the limit of zero quark mass, it follows that 
our final result in Eq.~(\ref{two-loop}) has the correct chiral limit.
 \item Lastly, we have explicitly verified in 
$\overline{MS}$ renormalization that the constant $C$ is 
absent, as must be the case.

\end{enumerate}

Yet more on this subject remains to be done.  This 
is especially true of the material composing 
Sect.~VIII, where an application of $SU(3)$-breaking 
sum rules to determine finite ${\cal O}(p^6)$ counterterms 
was discussed and additional points were raised.  
Future work will be needed to carefully analyze 
the axialvector sum rules, particularly the role of 
existing data to provide as precise a determination of the counterterms 
as experimental uncertainties allow.  We can, of course, 
combine the results of the present axialvector study with 
the vector results of Ref.~\cite{gk1} to study an even wider 
range of sum rules ({\it e.g.} as with the proposed determination 
of $R_{\rm A}$ discussed above).  On an even more ambitious level, 
our experience with such relations makes us optimistic 
about the possibility of describing a possible framework 
for interpreting chiral sum rules to {\it arbitrary} order 
in the chiral expansion.  Finally, it will be of interest 
to reconsider the phenomenological extraction of spectral functions 
such as $\rho_{{\rm A}33}^{(0)}[3\pi]$ and to 
stimulate experimental efforts to extract spectral 
function information involving non-pionic particles 
such as kaons and etas.

\acknowledgments 
We would like to thank J. Gasser, M. Knecht and J. Stern for 
many fruitful discussions on the subject discussed in this 
article. One of us (JK) acknowledges support from the 
Institut de Physique Nucl\'eaire, Orsay,\footnote{Laboratoire 
de Recherche des Universit\'es Paris XI et Paris VI, 
associ\'e au CNRS} where a substantial part of his contribution 
to this project was performed.  The research described here 
was supported in part by the National Science Foundation 
and by Schweizerischer Nationalfonds.


\eject
\appendix
\section{\bf Sunset Integrals}
In the following, we compile mathematical details related 
to the sunset amplitudes which appear in the two-loop analysis.  
First, we give integral expressions for the 
quantities occurring in the 1PI, vertex and self-energy 
amplitudes.  Then we write down the finite 
sunset amplitudes which remain after the singular parts 
have been identified.  Finally we discuss certain identities 
which relate various sunset integrals.  Additional work on sunset 
integrals can be found in Refs.~\cite{s1,bur96,bt97,betal97} for the equal 
mass case and in Refs.~\cite{s5,s6,s7,jet96} for the general mass case.
\subsection{\bf Definitions of Sunset Integrals}
For the sunset amplitudes containing unequal masses, we shall 
denote the mass occurring twice as `$M$' and the third mass 
as `$m$' ({\it e.g.} for ${\bar K}K\pi$ amplitudes, we have 
$M \to m_K$ and $m \to m_\pi$).  
The quantity ${\cal H}_{\mu\nu}$ appearing in the 1PI sunset 
amplitude of Eq.~(\ref{5}) is defined by the integral expression 
\beqa
& & F_0^2 {\cal H}_{\mu\nu}(q^2,m^2,M^2)
\equiv \int d{\tilde t}~
{(q - 3t)_\mu (q - 3t)_\nu \over
t^2 - m^2} \ \int d{\tilde b}~
{1 \over  (b^2 - M^2)\cdot ((Q-b)^2 - M^2)} \nonumber \\
& & \phantom{xxxxxxxxxxxx} = q_\mu q_\nu S - 3 q_\mu S_\nu 
- 3 q_\nu S_\mu + 9 S_{\mu\nu} \ \ , 
\label{ss3}
\eeqa
where $Q \equiv q - t$ and 
\beq
\{ S; S_\mu; S_{\mu\nu} \} 
\equiv \int d{\tilde t}~
{\{1; t_\mu; t_\mu t_\nu \}  \over t^2 - m^2} \ \int d{\tilde b}~
{1 \over  (b^2 - M^2)\cdot ((Q-b)^2 - M^2)} \ \ .
\label{se6aa}
\eeq
From covariance, we have 
\beqa
S_\mu (q^2,m^2,M^2) &\equiv& q_\mu {\bar S} (q^2,m^2,M^2) \ ,
\nonumber \\
S_{\mu\nu}(q^2,m^2,M^2) &\equiv& q_\mu q_\nu 
S_1 (q^2,m^2,M^2) + g_{\mu\nu} S_2 (q^2,m^2,M^2) \ \ ,
\label{sfcns}
\eeqa
Also appearing in Eq.~(\ref{5}) is ${\cal L}_{\mu\nu}$, defined by 
\beq
F_0^2 {\cal L}_{\mu\nu}(q^2,m^2,M^2) = \int d{\tilde t}~
{1 \over t^2 - m^2} \ \int d{\tilde b}~
{(Q - 2b)_\mu (Q - 2b)_\nu
\over  (b^2 - M^2)\cdot ((Q-b)^2 - M^2)} \ \ , 
\label{ss4}
\eeq
which can be expressed in the equivalent form 
\begin{eqnarray}
F_0^2 \lefteqn{{\cal L}_{\mu\nu} (q^2, m^2, M^2) \equiv
{1\over d-1}\bigg[ q_\mu q_\nu {\cal K}_1 (q^2, m^2, M^2)} 
\nonumber \\
& & \phantom{xxxxxx} + g_{\mu\nu} {\cal K}_2 (q^2, m^2, M^2) + 
4 g_{\mu\nu} \left( {3\over 2} - 
{4-d\over 2}\right)A(m^2)A(M^2)  \bigg] \ \ ,
\label{fcnia}
\end{eqnarray}
where ${\cal K}_1$ and ${\cal K}_2$ are given respectively in 
Eqs.~(\ref{k1}),(\ref{k2}) below. 

The vector-valued integrals $I_{1\mu}$ and $I_{2\mu}$ 
which contribute to the axialvector vertex function 
in Eq.~({\ref{6e}), are defined as 
\beqa
\lefteqn{I_{1\mu} \left(q^2;m^2 ;M^2;\Lambda\right) \equiv
q_\mu I_1 (q^2;m^2 ;M^2;\Lambda)} 
\nonumber \\
& & \phantom{xxxxxxx} = \int d{\tilde t}~{(q - 3t)_\mu 
\over t^2 - m^2}~ \int d{\tilde b}~
{ Q^2 - 2q\cdot t + 2 b\cdot (Q-b) + \Lambda 
\over  (b^2 - M^2)\cdot ((Q-b)^2 - M^2)} \ \ ,
\label{int1} \\
\lefteqn{I_{2\mu} \left(q^2;m^2 ;M^2\right) \equiv
q_\mu I_2 (q^2;m^2 ;M^2)} 
\nonumber \\
& & \phantom{xxxxxxx} = \int d{\tilde t}~{(q + t)^\nu 
\over t^2 - m^2}~ \int d{\tilde b}~
{ (Q - 2b)_\mu (Q - 2b)_\nu 
\over  (b^2 - M^2)\cdot ((Q-b)^2 - M^2)}\ \ .
\label{int2}
\eeqa

Finally, in the calculation of self-energies, 
there appear quantities $S$, $R$ and $U$.  The 
function $S$ is already defined in Eq.~(\ref{se6aa}), 
and we have for $R$ and $U$, 
\beqa
\lefteqn{R(q^2; m^2; M^2; \Lambda) \equiv} \nonumber \\
& & \phantom{xxxxxxx} \int {d{\tilde t}
\over t^2 - m^2}~ \int d{\tilde b}~
{\left[ (q - t)^2 - 2q\cdot t + 2 b\cdot (Q-b) + \Lambda \right]^2 
\over  (b^2 - M^2)\cdot ((Q-b)^2 - M^2)} \ \  , \label{se7} \\
\lefteqn{U(q^2; m^2; M^2) \equiv} \nonumber \\
& & \phantom{xxxxxxx} \int {d{\tilde t} \over t^2 - m^2}~ \int d{\tilde b}~
{\left( (q + t) \cdot (2b + t - q)\right)^2 
\over  (b^2 - M^2)\cdot ((Q-b)^2 - M^2)} \ \ .
\label{se7a} 
\eeqa

Analysis reveals that all the sunset contributions can be 
expressed in terms of the functions 
$S$, ${\bar S}$, $S_1$, $S_2$, ${\cal K}_1$ and ${\cal K}_2$.   
This is already evident from Eq.~(\ref{ss3}) for 
${\cal H}_{\mu\nu}$ and from Eq.~(\ref{fcnia}) for 
${\cal L}_{\mu\nu}$.  One can deduce the additional 
relations 
\beqa
\lefteqn{I_{1\mu} \left(q^2;m^2 ;M^2;\Lambda\right) =
-2 q_\mu \left[ 2 A^2 (M^2) + A(m^2) A(M^2) \right]} 
\nonumber \\
& & \phantom{xxxx} + \left( q_\mu S(q^2; m^2; M^2) 
- 3 S_\mu (q^2; m^2; M^2) \right)
\left[ 2 (q^2 + m^2 - M^2) + \Lambda \right] \nonumber \\
& & \phantom{xxxx} - 6 \left( q_\mu q_\nu S^{\nu} (q^2; m^2; M^2) 
- 3 q^\nu S_{\mu\nu} (q^2; m^2; M^2) \right)
\label{int1a} \\
\lefteqn{I_{2\mu} \left(q^2;m^2 ;M^2\right) =
q_\mu {2(3 \ +\ d-4)\over  d-1}  A(m^2) A(M^2)} 
\nonumber \\
& & \phantom{xxxx} + {2 F_0^2 \over d-1} q_\mu 
\left( q^2 {\cal K}_1(q^2; m^2; M^2) 
+ {\cal K}_2(q^2; m^2; M^2) \right) \ \ , 
\label{int2a}
\eeqa
as well as 
\beqa
\lefteqn{R(q^2; m^2; M^2; \Lambda) =} \nonumber \\
& & \phantom{xxxxxx} \Lambda^2 S(q^2; m^2; M^2) 
- 4 \Lambda \bigg[ -(q^2 + m^2 - M^2) 
S(q^2; m^2; M^2) \nonumber \\
& & \phantom{xxxxxx} + 3 q_\mu S^\mu (q^2; m^2; M^2) 
- A^2 (M^2) + A(M^2) A(m^2) \bigg] \label{se7b} \\
& & \phantom{xxxxxx} + 36 q_\mu q_\nu S^{\mu\nu} (q^2; m^2; M^2) 
+ 24 (M^2 - m^2 - q^2) q_\mu S^\mu (q^2; m^2; M^2) 
\nonumber \\
& & \phantom{xxxxxx} + 4 (q^2 + m^2 - M^2)^2 S(q^2; m^2; M^2) 
+ A^2(M^2) (4 m^2 - 12 q^2) 
\nonumber \\
& & \phantom{xxxxxx} + A(m^2) A(M^2) (8 M^2 - 6 q^2 - 6 m^2 ) 
\nonumber \\
\lefteqn{U(q^2; m^2; M^2) =} \nonumber \\
& & \phantom{xxxxxx} {4\over d-1}\bigg[ + (q^2 + m^2) \left( {3\over 2} + 
{d-4\over 2} \right) A(m^2) A(M^2) \nonumber \\
& & \phantom{xxxxxx} + q^2 \left( q^2 {\cal K}_1 (q^2; m^2; M^2) 
+ {\cal K}_2 (q^2; m^2; M^2) \right) \bigg] \ .
\label{se7c} 
\eeqa

The functions $S$, ${\bar S}$, $S_1$, $S_2$, ${\cal K}_1$ and 
${\cal K}_2$ can in turn each be written as the sum of terms 
(which diverge in the $d\to 4$ limit) proportional to 
gamma functions plus finite-valued functions 
$\{ Y_c^{(n)} (q^2, m^2, M^2) \}$ and 
$\{ Z_c^{(n)} (q^2, m^2, M^2) \}$.  Thus, we have 
\beqa 
& & S (q^2, m^2, M^2) = {\Gamma^2 (2 - d/2)\over (4\pi)^d}~(M^2)^{d-4} 
\nonumber \\
& &  \phantom{xxxx} \times {1\over d-2} \bigg[ - 2 \left( \bigg(
{m^2 \over M^2}\bigg)^{d/2 - 1} 
+ {1 \over d-3} \right) M^2 + {1\over 5 - d}m^2 \nonumber \\
& & \phantom{xxxx} + { 4 - d\over d(5-d)}q^2 \bigg] 
- Y^{(0)}_0 m^2 + (2 Y^{(1)}_0 - Y^{(0)}_0 )q^2 \ ,
\label{s} \\
& & {\bar S} (q^2, m^2, M^2) = {\Gamma^2 (2 - d/2) \over 
(4\pi)^d}~(M^2)^{d-4} \nonumber \\
& &  \phantom{xxxx} \times {1\over d} \left( -  
{2 \over d-3}M^2 - {4 - d \over
(5 - d)(d-2)} m^2 + { 4 - d\over (5-d)(d+2)}q^2 \right) \nonumber \\
& & \phantom{xxxx} + ( Y^{(0)}_0 - 2Y^{(1)}_0) m^2 
+ (3 Y^{(2)}_0 - 2 Y^{(1)}_0 )q^2 + Z_0^{(1)} \ , \label{smu} \\
& & S_1 (q^2, m^2, M^2) = {\Gamma^2 (2 - d/2) \over 
(4\pi)^d}~(M^2)^{d-4} \nonumber \\
& &  \phantom{xxxx} \times {1 \over d + 2} \left( -  {2 \over
d-3}M^2 - {4 - d \over d (5 - d)} m^2 + { 4 - d\over (d+4)(5-d)}q^2
\right) \nonumber \\
& & \phantom{xxxx} + ( 2Y^{(1)}_0 - 3Y^{(2)}_0) m^2 
+ (4 Y^{(3)}_0 - 3Y^{(2)}_0 )q^2 + 2 Z_0^{(2)} \label{s1} \\
& & S_2 (q^2, m^2, M^2) = {\Gamma^2 (2 - d/2) \over 
(4\pi)^d}~(M^2)^{d-4} \nonumber \\
& &  \phantom{xxxx} \times \bigg[ \bigg( - {2\over d(d-2)}\left( 
\bigg( {m^2 \over M^2}\bigg)^{d/2} + {2\over d-2} \right) M^4 - 
{2\over d(d-2)(d-3)}m^2 M^2 \nonumber \\
& & \phantom{xxxx} + {2\over d(d-3)(d+2)}q^2 M^2 
+ {1\over d(5-d)(d-2)} m^4 \nonumber \\
& & \phantom{xxxx} + {2(4-d)\over d(5-d)(d^2 - 4)}q^2
m^2 + {d-4\over d(5-d)(d+2)(d+4)}q^4 \bigg) \bigg] \nonumber \\
& & \phantom{xxxx} + { 1 \over 2} \bigg( ( Y^{(1)}_0 - Y^{(0)}_0) m^4 
+ (2 Y^{(3)}_0 - 3Y^{(2)}_0 + Y^{(1)}_0 ) q^4 \nonumber \\
& & \phantom{xxxx} + ( 4Y^{(1)}_0 - 3Y^{(2)}_0 - Y^{(0)}_0 ) m^2 q^2 
-  Z_0^{(1)} m^2 + ( 2 Z_0^{(2)} -  Z_0^{(1)} ) q^2 \bigg) \ , 
\label{smunu}
\eeqa
and 
\beqa
& & {\cal K}_1 (q^2, m^2, M^2) = 
{\Gamma^2 (2 - d/2) \over (4\pi)^d}~(M^2)^{d-4}~ 
{d - 1 \over d - 2}\bigg[ - { 16 \over d(d-3)(5-d)(d+2)}M^2 \nonumber \\
& &  \phantom{xxxx} + \left( - {2\over 3} 
\bigg({m^2 \over M^2}\bigg)^{d/2 - 2} + 
{24  \over d(5-d)(7-d)(d+2)}\right) m^2 \nonumber \\
& & \phantom{xxxx} + {24(4-d) \over d(7-d)(5-d)(d+2)(d+4)}q^2 \bigg]
+ ( 6 Y^{(1)}_1 - 3 Y^{(2)}_1 - 3 Y^{(0)}_1 ) m^2 
\nonumber \\
& & \phantom{xxxx} + ( 4 Y^{(3)}_1 - 9 Y^{(2)}_1 + 6 Y^{(1)}_1 - Y^{(0)}_1 ) q^2 
+ 2 Z_1^{(2)} - 2 Z_1^{(1)} 
\label{k1} \\
& & {\cal K}_2 (q^2, m^2, M^2) =  {\Gamma^2 (2 - d/2) \over 
(4\pi)^d}~(M^2)^{d-4}~ {d-1\over d(d-2)}~\bigg[  
{2(d-1)\over (d-3)(5-d)}m^2 M^2 \nonumber \\
& &  \phantom{xxxx} -  {4(d-1)\over (d-2)(d-3)} M^4 
+ \left(  {2(d-1)\over 3}\bigg( {m^2 \over M^2}\bigg)^{d/2-2} 
- {d-1\over (5-d)(7-d)} \right)m^4 \nonumber \\
& & \phantom{xxxx} + \left( {2d\over 3} \bigg( {m^2 \over M^2}\bigg)^{d/2 - 2} 
+ {2(d^2 -5d - 8) \over (5-d)(7-d)(d+2)} \right)q^2 m^2 
\nonumber \\
& & \phantom{xxxx} + {2(6 + 3d - d^2) \over (d-3)(5-d)(d+2)} q^2 M^2 
+ {(4-d)(d^2 - 3d -22)\over (5-d)(7-d)(d+2)(d+4)}q^4  \bigg] 
\nonumber \\
& & \phantom{xxxx} + \left( - 7 Y^{(3)}_1 + {27\over 2} Y^{(2)}_1 
- {15\over 2} Y^{(1)}_1 + Y^{(0)}_1 \right) q^4 \nonumber \\
& & \phantom{xxxx} + \left( {15\over 2} Y^{(2)}_1 - 12 Y^{(1)}_1 + {9\over 2}
Y^{(0)}_1 \right) m^2 q^2 + {3\over 2}\left( Y^{(0)}_1  
- Y^{(1)}_1 \right) m^4 
\nonumber \\
& & \phantom{xxxx} + \left( {7\over 2} Z_1^{(1)} - 5 Z_1^{(2)} \right) q^2 
+ {3\over 2} Z_1^{(1)}m^2 \ . 
\label{k2}
\eeqa
For the sake of simplicity, we have omitted the arguments of 
the $\{ Y_c^{(n)} \}$ and the $\{ Z_c^{(n)} \}$.  We have verified
that Eq.~(\ref{s}) agrees in the equal mass limit with the explicit 
expression appearing in Ref.~\cite{betal97}.
\subsection{\bf The Finite Sunset Integrals}
Having identified the singular parts of the sunset functions 
by expanding these quantities in a Laurent series about 
$d = 4$, one can express the finite-valued functions 
$\{ Y_c^{(n)} \}$ and $\{ Z_c^{(n)} \}$ which remain 
by means of integral representations, 
\beq
Y^{(n)}_c \equiv {1 \over (16\pi^2)^2} 
\int_{4M^2}^\infty {d\sigma \over \sigma} 
\left( 1 - {4M^2 \over \sigma}\right)^{1/2 + c}
\int_0^1 dx\ x^n ~ \ln(1 + \Delta g) \ \ ,
\label{fn1}
\eeq
and 
\beq
Z^{(n)}_c \equiv {1 \over (16\pi^2)^2} 
\int_{4M^2}^\infty d\sigma~ \left( 1 - {4M^2 \over 
\sigma}\right)^{1/2 + c} \int_0^1 dx\ x^n ~ \left( \ln(1 + \Delta g) - 
\Delta g \right) 
\label{fn2}
\eeq
where 
\beq
\Delta g \equiv \left( {m^2\over x} - q^2 \right){1-x\over \sigma} \ \ .
\label{fn3}
\eeq
For convenience, we shall introduce the dimensionless variables 
\beq
{\bar q}^2 \equiv {q^2 \over 4M^2} \qquad {\rm and} \qquad r^2 \equiv 
{m^2 \over 4M^2} \ \ , 
\label{fn4}
\eeq
and likewise work with the {\it reduced functions} 
${\bar Y}^{(n)}_c$ and ${\bar Z}^{(n)}_c$, 
\beqa
Y^{(n)}_c ({\bar q}^2, r^2)  &\equiv& {1 \over (16\pi^2)^2} 
{\bar Y}^{(n)}_c ({\bar q}^2, r^2)  \nonumber \\
Z^{(n)}_c ({\bar q}^2, r^2)  &\equiv& {4M^2 \over (16\pi^2)^2} 
{\bar Z}^{(n)}_c ({\bar q}^2, r^2)  \label{fn7a}
\eeqa
One is allowed to express such finite quantities 
in terms of the physical meson masses, and it is unerstood 
we do so in the remainder of this section.  For the six flavour 
configurations which can contribute to the sunset 
amplitude, the parameter $r^2$ takes on the numerical values 
\beq
r^2 = \left\{ \begin{array}{ll} 
0.016 & (\eta\eta\pi) \\ 
0.020 & ({\bar K}K\pi) \\ 
0.25 & (3\pi, 3\eta) \\ 
0.31 & ({\bar K}K\eta) \\ 
3.82 & (\pi\pi\eta) \ . 
\end{array} \right. 
\label{fn4a}
\eeq

\subsubsection{\bf Behaviour at $r^2 = 0$ and Near $q^2 = 0$}
In the $r^2 = 0$ limit ({\it i.e.} $m^2 = 0$), analytic expressions 
can be obtained for ${\bar Y}^{(n)}_c$ and ${\bar Z}^{(n)}_c$ 
\beq
{\bar Y}^{(n)}_c ({\bar q}^2 , 0) = - \sum_{k=1}^\infty ~
{B(k+1;n+1)~B(k;a + 3/2) \over k} ~{\bar q}^{2k} \ \ ,
\label{fn8}
\eeq
and 
\beq
{\bar Z}^{(n)}_c ({\bar q}^2 , 0) = - \sum_{k=2}^\infty ~
{B(k+1;n+1)~B(k-1;a + 3/2) \over k} ~{\bar q}^{2k} \ \ ,
\label{fn9}
\eeq
where $B(m;n)$ denotes the Euler beta function.  Observe in the 
summations that the indices begin at $k=1$ for ${\bar Y}^{(n)}_c$ 
and at $k=2$ for ${\bar Z}^{(n)}_c$, {\it i.e.} that 
\beq
{\bar Y}^{(n)}_c (0,0) \ = \ {\bar Z}^{(n)}_c (0,0) \ = \  
{\bar Z}^{(n)'}_c (0,0) \ = \  0 \ \ .
\label{fn9a}
\eeq
For the more general case of nonzero $r^2$ but small $q^2$, 
it is useful to employ a power series 
\beqa
{\bar Y}^{(n)}_c ({\bar q}^2, r^2)  &=& 
{\bar Y}^{(n)}_c (0, r^2)  + {\bar Y}^{(n)'}_c (0, r^2) {\bar q}^2   
+ {1\over 2} {\bar Y}^{(n)''}_c (0, r^2) {\bar q}^4 + \dots    
\nonumber \\
{\bar Z}^{(n)}_c ({\bar q}^2, r^2)  &=& 
{\bar Z}^{(n)}_c (0, r^2)  + {\bar Z}^{(n)'}_c (0, r^2) {\bar q}^2   
+ {1\over 2} {\bar Z}^{(n)''}_c (0, r^2) {\bar q}^4 + \dots    \ \ .
\eeqa
For nonzero $r^2$, one can obtain numerical values for the 
above $q^2 = 0$ derivatives of ${\bar Y}^{(n)}_c ({\bar q}^2, r^2)$ 
and ${\bar Z}^{(n)}_c ({\bar q}^2, r^2)$.  Of course, the integral 
representations of Eqs.~(\ref{fn1}),(\ref{fn2}) allow also for 
a straightforward numerical determination of the real part 
of the sunset amplitudes for arbitrary $q^2$.  However, some care 
must be taken to obtain accurate values for $q^2$ close to 
or above three-particle thresholds.  

\subsubsection{\bf Imaginary Parts}

For ${\bar q}^2 <1$, the finite sunset amplitudes are real-valued.
However, $Y^{(n)}_c$ and $Z^{(n)}_c$ have a branch point singularity 
at ${\bar q}^2 = (1 + r)^2$ (corresponding to $q^2 = (2M + m)^2$) and 
become complex-valued for ${\bar q}^2 > (1 + r)^2$.  We shall be 
concerned here with determining the imaginary parts of these
quantities.

Consider first the integral $X^{(n)}({\bar q}^2,r^2)$ defined by 
\beq
X^{(n)}({\bar q}^2,r^2) \equiv \int_0^1 dx\ x^n ~ \ln(1 + \Delta g) \ \ ,
\label{im6}
\eeq
which can be rewritten as 
\beqa
\lefteqn{X^{(n)}({\bar q}^2,r^2) =} \nonumber \\
& & \int_0^1 dx\ x^n ~ \bigg[ 
\ln\left( x^2 + x\bigg( {1\over u{\bar q}^2} - 1 - 
{r^2 \over {\bar q}^2}\bigg) + {r^2 \over {\bar q}^2} \right) 
+ \ln(u{\bar q}^2/x) \bigg] \nonumber \\ 
&=& \int_0^1 dx\ x^n ~ \bigg[ \ln \left((x - x_+)(x - x_-)\right) 
+ \ln(u{\bar q}^2/x) \bigg] \ \ ,
\label{im7}
\eeqa
where $x_\pm$ are given by 
\beq
x_\pm = {1\over 2}\left[ 1 - {1\over u{\bar q}^2} + 
{r^2 \over {\bar q}^2} \pm \sqrt{\bigg( 1 - {1\over u{\bar q}^2} + 
{r^2 \over {\bar q}^2}\bigg)^2 - {4 r^2 \over {\bar q}^2} }~\right] \ \ .
\label{im8}
\eeq

The imaginary part of $X^{(n)}({\bar q}^2,r^2)$ will occur when 
the argument of the first logarithm in the above becomes negative, 
\beqa
{\cal I}m~X^{(n)}({\bar q}^2,r^2) &=& \int_0^1 dx\ x^n ~ 
{\cal I}m~ \ln \left((x - x_+)(x - x_-)\right) \nonumber \\
&=& - \pi~\int_{x_-}^{x_+} dx\ x^n \nonumber \\
&=& - {\pi \over n+1} \left( x_+^{n+1} - x_-^{n+1} \right) \ \ ,
\label{im9}
\eeqa
so that 
\beqa
{\cal I}m~{\bar Y}^{(n)}_c ({\bar q}^2,r^2) &=& - {\pi \over n+1} 
\int_{u_0}^1 {du \over u} (1 - u)^{1/2 + a} 
\left( x_+^{n+1} - x_-^{n+1} \right) \ \ ,
\label{im10}
\eeqa
where 
\beq
u_0 = {1 \over \left( \sqrt{ {\bar q}^2 } - \sqrt{ r^2 } \right)^2 } \ \ . 
\label{im11}
\eeq
The lower limit $u_0$ on the $u$-integral is simply a reflection 
of the branch point occurring in the sunset amplitude at 
$q^2 = (2M + m)^2$.  Proceeding in like manner leads to 
the following formula for ${\cal I}m~Z^{(n)}_c$, 
\beqa
{\cal I}m~{\bar Z}^{(n)}_c ({\bar q}^2,r^2) &=& - {\pi \over n+1} 
\int_{u_0}^1 {du \over u^2} (1 - u)^{1/2 + a} 
\left( x_+^{n+1} - x_-^{n+1} \right) \ \ .
\label{im12}
\eeqa

\subsection{\bf Identities}

Given the set of sunset integrals $\{ S; S_\mu; S_{\mu\nu} \}$, 
it is not difficult to infer the following `trace identity', 
\beq
S^\mu_\mu (q^2, m^2, M^2) = m^2 S (q^2, m^2, M^2) + A^2 (M^2) \ \ ,
\label{tr3}
\eeq
which is valid for arbitrary kinematics.  

It turns out that several more identities become derivable 
in the equal mass limit of $SU(3)$ symmetry.  This is a 
consequence of the symmetry constraint that the isospin and 
hypercharge results agree.  Indeed, their direct comparison 
serves to check the correctness of the calculation.  
Interestingly, the identities discovered in the $SU(3)$ limit 
are typically not at all {\it a priori} obvious.  Below, we list 
and indicate the source of relations:
\begin{enumerate}
\item Relating ${\bar S}$ to $S$:
\beq
{\bar S} (q^2, m^2, m^2) = {1 \over 3} S (q^2, m^2, m^2) \ \ .
\label{tr9}
\eeq
\item 1PI amplitudes:
\beq
{\cal H}_{\mu\nu} (q^2, m^2 , m^2 ) = 
3 {\cal L}_{\mu\nu} (q^2, m^2 , m^2 ) \ \ .
\label{con1}
\eeq
\item Vertex functions:
\beqa
I_{1\mu}(q^2;m^2;m^2;\Lambda) &=& I_{1\mu}(q^2;m^2;m^2;0) 
\nonumber \\
I_{1\mu}(q^2;m^2;m^2;0) &=& 3 I_{2\mu}(q^2;m^2;m^2;0) 
\label{con2}
\eeqa
\item Self-energies:
\beqa
R(q^2; m^2; m^2; \Lambda) &=& R(q^2; m^2; m^2; 0) + 
\Lambda^2 S(q^2;m^2;m^2) \nonumber \\
R(q^2; m^2; m^2; 0) &=& 3~ U(q^2; m^2; m^2) \ \ . 
\label{con3}
\eeqa
\end{enumerate}
\section{Compendium of Finite Results}
In this Appendix we complete the compilation begun in Sect.~VI 
of finite results in our two-loop calculation.  We list in 
turn expressions for the hypercharge polarization functions, 
then the pion and eta decay constants and finally the 
pion and eta masses.  The Appendix concludes with a brief 
summary relating our ${\overline\lambda}$-subtraction 
renormalization with the $\overline{MS}$ scheme of 
Ref.~\cite{betal97}.

\subsection{Hypercharge Polarization Functions}
First we display the corresponding hypercharge results, 
beginning with the remnant piece of the polarization 
function ${\hat \Pi}_8^{(1)}$, 
\begin{eqnarray}
& & F_\pi^2 {\hat \Pi}_{8,{\rm rem}}^{(1)} =
{M_\pi^2 \over \pi^4} \left[ {13 \over 6144} - {C \over 512} + 
{1\over 512} \log{M_K^2 \over \mu^2} \right]
\nonumber \\ 
& & + {M_K^2 \over \pi^4} \left[ - {1 \over 4096} - {9~C \over 1024} 
+ \log{M_K^2 \over \mu^2} \left( {9 \over 1024} - 
{3\pi^2 \over 4} L_{10}^{(0)} - {3 \over 1024} \log{M_K^2 \over \mu^2} 
\right) \right] \nonumber \\ 
& & + {q^2 \over \pi^4} \left[ - {35 \over 32768} + {3~C \over 4096} 
- {3 \over 4096} \log{M_K^2 \over \mu^2} \right] \ \ ,
\label{fin2b1a}
\end{eqnarray}
and then the counterterm contribution, 
\begin{equation}
{\hat \Pi}_{8,{\rm CT}}^{(1)} (q^2) = 
- {q^2 \over F_\pi^2} P_{\rm A} 
- {4 M_\pi^2 \over F_\pi^2} \left( R_{\rm A} - {1\over 3} Q_{\rm A} 
\right) - {8 M_K^2 \over F_\pi^2} 
\left( R_{\rm A} + {2\over 3} Q_{\rm A} \right) \ .
\label{fin2b1c} 
\end{equation}
As was explained in Sect.~VI, the $L_{10}^{(-1)}$ dependence 
from the polarizations ${\hat \Pi}_{3,8,{\rm rem}}^{(1)}$
has been removed via the definitions of $P_{\rm A},Q_{\rm A},
R_{\rm A}$ given in Eq.~(\ref{ctA}).  

Concluding with the part from the finite functions, we have 
\begin{eqnarray}
& & {\hat \Pi}_{8,{\rm YZ}}^{(1)} (q^2) = 
{1\over 2} {\cal H}^{qq}_{\rm YZ} (q^2, M_\pi^2, M_K^2) + 
{1\over 2} {\cal H}^{qq}_{\rm YZ} (q^2, M_\eta^2, M_K^2) - 
R_{8,{\rm YZ}} \ \ ,
\label{fin2b1b}
\end{eqnarray}
where, making use of the notation established in Eq.~(\ref{aux}), 
\begin{eqnarray}
& & R_{8,{\rm YZ}} (q^2) = 
{1\over 2(q^2 - M_\eta^2)} 
\bigg[ {\overline I}_{1,{\rm YZ}} (q^2 ; M_\pi^2; M_K^2 ; {2(M_\pi^2 - M_K^2)\over 3}) 
\nonumber \\
& & + {\overline I}_{1,{\rm YZ}} 
(q^2 ; M_\eta^2; M_K^2 ; {2(3 M_K^2 - M_\pi^2)\over 3}) \bigg] 
- {1\over (q^2 - M_\eta^2)^2}\bigg[ 
{M_\pi^4 \over 6} {\breve S}_{\rm YZ} (q^2 ; M_\eta^2 ; M_\pi^2) 
\nonumber \\ 
& & + {(16 M_K^2 - 7 M_\pi^2 )^2 \over 486} 
{\breve S}_{\rm YZ} (q^2 ; M_\eta^2 ; M_\eta^2) 
+ {1\over 8} {\breve R}(q^2 ; M_\pi^2 ; M_K^2; 
{2 (M_\pi^2 - M_K^2)\over 3}) 
\nonumber \\ 
& & + {1\over 8} {\breve R}(q^2 ; M_\eta^2 ; M_K^2; 
{2 (3 M_K^2 - M_\pi^2)\over 3}) 
\bigg] \ \ ,
\label{rmdr3a} 
\end{eqnarray}
and at $q^2 = 0$, we find 
\begin{equation}
F_\pi^2 {\hat \Pi}_{8,{\rm YZ}}^{(1)} (0) = 5.494 \times 
10^{-6}~{\rm GeV}^2 \ \ .
\label{val188}
\end{equation}
The last of the polarization functions is ${\hat \Pi}_8^{(0)}$, for 
which the remnant piece is 
\begin{eqnarray}
\lefteqn{F_\pi^2 {\hat \Pi}_{8,{\rm rem}}^{(0)} (q^2) =} \nonumber \\ 
& & {M_K^4 \over \pi^4} \left( - {2971 \over 497664} 
+ {307~C \over 62208} - {1\over 256} \log{M_K^2 \over \mu^2} 
- {1\over 972} \log{M_\eta^2 \over \mu^2} \right)
\nonumber \\ 
& & + {M_\pi^2 M_K^2 \over \pi^4} \left( 
{8347 \over 995328} - {787~C \over 124416} 
+ {25\over 4608} \log{M_K^2 \over \mu^2} 
+ {7\over 7776} \log{M_\eta^2 \over \mu^2} \right)
\nonumber \\ 
& & {M_\pi^4 \over \pi^4}\left( - {28123 \over 7962624}
+ {2707~C \over 995328} - {1\over 3072} \log{M_\pi^2 \over \mu^2} 
- {49\over 248832} \log{M_\eta^2 \over \mu^2} \right.
\nonumber \\ 
& & \left. \phantom{xxxxx}
- {9\over 4096} \log{M_K^2 \over \mu^2} \right) \ \ .
\label{fin2b2a}
\end{eqnarray}
Next comes the counterterm contribution, 
\begin{eqnarray}
& & {\hat \Pi}_{8,{\rm CT}}^{(0)} =
M_\pi^4 \left( - 4 B_{11}^{(0)} + {32\over 3} B_{13}^{(0)} \right) 
\label{fin2b2c} \\ 
& & \phantom{xxxxx} + M_\pi^2 M_K^2 {32 \over 3} 
\left( B_{11}^{(0)} - 2 B_{13}^{(0)} \right) 
+ M_K^4 {32 \over 3} \left( - B_{11}^{(0)} 
+ B_{13}^{(0)} \right) \bigg] \ \ , 
\nonumber 
\end{eqnarray}
followed finally by the piece from the finite functions, 
\begin{eqnarray}
\lefteqn{F_\pi^2 {\hat \Pi}_{8,{\rm YZ}}^{(0)} (q^2) = 
{9\over 2} S_{2,{\rm YZ}} (q^2, M_\pi^2, M_K^2) + 
{9\over 2} S_{2,{\rm YZ}} (q^2, M_\eta^2, M_K^2)} 
\nonumber \\
& & + q^2 \bigg( {1\over 2} {\cal H}^{qq}_{\rm YZ} (q^2, M_\pi^2, M_K^2) + 
{1\over 2} {\cal H}^{qq}_{\rm YZ} (q^2, M_\eta^2, M_K^2) 
- R_{8,{\rm YZ}} (q^2)\bigg) 
\nonumber \\
& & - 2 \bigg[ {1\over 4}I_{1,{\rm YZ}} 
\left(M_\eta^2;M_\pi^2;M_K^2;{2\over 3}(M_\pi^2 - M_K^2)\right) 
\nonumber \\
& & + {1\over 4}I_{1,{\rm YZ}} 
\left(M_\eta^2;M_\eta^2;M_K^2;{2\over 3}(3 M_K^2 - M_\pi^2)\right) 
\label{fin2b2b} \\
& & - {M_\pi^4 \over 12} S'_{\rm YZ} (M_\eta^2 ; M_\eta^2 ; M_\pi^2 ) -
{(16 M_K^2 - 7 M_\pi^2)^2 \over 972} S'_{\rm YZ} 
(M_\eta^2 ; M_\eta^2 ; M_\eta^2 ) 
\nonumber \\
& & - {1\over 16} R'_{\rm YZ} 
(M_\eta^2 ; M_\pi^2 ; M_K^2 ; {2 \over 3}(M_\pi^2 - M_K^2)) 
\nonumber \\
& & - {1\over 16} R'_{\rm YZ} (M_\eta^2 ; M_\eta^2 ; M_K^2 ; 
{2\over 3}(3 M_K^2 - M_\pi^2)) 
\bigg] \nonumber 
\ \ .
\nonumber 
\end{eqnarray}
The numerical value of the YZ-part at $q^2 = 0$ is 
\begin{equation}
F_\pi^2 {\hat \Pi}_{8,{\rm YZ}}^{(0)} (0) = - 1.753 \times 
10^{-6}~{\rm GeV}^4 \ \ .
\label{val088}
\end{equation}
\subsection{\bf Meson Decay Constants}
Before displaying explicit forms for $F^{(4)}_{\pi,\eta}$, 
we must (i) implement the procedure ({\it cf} Sect.~VI) for 
removing all contributions from the $\{ L_\ell^{(-1)}\}$ 
counterterms and (ii) re-express all 
one-loop mass and decay constant corrections in terms of 
one-loop renormalized quantities.  

There are {\it a priori} seven of the ${\cal O}(p^4)$ counterterms 
$L_\ell^{(-1)}\ (\ell = 1,\dots,6,8)$ 
which contribute to the decay constants 
$F^{(4)}_{(\pi,\eta),{\rm rem}}$.  Analogous to the procedure 
followed for the polarization functions 
${\hat \Pi}_{(3,8),{\rm rem}}^{(1)}$, we remove the 
$\{ L_\ell^{(-1)}\}$ dependence by defining the following 
four effective ${\cal O}(p^4)$ counterterms, 
\beqa
{\tilde B}_1 &\equiv& F_\pi^2 \left(
A^{(0)} - 3 E^{(0)} \right) + 
{28 \over 3}L_1^{(-1)} + {34 \over 3}L_2^{(-1)} 
+ {25 \over 3}L_3^{(-1)} 
\nonumber \\ 
& & \phantom{x} - {26 \over 3} L_4^{(-1)} + 
{8 \over 3} L_5^{(-1)} + 12 L_6^{(-1)} - 12 L_8^{(-1)} \ ,
\nonumber \\ 
{\tilde B}_2 &\equiv& F_\pi^2 \left(
B^{(0)} - 2 E^{(0)}\right) + 
{32 \over 9}L_1^{(-1)} + {8 \over 9}L_2^{(-1)} 
+ {8 \over 9}L_3^{(-1)} 
\nonumber \\ 
& & \phantom{x} - {106 \over 9} L_4^{(-1)} + 
{22 \over 9} L_5^{(-1)} + 20 L_6^{(-1)} \ ,
\nonumber \\ 
{\tilde B}_3 &\equiv& F_\pi^2 \left(
C^{(0)} + E^{(0)} \right) - 
{28 \over 9}L_1^{(-1)} - {34 \over 9}L_2^{(-1)} 
- {59 \over 18}L_3^{(-1)} 
\label{effdc}  \\ 
& & \phantom{x} + {26 \over 9} L_4^{(-1)} - 3 
L_5^{(-1)} - 4 L_6^{(-1)} + 6 L_8^{(-1)} \ ,
\nonumber \\ 
{\tilde B}_4 &\equiv& F_\pi^2 D^{(0)} 
- {104 \over 9}L_1^{(-1)} - {26 \over 9}L_2^{(-1)} 
- {61 \over 18}L_3^{(-1)} 
\nonumber \\ 
& & \phantom{x} + {34 \over 9} L_4^{(-1)} - 
L_5^{(-1)} + 4 L_6^{(-1)} + 2 L_8^{(-1)} \ \ .
\nonumber 
\eeqa

It is natural to express the one-loop mass and decay constant 
corrections in terms of {\it one-loop renormalized quantities}.  
Thus, in the one-loop expressions for the decay constants and 
masses, we shall replace the tree-level parameters by their 
one-loop corrected counterparts, 
\beq
m_i^2 \to M_i^2 + \Delta_i \ , \qquad 
F_0 \to F_i + \delta_i \ ,\ \qquad (i = \pi, K, \eta) \ .
\label{mcorr}
\eeq
The quantities $\Delta_i , \delta_i \ (i = \pi, K, \eta)$ are 
compiled in Ref.~\cite{gk1} (also see Eq.~(\ref{fm2}) of this paper).  
Thus, we write the decay constant relations for $F_\pi$ 
and $F_\eta$ through one-loop as 
\beqa
F_\pi &=& F_0 + {1\over F_\pi} \bigg[ 
4 L_4^{(0)} \left( M_\pi^2 + 2 M_K^2 \right) + 4 L_5^{(0)} M_\pi^2 
\nonumber \\
& & - {M_\pi^2 \over 16 \pi^2} \ln {M_\pi^2 \over \mu^2} 
- {M_K^2 \over 32 \pi^2} \ln {M_K^2 \over \mu^2} \bigg] + \dots 
\nonumber \\
F_\eta &=& F_0 + {1\over F_\pi} \bigg[ 
4 L_4^{(0)} \left( M_\pi^2 + 2 M_K^2 \right) 
+ 4 L_5^{(0)} M_\eta^2 
\nonumber \\
& & - {3 M_K^2 \over 32 \pi^2} \ln {M_K^2 \over \mu^2} \bigg] + \dots \ ,
\label{redc}
\eeqa
and the mass relations for $M_\pi^2$, $M_K^2$ and $M_\eta^2$ 
through one-loop as 
\beqa 
M_\pi^2 &=& m_\pi^2 + {M_\pi^2 \over F_\pi^2} \bigg[
- 8 \left( L_4^{(0)} - 2 L_6^{(0)} \right)
\left( M_\pi^2 + 2 M_K^2 \right)
- 8 \left( L_5^{(0)} - 2 L_8^{(0)} \right) M_\pi^2 
\nonumber \\
& & + {M_\pi^2 \over 32 \pi^2} \ln {M_\pi^2 \over \mu^2} 
- {M_\eta^2 \over 96 \pi^2} \ln {M_\eta^2 \over \mu^2} \bigg] 
+ \dots \ ,
\nonumber \\
M_K^2 &=& m_K^2 + {M_K^2 \over F_\pi^2} \bigg[ 
- 8 \left( L_4^{(0)} - 2 L_6^{(0)} \right) 
\left( M_\pi^2 + 2 M_K^2 \right) 
- 8 \left( L_5^{(0)} - 2 L_8^{(0)} \right) M_K^2 
\nonumber \\
& & + {M_\eta^2 \over 48 \pi^2} \ln {M_\eta^2 \over \mu^2} 
\bigg] + \dots \ ,
\nonumber \\
M_\eta^2 &=& m_\eta^2 + {M_\eta^2 \over F_\pi^2} \bigg[ 
- 8 \left( L_4^{(0)} - 2 L_6^{(0)} \right) 
\left( M_\pi^2 + 2 M_K^2 \right) 
- 8 \left( L_5^{(0)} - 2 L_8^{(0)} \right) M_\eta^2 
\nonumber \\
& & + {M_K^2 \over 16 \pi^2} \ln {M_K^2 \over \mu^2} 
- {M_\eta^2 \over 24 \pi^2} \ln {M_\eta^2 \over \mu^2} 
\bigg] 
\nonumber \\
& & + {M_\pi^2 \over F_\pi^2} \bigg[ 
- {M_\pi^2 \over 32 \pi^2} \ln {M_\pi^2 \over \mu^2} 
+ {M_K^2 \over 48 \pi^2} \ln {M_K^2 \over \mu^2} 
+ {M_\eta^2 \over 96 \pi^2} \ln {M_\eta^2 \over \mu^2} \bigg] 
\nonumber \\
& & + {128~(M_K^2 - M_\pi^2)^2 \over 9~F_\pi^2} 
\left( 3 L_7^{(0)} + L_8^{(0)} \right) \ \ .
\label{rems}
\eeqa
The expressions in Eqs.~(\ref{redc})-(\ref{rems}) constitute 
our {\it conventions} for these quantities.  In adopting 
our convention, we have made two kinds of choices: 
\begin{enumerate}
\item The prefactors $1/F_0$ (for decay constants) and 
$1/F_0^2$ (for masses) have been replaced respectively by 
$1/F_\pi$ and $1/F_\pi^2$ since $F_\pi$ is the most accurately 
determined decay constant.  
\item We have employed $M_\eta^2$ explicitly throughout rather than 
use the GMO relation to replace it.
\end{enumerate}

A consequence of the procedure just described is to 
introduce additional `spill-over' corrections at two-loop order to 
the set of decay constants and masses which were calculated 
earlier in the paper.  Such contributions 
depend on the choice of convention discussed above.  
It is understood that the two-loop decay constants and masses 
($F^{(4)}_{\pi,{\rm rem}}$, $F^{(4)}_{\eta,{\rm rem}}$, 
$M^{(6)2}_{\pi,{\rm rem}}$, $M^{(6)2}_{\eta,{\rm rem}}$)
listed in the remainder of this appendix contain these 
spill-over corrections.  Finally, for convenience we use the 
GMO relation in two-loop contributions to eliminate all factors 
of the eta mass not occurring inside logarithms.  Any error 
thereby made would appear in higher orders.  

We have then for the pion decay constant, 
\begin{eqnarray}
& & F_\pi^3 F^{(4)}_{\pi,{\rm rem}} = 
M_\pi^4 \left( - {283 \over 589824 \pi^4} - 
{547~C \over 73728 \pi^4} 
+ {1\over \pi^2} \left[ - {1 \over 8}L_1^{(0)} 
- {37 \over 144}L_2^{(0)} - {7 \over 108} L_3^{(0)} \right] 
\right. 
\nonumber \\
& & \left. \phantom{xxxxxx} 
+ 8 L_4^{(0)} \left[ 5 L_4^{(0)} + 10 L_5^{(0)} - 8 L_6^{(0)} - 
8 L_8^{(0)} \right] + 8 L_5^{(0)} \left[ 5 L_5^{(0)} - 
8 L_6^{(0)} - 8 L_8^{(0)} \right] \right.
\nonumber \\ 
& & \left. \phantom{xxxxxx} + {1\over \pi^4} \log{M_K^2 \over \mu^2}
\left[ - {3 \over 8192} - {1 \over 12288} \log{M_K^2 \over \mu^2} \right]
\right. 
\nonumber \\ 
& & \left. \phantom{xxxxxx} 
+ {1\over \pi^2} \log{M_\eta^2 \over \mu^2}
\left[ -{1 \over 36864\pi^2} + {1 \over 18} L_1^{(0)} + 
{1\over 72} L_2^{(0)} + {1 \over 72} L_3^{(0)} 
- {1 \over 24} L_4^{(0)} 
\right. \right.
\nonumber \\ 
& & \left. \left. \phantom{xxxxxx} 
- {1 \over 12288 \pi^2} \log{M_K^2 \over \mu^2} \right] 
+ {1\over \pi^2} \log{M_\pi^2 \over \mu^2}
\left[ {119 \over 12288 \pi^2} + {7 \over 4} L_1^{(0)} + 
L_2^{(0)} \right. \right. 
\nonumber \\ 
& & \left. \left. \phantom{xxxxxx} 
+ {7 \over 8} L_3^{(0)} - {9 \over 8} L_4^{(0)} - {3 \over 4} L_5^{(0)} 
+ {1 \over 4096 \pi^2} \log{M_K^2 \over \mu^2}
+ {1 \over 1024 \pi^2} \log{M_\pi^2 \over \mu^2} \right] \right) 
\nonumber \\ 
& & + M_\pi^2 M_K^2 \left( {101 \over 24576 \pi^4} - 
{19~C \over 6144 \pi^4} + 32 L_4^{(0)} 
\left[ 5 L_4^{(0)} + 3 L_5^{(0)}\right] 
- 128 L_6^{(0)} \left[ 2 L_4^{(0)} + L_5^{(0)}\right] \right. 
\nonumber \\ 
& & \left. \phantom{xxxxxx} + {1\over \pi^2} \left[ 
{1 \over 18}L_2^{(0)} + {1 \over 54}L_3^{(0)} 
\right] + {1\over \pi^2} \log{M_K^2 \over \mu^2}
\left[ {13 \over 6144 \pi^2} 
- {1 \over 8}L_4^{(0)} - {3 \over 8}L_5^{(0)} 
\right.\right.
\nonumber \\ 
& & \left. \left. \phantom{xxxxxx} 
+ {7 \over 6144 \pi^2} \log{M_K^2 \over \mu^2} 
+ {1 \over 512\pi^2} \log{M_\pi^2 \over \mu^2}\right] \right. 
\nonumber \\ 
& & \left. \phantom{xxxxxx} 
+ {1\over \pi^2} \log{M_\eta^2 \over \mu^2}
\left[ {1 \over 1536 \pi^2} - {4 \over 9} L_1^{(0)} - 
{1 \over 9} L_2^{(0)} - {1 \over 9} L_3^{(0)} \right. \right.
\label{4a1} \\ 
& & \left. \left. \phantom{xxxxxx} + {1 \over 3} L_4^{(0)} 
+ {1 \over 1536\pi^2} \log{M_K^2 \over
\mu^2} \right] - {1\over 2\pi^2} L_4^{(0)} 
\log{M_\pi^2 \over \mu^2} \right) 
\nonumber \\ 
& & + M_K^4 \left( - {91 \over 24576 \pi^4} - 
{43~C \over 6144 \pi^4} + 32 L_4^{(0)} 
\left[ 5 L_4^{(0)} + 2 L_5^{(0)} - 8 L_6^{(0)} - 4 L_8^{(0)} \right] 
\right. 
\nonumber \\ 
& & \left. \phantom{xxxxxx} + {1\over \pi^2} \left[ 
- {13 \over 36}L_2^{(0)} - {43 \over 432}L_3^{(0)} 
\right] + {1\over \pi^2} \log{M_K^2 \over \mu^2}
\left[ {1 \over 96 \pi^2} + 2 L_1^{(0)} 
\right. \right. 
\nonumber \\ 
& & \left. \left. \phantom{xxxxxx} + 
{1 \over 2}L_2^{(0)} + {5 \over 8} L_3^{(0)} - {5 \over 4}L_4^{(0)} 
- {1\over 768 \pi^2} \log{M_K^2 \over \mu^2} \right] \right. 
\nonumber \\ 
& & \left. \phantom{xxxxxx} 
+ {1\over \pi^2} \log{M_\eta^2 \over \mu^2}
\left[ -{1 \over 768 \pi^2} + {8 \over 9} L_1^{(0)} + 
{2 \over 9} L_2^{(0)} + {2 \over 9} L_3^{(0)} \right. \right.
\nonumber \\ 
& & \left. \left. \phantom{xxxxxx} - {2 \over 3} L_4^{(0)} 
- {1 \over 768\pi^2} \log{M_K^2 \over \mu^2}\right]
\right) \ ,
\nonumber 
\end{eqnarray}
and the corresponding counterterm amplitude is 
\begin{equation}
F_\pi^3 F^{(4)}_{\pi,{\rm CT}} = M_\pi^4 \left( 2 {\tilde B}_1 
- 2 {\tilde B}_2 - 4 {\tilde B}_4 \right) - 
4 M_\pi^2 M_K^2 {\tilde B}_2 - 8 M_K^4 {\tilde B}_4 \ \ .
\label{4a3}
\end{equation}
Finally, we find for the YZ contribution 
\begin{eqnarray}
& & F_\pi^3 F^{(4)}_{\pi,{\rm YZ}} = 
{2\over 9} I_{1,{\rm YZ}}(M_\pi^2; M_\pi^2; 
M_\pi^2; M_\pi^2;)  
+ {1\over 36} I_{1,{\rm YZ}}(M_\pi^2; M_\pi^2; M_K^2;
2(M_\pi^2 + M_K^2)) 
\nonumber \\ 
& & + {1\over 12} I_{1,{\rm YZ}}(M_\pi^2; M_\eta^2; 
M_K^2; {2\over 3}(M_\pi^2 - M_K^2))  
+ {1\over 2} I_{2,{\rm YZ}}(M_\pi^2; M_\pi^2; M_K^2)  
\nonumber \\
& & - {1\over 2}\left[ {M_\pi^4 \over 6} S_{\rm YZ}' 
(M_\pi^2, M_\pi^2, M_\pi^2) 
+ {M_\pi^4 \over 18} S_{\rm YZ}' (M_\pi^2, M_\pi^2, M_\eta^2) 
\right. \label{4a2} \\
& & \left. + {1\over 9} R_{\rm YZ}'(M_\pi^2; M_\pi^2; 
M_\pi^2; M_\pi^2) 
+ {1\over 72} R_{\rm YZ}'(M_\pi^2; M_\pi^2; M_K^2;
2(M_\pi^2 + M_K^2)) \right.
\nonumber \\ 
& & \left. + {1\over 24} R_{\rm YZ}'(M_\pi^2; M_\eta^2; 
M_K^2; {2\over 3}(M_\pi^2 - M_K^2))  
+ {1\over 4} U_{\rm YZ}'(M_\pi^2; M_\pi^2; M_K^2) \right]
\nonumber 
\end{eqnarray}

The eta decay constant is treated analogously and one 
finds for the remnant contribution, 
\begin{eqnarray}
& & F_\pi^3 F^{(4)}_{\eta,{\rm rem}} = 
M_\pi^4 \left( {139903 \over 15925248 \pi^4} - 
{5137~C \over 1990656\pi^4} + 8 L_4^{(0)} \left[ 5 L_4^{(0)} 
- 8 L_6^{(0)} - 8 L_8^{(0)} \right] \right. 
\nonumber \\
& & \left. \phantom{xxxxxx} 
+ {8 \over 3} L_5^{(0)} \left[ 14 L_4^{(0)} - L_5^{(0)} 
+ 8 L_6^{(0)} - 64 L_7^{(0)} - 24 L_8^{(0)} \right] \right. 
\nonumber \\
& & \left. \phantom{xxxxxx} + {1\over \pi^2} \left[ - {1 \over 72}L_1^{(0)} 
- {29 \over 144}L_2^{(0)} - {5 \over 72} L_3^{(0)} \right] \right.
\nonumber \\ 
& & \left. \phantom{xxxxxx} + {1\over \pi^4} \log{M_K^2 \over \mu^2}
\left[ {49 \over 8192} + {5 \over 4096} \log{M_K^2 \over \mu^2} \right]
\right. 
\nonumber \\ 
& & \left. \phantom{xxxxxx} 
+ {1\over \pi^2} \log{M_\eta^2 \over \mu^2}
\left[ -{145 \over 995328 \pi^2} + {1 \over 12} L_1^{(0)} + 
{1\over 12} L_2^{(0)} + {1 \over 24} L_3^{(0)} \right. \right.
\nonumber \\ 
& & \left. \left. \phantom{xxxxxx} - {1 \over 24} L_4^{(0)} 
- {1 \over 4096 \pi^2} \log{M_K^2 \over \mu^2}
\right] \right. 
\nonumber \\ 
& & \left. \phantom{xxxxxx} 
+ {1\over \pi^2} \log{M_\pi^2 \over \mu^2}
\left[ -{25 \over 12288 \pi^2} + {3 \over 2} L_1^{(0)} + 
{3\over 8} L_2^{(0)} + {3 \over 8} L_3^{(0)} \right. \right.
\nonumber \\ 
& & \left. \left. \phantom{xxxxxx} - {9 \over 8} L_4^{(0)} 
+ {1 \over 12} L_5^{(0)} 
- {9 \over 4096 \pi^2} \log{M_K^2 \over \mu^2}
\right] \right) 
\nonumber \\ 
& & + M_\pi^2 M_K^2 \left( - {3443 \over 248832 \pi^4} - 
{127~C \over 497664 \pi^4} \right. 
\nonumber \\ 
& & \left. \phantom{xxxxxx} + 32 L_4^{(0)} 
\left[ 5 L_4^{(0)} - 8 L_6^{(0)} \right] + {32 \over 3} 
L_5^{(0)} \left[ 5 L_4^{(0)} - 4 L_6^{(0)} + 32 L_7^{(0)} + 16 L_8^{(0)} 
\right] \right. 
\nonumber \\ 
& & \left. \phantom{xxxxxx} + {1\over \pi^2} \left[ 
{1 \over 9}L_1^{(0)} 
+ {1 \over 9}L_2^{(0)} + {1 \over 18}L_3^{(0)} \right. \right]
\nonumber \\ 
& & \left. \phantom{xxxxxx} + {1\over \pi^2} 
\log{M_\pi^2 \over \mu^2} \left[ - {1\over 2} L_4^{(0)} 
- {1\over 3} L_5^{(0)} + 
{3\over 512 \pi^2} \log{M_K^2 \over \mu^2} \right] \right.
\nonumber \\ 
& & \left. \phantom{xxxxxx} + {1\over \pi^2} \log{M_K^2 \over \mu^2}
\left[ -{41 \over 18432 \pi^2} 
- {1 \over 8}L_4^{(0)} - {5 \over 24}L_5^{(0)} 
- {5\over 2048 \pi^2} \log{M_K^2 \over \mu^2} \right] \right. 
\nonumber \\ 
& & \left. \phantom{xxxxxx} 
+ {1\over \pi^2} \log{M_\eta^2 \over \mu^2}
\left[ {187 \over 124416 \pi^2} - {2 \over 3} L_1^{(0)} - 
{2 \over 3} L_2^{(0)} - {1 \over 3} L_3^{(0)} \right. \right.
\nonumber \\ 
& & \left. \left. \phantom{xxxxxx} + {1 \over 3} L_4^{(0)} 
+ {1 \over 512 \pi^2} \log{M_K^2 \over \mu^2}\right]
\right) 
\label{4b1} \\ 
& & + M_K^4 \left( {11531 \over 1990656\pi^4} - 
{7303~C \over 497664\pi^4} + {1\over \pi^2} 
\left[ - {2 \over 9}L_1^{(0)} - {17 \over 36}L_2^{(0)} 
- {19 \over 144}L_3^{(0)} \right] \right.
\nonumber \\ 
& & \left. \phantom{xxxxxx} 
+ 32 L_4^{(0)} \left[ 5 L_4^{(0)} - 8 L_6^{(0)} 
- 4 L_8^{(0)} \right] \right. 
\nonumber \\ 
& & \left. \phantom{xxxxxx} 
+ {64 \over 3} L_5^{(0)} 
\left[ 7 L_4^{(0)} + 2 L_5^{(0)} 
- 8 L_6^{(0)} - 8 L_7^{(0)} - 8 L_8^{(0)} \right] \right. 
\nonumber \\ 
& & \left. \phantom{xxxxxx} + {1\over \pi^2} \log{M_K^2 \over \mu^2}
\left[ {11 \over 512 \pi^2} + 2 L_1^{(0)} + 
{1 \over 2}L_2^{(0)} + {7 \over 8} L_3^{(0)} 
- {5 \over 4}L_4^{(0)} \right. \right. 
\nonumber \\ 
& & \left. \left. \phantom{xxxxxx} - {2 \over 3} L_5^{(0)} 
+ {1\over 512 \pi^2} \log{M_K^2 \over \mu^2} \right] 
+ {1\over \pi^2} \log{M_\eta^2 \over \mu^2}
\left[ -{211 \over 62208 \pi^2} + {4 \over 3} L_1^{(0)} \right. \right.
\nonumber \\ 
& & \left. \left. \phantom{xxxxxx} + {4 \over 3} L_2^{(0)} 
+ {2 \over 3} L_3^{(0)} - {2 \over 3} L_4^{(0)} 
- {1 \over 256\pi^2} \log{M_K^2 \over \mu^2}\right]
\right) \ ,
\nonumber 
\end{eqnarray}
whereas the counterterm and YZ amplitudes are respectively 
\begin{eqnarray}
& & F_\pi^3 F^{(4)}_{\eta,{\rm CT}} = {2\over 3} 
M_\pi^4 \left( 3 {\tilde B}_1 + {\tilde B}_2 
+ 8 {\tilde B}_3 - 6 {\tilde B}_4 \right) 
\nonumber \\ 
& & - {4\over 3} M_\pi^2 M_K^2 \left( 4 {\tilde B}_1 + 
{\tilde B}_2 + 8 {\tilde B}_3 \right) 
+ {8\over 3} M_K^4 \left( 2 {\tilde B}_1 - 2
{\tilde B}_2 + 2 {\tilde B}_3 - 3 {\tilde B}_4 \right)
\ \ ,
\label{4b3}
\end{eqnarray}
and 
\begin{eqnarray}
\lefteqn{F_\pi^3 F^{(4)}_{\eta,{\rm YZ}} =} 
\nonumber \\
& & {1\over 4} I_{1,{\rm YZ}}(M_\eta^2; M_\pi^2; 
M_K^2; {2\over 3}(M_\pi^2 - M_K^2))  \nonumber \\
& & + {1\over 4} I_{1,{\rm YZ}}(M_\eta^2; M_\eta^2; M_K^2;
{2\over 3}(3 M_K^2 - M_\pi^2)) 
\nonumber \\ 
& & - {1\over 2}\left[ {M_\pi^4 \over 6} S_{\rm YZ}' 
(M_\eta^2, M_\eta^2, M_\pi^2) 
+ {(16 M_K^2 - 7 M_\pi^2)^2 \over 486} 
S_{\rm YZ}' (M_\eta^2, M_\eta^2, M_\eta^2) \right.
\nonumber \\
& & \left. + {1\over 8} R_{\rm YZ}'(M_\eta^2; M_\pi^2; 
M_K^2; {2\over 3}(M_\pi^2 - M_K^2))  \right. 
\nonumber \\
& & \left. + {1\over 8} R_{\rm YZ}'(M_\eta^2; M_\eta^2; M_K^2;
{2\over 3}(3 M_K^2 - M_\pi^2)) \right] \ \ .
\label{4b2}
\end{eqnarray}
\subsection{\bf Meson Masses}
In order to remove the seven $L_\ell^{(-1)}$ 
${\cal O}(p^4)$ counterterms ($\ell = 1,\dots,8$) 
from $ M^{(6)2}_{(\pi,\eta),{\rm rem}}$, we define the 
following five effective ${\cal O}(p^6)$ counterterms,  
\beqa
{\tilde B}_5 &\equiv& F_\pi^2 \left(
B_3^{(0)} + {1\over 6} F^{(0)} + B_4^{(0)} + B_5^{(0)} 
+ 3 B_7^{(0)} \right) 
\nonumber \\ 
& & + {20 \over 3}L_4^{(-1)} + {23 \over 3}L_5^{(-1)} 
- {40 \over 3}L_6^{(-1)} 
- 40 L_7^{(-1)} - {86 \over 3} L_8^{(-1)} \ ,
\nonumber \\
{\tilde B}_6 &\equiv& {1\over 648} \bigg[ F_\pi^2 \left( 648 
B_6^{(0)} - 36 F^{(0)} - 216 B_4^{(0)} - 216 B_5^{(0)} 
- 648 B_7^{(0)} \right) 
\nonumber \\
& & + 3168 L_4^{(-1)} + 24 L_5^{(-1)} - 6336 L_6^{(-1)} + 
9792 L_7^{(-1)} + 3216 L_8^{(-1)} \bigg] \ ,
\nonumber \\ 
{\tilde B}_7 &\equiv& F_\pi^2 \left( B_{14}^{(0)} 
- {3\over 2} B_4^{(0)} - {3\over 2} B_5^{(0)} 
- {9\over 2} B_7^{(0)} \right) 
\nonumber \\ 
& & - 8 L_6^{(-1)} + 64 L_7^{(-1)} + {62 \over 3} L_8^{(-1)} \ ,
\nonumber \\ 
{\tilde B}_8 &\equiv& F_\pi^2 \left( B_{15}^{(0)} 
+ {1\over 3} F^{(0)} + B_4^{(0)} 
+ 2 B_5^{(0)} + 3 B_7^{(0)} \right) 
\nonumber \\ 
& & - 2 L_5^{(-1)} + {16 \over 3}L_6^{(-1)} 
- 72 L_7^{(-1)} - 24 L_8^{(-1)} 
\nonumber \\ 
{\tilde B}_9 &\equiv& F_\pi^2 \left( B_{16}^{(0)} 
- {1\over 6} F^{(0)} - 2 B_4^{(0)} - B_5^{(0)} - 3 B_7^{(0)} \right) 
\nonumber \\ 
& & + L_5^{(-1)} + {152 \over 3} L_7^{(-1)} + {62 \over 3} L_8^{(-1)}
\ \ .
\label{effmm}
\eeqa

Beginning with the pion squared-mass, we have 
\begin{eqnarray}
& & F_\pi^4 M^{(6)2}_{\pi,{\rm rem}} = 
M_\pi^6 \left( - {3689 \over 884736 \pi^4} + 
{1403~C \over 110592 \pi^4} \right.
\nonumber \\ 
& & \left. \phantom{xxxxxx} + {1\over \pi^2} \left[ {1 \over 4}L_1^{(0)} 
+ {37 \over 72}L_2^{(0)} + {7 \over 54} L_3^{(0)} \right] \right.
\nonumber \\ 
& & \left. \phantom{xxxxxx} 
+ 128 L_4^{(0)} \left[ - L_4^{(0)} - 2 L_5^{(0)}
+ 4 L_6^{(0)} + 4 L_8^{(0)} \right]
\right. \nonumber \\ 
& & \left. \phantom{xxxxxx} 
+ 128 L_5^{(0)} \left[ - L_5^{(0)} + 4 L_6^{(0)} + 4 L_8^{(0)} \right]
\right. \nonumber \\ 
& & \left. \phantom{xxxxxx} 
- 512 L_6^{(0)} \left[ L_6^{(0)} + 2 L_8^{(0)} \right] 
- 512 L_8^{(0)2} 
\right. \nonumber \\ 
& & \left. \phantom{xxxxxx} 
+ {1\over \pi^2} \log{M_\eta^2 \over \mu^2}
\left[ -{13 \over 55296\pi^2} - {1 \over 9} L_1^{(0)} -
{1\over 36} L_2^{(0)} - {1 \over 36} L_3^{(0)} 
+ {1 \over 6} L_4^{(0)} \right. \right.
\nonumber \\ 
& & \left. \left. \phantom{xxxxxx} + {2 \over 27} L_5^{(0)} 
- {2 \over 9} L_6^{(0)} + {4 \over 9} L_7^{(0)} + {1 \over 18} L_8^{(0)} 
- {31 \over 165888 \pi^2} \log{M_\eta^2 \over \mu^2} \right] \right. 
\nonumber \\ 
& & \left. \phantom{xxxxxx} 
+ {1\over \pi^2} \log{M_\pi^2 \over \mu^2}
\left[ -{281 \over 18432 \pi^2} - {7 \over 2} L_1^{(0)} - 2 
L_2^{(0)} - {7 \over 4} L_3^{(0)} + {9 \over 2} L_4^{(0)} 
\right. \right.
\nonumber \\ 
& & \left. \left. \phantom{xxxxxx} + 3 L_5^{(0)} 
- 8 L_6^{(0)} - {11 \over 2} L_8^{(0)} 
+ {5 \over 2048 \pi^2} \log{M_\pi^2 \over \mu^2}
\right. \right. \nonumber \\ 
& & \left. \left. \phantom{xxxxxx} 
- {1 \over 3072 \pi^2} \log{M_\eta^2 \over \mu^2} \right] 
+ {1\over \pi^4} \log{M_K^2 \over \mu^2} \left[ -{35 \over 36864}
- {13 \over 18432 } \log{M_K^2 \over \mu^2} 
\right. \right. \nonumber \\ 
& & \left. \left. \phantom{xxxxxx} 
+ {3 \over 2048} \log{M_\pi^2 \over \mu^2} 
- {1 \over 18432} \log{M_\eta^2 \over \mu^2} 
\right] \right)
\label{3a1} \\ 
& & + M_\pi^4 M_K^2 \left( -{49 \over 55296 \pi^4} + 
{47~C \over 13824 \pi^4} + {1\over \pi^2} \left[ 
- {1 \over 9}L_2^{(0)} - {1 \over 27}L_3^{(0)} \right] \right. 
\nonumber \\ 
& & \left. \phantom{xxxxxx} 
+ 128 L_4^{(0)} \left[ - 4 L_4^{(0)} - 3 L_5^{(0)} 
+ 16 L_6^{(0)} + 6 L_8^{(0)} \right]
\right. \nonumber \\ 
& & \left. \phantom{xxxxxx} 
+ 256 L_6^{(0)} \left[ 3 L_5^{(0)} - 8 L_6^{(0)} - 6 L_8^{(0)} \right] 
\right. \nonumber \\ 
& & \left. \phantom{xxxxxx} 
+ {1\over \pi^2} \log{M_K^2 \over \mu^2}
\left[ - {1 \over 576 \pi^2} 
+ {1 \over 2} L_4^{(0)} + {1 \over 2} L_5^{(0)} - L_6^{(0)} 
- L_8^{(0)} - {5 \over 4608 \pi^2} \log{M_K^2 \over \mu^2} \right] 
\right. \nonumber \\ 
& & \left. \phantom{xxxxxx} 
+ {1\over \pi^2} \log{M_\eta^2 \over \mu^2} \left[ - {7 \over 6912 \pi^2} 
+ {8 \over 9} L_1^{(0)} + {2 \over 9} L_2^{(0)} + {2 \over 9} L_3^{(0)} 
\right. \right.
\nonumber \\ 
& & \left. \left. \phantom{xxxxxx} 
- {7 \over 6} L_4^{(0)} - {10 \over 27} L_5^{(0)} 
+ {13 \over 9} L_6^{(0)} - {20 \over 9} L_7^{(0)} 
- {2 \over 9} L_8^{(0)} 
+ {7 \over 20736\pi^2} \log{M_\eta^2 \over \mu^2} \right]
\right. 
\nonumber \\ 
& & \phantom{xxxxxx} \left. 
+ {1\over \pi^2} \log{M_\pi^2 \over \mu^2}
\left[ {5 \over 2} L_4^{(0)} - 5 L_6^{(0)} 
+ {5 \over 2304 \pi^2} \log{M_\eta^2 \over \mu^2} 
\right] \right) 
\nonumber \\ 
& & + M_\pi^2 M_K^4 \left( {91 \over 12288 \pi^4} + 
{43~C \over 3072 \pi^4} + {1\over \pi^2} \left[ 
{13 \over 18}L_2^{(0)} + {43 \over 216}L_3^{(0)} \right]
\right. \nonumber \\ 
& & \left. \phantom{xxxxxx} 
+ 128 L_4^{(0)} \left[ - 4 L_4^{(0)} - L_5^{(0)} 
+ 16 L_6^{(0)} + 2 L_8^{(0)} \right] 
\right. \nonumber \\ 
& & \left. \phantom{xxxxxx} 
+ 256 L_6^{(0)} \left[ L_5^{(0)} - 8 L_6^{(0)} - 2 L_8^{(0)} \right] 
\right. \nonumber \\ 
& & \left. \phantom{xxxxxx} 
+ {1\over \pi^2} \log{M_K^2 \over \mu^2}
\left[ - {1 \over 48 \pi^2} 
- 4 L_1^{(0)} - L_2^{(0)} - {5 \over 4}L_3^{(0)} 
+ 5 L_4^{(0)} \right.\right.
\nonumber \\ 
& & \left. \left. \phantom{xxxxxx} + L_5^{(0)} 
- 6 L_6^{(0)} - 2 L_8^{(0)} 
+ {11 \over 3072 \pi^2} \log{M_K^2 \over \mu^2} 
\right] \right. 
\nonumber \\ 
& & \left. \phantom{xxxxxx} 
+ {1\over \pi^2} \log{M_\eta^2 \over \mu^2}
\left[ {1 \over 384 \pi^2} - {16 \over 9} L_1^{(0)} - 
{4 \over 9} L_2^{(0)} - {4 \over 9} L_3^{(0)} + 2 L_4^{(0)} 
\right. \right.
\nonumber \\ 
& & \left. \left. \phantom{xxxxxx} + {8 \over 27} L_5^{(0)} 
- {20 \over 9} L_6^{(0)} + {16 \over 9} L_7^{(0)} 
+ {1 \over 1152\pi^2} \log{M_K^2 \over \mu^2} 
- {1 \over 10368 \pi^2} \log{M_\eta^2 \over \mu^2} \right]\right) \ ,
\nonumber
\end{eqnarray}

\begin{eqnarray}
\lefteqn{F_\pi^4 M^{(6)2}_{\pi,{\rm CT}} =} \nonumber \\ 
& & 2 M_\pi^6 \left( - 2{\tilde B}_1 + 2 {\tilde B}_2 + 
4 {\tilde B}_4 + 5 {\tilde B}_5 + 3 {\tilde B}_6 + 
4 {\tilde B}_7 - 2 {\tilde B}_9 \right) \ ,
\nonumber \\ 
& & + 8 M_\pi^4 M_K^2 \left( {\tilde B}_2 + 3 {\tilde B}_6 
- {\tilde B}_9 \right) \ ,
\nonumber \\ 
& & + 8 M_\pi^2 M_K^4 \left( 2 {\tilde B}_4 + {\tilde B}_5 
+ 3 {\tilde B}_6 \right) \ ,
\label{3a3}
\end{eqnarray}
and 
\begin{eqnarray}
\lefteqn{F_\pi^4 M^{(6)2}_{\pi,{\rm YZ}} =} 
\nonumber \\ 
& & - {M_\pi^4 \over 6} S_{\rm YZ} (M_\pi^2, M_\pi^2, M_\pi^2) 
- {M_\pi^4 \over 18} S_{\rm YZ} (M_\pi^2, M_\pi^2, M_\eta^2) 
- {1\over 4} U_{\rm YZ}(M_\pi^2, M_\pi^2, M_K^2) \nonumber \\
& & - {1\over 9} R_{\rm YZ}(M_\pi^2; M_\pi^2; M_\pi^2; M_\pi^2) 
- {1\over 72} R_{\rm YZ}(M_\pi^2; M_\pi^2; M_K^2;2(M_\pi^2 + M_K^2)) 
\nonumber \\
& & - {1\over 24} R_{\rm YZ}(M_\pi^2; M_\eta^2; M_K^2;{2\over 3}
(M_\pi^2 - M_K^2)) 
\label{3a2}
\end{eqnarray}

Likewise, we find for the eta squared-mass, 
\begin{eqnarray}
& & F_\pi^4 M^{(6)2}_{\eta,{\rm rem}} = 
M_\pi^6 \left( - {13405 \over 7962624 \pi^4} + 
{14005~C \over 2985984 \pi^4} \right. 
\nonumber \\
& & \left. \phantom{xxxxxx} + {128\over 3} L_4^{(0)} 
\left[ L_4^{(0)} + {2 \over 3}L_5^{(0)} 
- 4 L_6^{(0)} + 24 L_7^{(0)} + 8 L_8^{(0)} \right] 
\right. \nonumber \\
& & \left. \phantom{xxxxxx} + {128\over 9} L_5^{(0)} 
\left[ - {1 \over 3}L_5^{(0)} 
- 4 L_6^{(0)} + 64 L_7^{(0)} + {64\over 3} L_8^{(0)} \right] 
\right. \nonumber \\
& & \left. \phantom{xxxxxx} + {512\over 3} L_6^{(0)} 
\left[ L_6^{(0)} - 12 L_7^{(0)} - 4 L_8^{(0)} \right] 
\right. \nonumber \\
& & \left. \phantom{xxxxxx} - {512\over 9} L_8^{(0)} 
\left[ 16 L_7^{(0)} + 5 L_8^{(0)} \right] 
\right. \nonumber \\
& & \left. \phantom{xxxxxx} + {1\over \pi^2} \left[ - 
{1 \over 108}L_1^{(0)} - {29 \over 216}L_2^{(0)} 
- {5 \over 108} L_3^{(0)} \right] 
\right. \nonumber \\ 
& & \left. \phantom{xxxxxx} 
+ {1\over \pi^2} \log{M_\eta^2 \over \mu^2}
\left[ -{1049 \over 1492992 \pi^2} + {1 \over 18} L_1^{(0)} + 
{1\over 18} L_2^{(0)} + {1 \over 36} L_3^{(0)} \right. \right.
\nonumber \\ 
& & \left. \left. \phantom{xxxxxx} - {1 \over 9} L_4^{(0)} -
{10 \over 81} L_5^{(0)} + {7 \over 27} L_6^{(0)} 
+ {16 \over 27} L_7^{(0)} + {17 \over 54} L_8^{(0)} 
\right. \right. 
\nonumber \\ 
& & \left. \left. \phantom{xxxxxx} 
- {67 \over 497664 \pi^2} \log{M_\eta^2 \over \mu^2} 
- {5 \over 6144 \pi^2} \log{M_K^2 \over \mu^2} 
+ {37 \over 27648 \pi^2} \log{M_\pi^2 \over \mu^2} \right] \right. 
\nonumber \\ 
& & \left. \phantom{xxxxxx} 
+ {1\over \pi^2} \log{M_\pi^2 \over \mu^2}
\left[ -{1 \over 2048 \pi^2} + L_1^{(0)} + {1\over 4}
L_2^{(0)} + {1 \over 4} L_3^{(0)} - {5 \over 3} L_4^{(0)} 
\right. \right.
\nonumber \\ 
& & \left. \left. \phantom{xxxxxx} + {1 \over 9} L_5^{(0)} 
+ {7 \over 3} L_6^{(0)} 
- 12 L_7^{(0)} - {85 \over 18} L_8^{(0)} 
+ {95 \over 18432 \pi^2} \log{M_\pi^2 \over \mu^2} \right] 
\right. \nonumber \\ 
& & \left. \phantom{xxxxxx} 
+ {1 \over \pi^4} \log{M_K^2 \over \mu^2} \left[ 
- {11 \over 4096} + {1 \over 6144} \log{M_K^2 \over \mu^2} 
+ {1 \over 2048} \log{M_\pi^2 \over \mu^2} \right] \right) 
\nonumber \\ 
& & + M_\pi^4 M_K^2 \left( -{2597 \over 221184 \pi^4} - 
{263~C \over 248832 \pi^4} \right. \nonumber \\ 
& & \left. \phantom{xxxxxx} + {128 \over 3} L_5^{(0)} 
\left[ - L_4^{(0)} + {2 \over 3} L_5^{(0)} + 2 L_6^{(0)} 
- 16 L_7^{(0)} - 4 L_8^{(0)} \right] \right.
\nonumber \\ 
& & \phantom{xxxxxx} \left.
+ {256 \over 3} L_8^{(0)} \left[ - L_4^{(0)} 
+ 2 L_6^{(0)} - 16 L_7^{(0)} - 8 L_8^{(0)} \right] \right. 
\nonumber \\ 
& & \left. \phantom{xxxxxx} + {1\over \pi^2} \left[ 
{1 \over 9}L_1^{(0)} + {11 \over 18}L_2^{(0)} 
+ {2 \over 9}L_3^{(0)} \right] \right. 
\nonumber \\ 
& & \left. \phantom{xxxxxx} + 
{1\over \pi^2} \log{M_K^2 \over \mu^2} \left[ 
- {373\over 27648 \pi^2} 
- {1 \over 6}L_4^{(0)} + {5 \over 18} L_5^{(0)} 
+ {1 \over 3}L_6^{(0)} - {16 \over 3} L_7^{(0)} 
\right. \right. 
\nonumber \\ 
& & \left. \left. \phantom{xxxxxx} 
- {7 \over 3}L_8^{(0)} 
+ {5 \over 1536 \pi^2} \log{M_\pi^2 \over \mu^2} 
- {7\over 2304 \pi^2} \log{M_K^2 \over \mu^2} \right]
\right. \nonumber \\ 
& & \left. \phantom{xxxxxx} 
+ {1\over \pi^2} \log{M_\eta^2 \over \mu^2}
\left[  {757 \over 124416 \pi^2} - {2 \over 3} L_1^{(0)} 
- {2 \over 3} L_2^{(0)} - {1 \over 3} L_3^{(0)} 
+ {5 \over 6} L_4^{(0)} 
\right. \right. \nonumber \\ 
& & \left. \left. \phantom{xxxxxx} 
+ {22 \over 27} L_5^{(0)} - {19 \over 9} L_6^{(0)} 
- {16 \over 3} L_7^{(0)} - {70 \over 27} L_8^{(0)} 
\right. \right. \nonumber \\ 
& & \phantom{xxxxxx} \left. \left. 
+ {97 \over 13824 \pi^2} \log{M_K^2 \over \mu^2} 
+ {11 \over 10368 \pi^2} \log{M_\eta^2 \over \mu^2} \right] \right. 
\label{3b1} \\ 
& & \phantom{xxxxxx} \left. 
+ {1\over \pi^2} \log{M_\pi^2 \over \mu^2}
\left[ {7 \over 1536 \pi^2} - 4  L_1^{(0)} 
- L_2^{(0)} - L_3^{(0)} + {9 \over 2} L_4^{(0)} 
\right. \right. 
\nonumber \\ 
& & \phantom{xxxxxx} \left. \left.
- {8 \over 9} L_5^{(0)} -  5 L_6^{(0)} + {52 \over 3} L_7^{(0)} 
+ {68 \over 9} L_8^{(0)} 
\right. \right. 
\nonumber \\ 
& & \phantom{xxxxxx} \left. \left.
- {53 \over 6912 \pi^2} \log{M_\eta^2 \over \mu^2} 
- {1\over 2304 \pi^2}\log{M_\pi^2 \over \mu^2} \right] \right) 
\nonumber \\ 
& & + M_\pi^2 M_K^4 \left( {6887 \over 331776 \pi^4} - 
{883~C \over 82944 \pi^4} 
\right. \nonumber \\ 
& & \left. \phantom{xxxxxx} 
+ 128 L_4^{(0)} \left[ - 4 L_4^{(0)} - L_5^{(0)} 
+ 16 L_6^{(0)} - 24 L_7^{(0)} - {26 \over 3} L_8^{(0)} \right] \right. 
\nonumber \\ 
& & \left. \phantom{xxxxxx} 
+ 256 L_5^{(0)} \left[ L_6^{(0)} 
- {16 \over 3} L_7^{(0)} - {32 \over 9} L_8^{(0)} \right] \right. 
\nonumber \\ 
& & \left. \phantom{xxxxxx} 
+ 512 L_6^{(0)} \left[ - 4 L_6^{(0)} 
+ 12 L_7^{(0)} + {13 \over 3} L_8^{(0)} \right] \right. 
\nonumber \\ 
& & \left. \phantom{xxxxxx} 
+ {8192 \over 3} L_8^{(0)} \left[ 
2 L_7^{(0)} + L_8^{(0)} \right] \right. 
\nonumber \\ 
& & \left. \phantom{xxxxxx} + {1\over \pi^2} \left[ -{4 \over 9}L_1^{(0)} 
- {11 \over 18}L_2^{(0)} - {17 \over 72}L_3^{(0)} \right] \right.
\nonumber \\ 
& & \left. \phantom{xxxxxx} + {1\over \pi^2} \log{M_K^2 \over \mu^2}
\left[  {59 \over 1728 \pi^2} + {4 \over 3}L_1^{(0)} + {1 \over 3}L_2^{(0)} 
+ {7 \over 12}L_3^{(0)} \right.\right.
\nonumber \\ 
& & \left. \left. \phantom{xxxxxx} 
- {1 \over 3} L_4^{(0)} 
- {11 \over 9} L_5^{(0)} 
- {2 \over 3} L_6^{(0)} + {40 \over 3}L_7^{(0)} 
+ {74 \over 9} L_8^{(0)} 
\right.  \right. 
\nonumber \\ 
& & \left. \left. \phantom{xxxxxx} 
- {1 \over 96 \pi^2} \log{M_\pi^2 \over \mu^2} 
+ {139 \over 27648 \pi^2} \log{M_K^2 \over \mu^2} 
\right] \right.
\nonumber \\ 
& & \left. \phantom{xxxxxx} + {1\over \pi^2} \log{M_\pi^2 \over \mu^2}
\left[ {8 \over 3} L_4^{(0)} + {16 \over 9} L_5^{(0)} 
- {16 \over 3} L_6^{(0)} - {16 \over 3} L_7^{(0)} 
- {16 \over 3} L_8^{(0)} 
\right] \right.
\nonumber \\ 
& & \left. \phantom{xxxxxx} + {1\over \pi^2} \log{M_\eta^2 \over \mu^2}
\left[ - {193 \over 10368 \pi^2} 
+ {8 \over 3} L_1^{(0)} + {8 \over 3} L_2^{(0)} + {4 \over 3} L_3^{(0)} 
\right. \right.
\nonumber \\ 
& & \left. \left. \phantom{xxxxxx} 
- 2 L_4^{(0)} - {40 \over 27} L_5^{(0)} 
+ {52 \over 9} L_6^{(0)} + {128 \over 9} L_7^{(0)} + {176 \over 27} L_8^{(0)} 
\right. \right.
\nonumber \\ 
& & \phantom{xxxxxx} \left. \left. 
+ {1 \over 108\pi^2} \log{M_\pi^2 \over \mu^2} 
- {73 \over 3456\pi^2} \log{M_K^2 \over \mu^2} 
- {1 \over 384\pi^2} \log{M_\eta^2 \over \mu^2} \right]\right) 
\nonumber \\ 
& & + M_K^6 \left( - {2401 \over 248832 \pi^4} + 
{6923~C \over 186624 \pi^4} 
+ {1\over \pi^2} \left[ {16 \over 27}
L_1^{(0)} + {34 \over 27}L_2^{(0)}  
+ {19 \over 54}L_3^{(0)} \right] \right.
\nonumber \\ 
& & \left. \phantom{xxxxxx} + {512 \over 3} L_4^{(0)} 
\left[ - 4 L_4^{(0)} - {11 \over 3}L_5^{(0)} 
+ 16 L_6^{(0)} + 12 L_7^{(0)} + 14 L_8^{(0)} \right] 
\right. \nonumber \\
& & \left. \phantom{xxxxxx} + {1024\over 27} L_5^{(0)} 
\left[ - 4 L_5^{(0)} 
+ 33 L_6^{(0)} + 30 L_7^{(0)} + 34 L_8^{(0)} \right] 
\right. \nonumber \\
& & \left. \phantom{xxxxxx} - {2048 \over 3} L_6^{(0)} 
\left[ 4 L_6^{(0)} + 6 L_7^{(0)} + 7 L_8^{(0)} \right] 
\right. \nonumber \\
& & \left. \phantom{xxxxxx} - {4096 \over 9} L_8^{(0)} 
\left[ 7 L_7^{(0)} + 5 L_8^{(0)} \right] 
\right. \nonumber \\
& & \left. \phantom{xxxxxx} + {1\over \pi^2} \log{M_K^2 \over \mu^2}
\left[ -{59 \over 864 \pi^2} - {16\over 3} L_1^{(0)} -
{4 \over 3}L_2^{(0)} - {7 \over 3} L_3^{(0)} 
+ 8 L_4^{(0)} + {40 \over 9} L_5^{(0)} \right. \right.
\nonumber \\ 
& & \left. \left. \phantom{xxxxxx} 
- {32 \over 3} L_6^{(0)} 
- 8 L_7^{(0)} - {104 \over 9} L_8^{(0)} 
- {31 \over 6912 \pi^2} \log{M_K^2 \over \mu^2} 
+ {7\over 288 \pi^2} \log{M_\eta^2 \over \mu^2} \right] \right. 
\nonumber \\ 
& & \left. \phantom{xxxxxx} 
+ {1\over \pi^2} \log{M_\eta^2 \over \mu^2}
\left[ {515 \over 23328 \pi^2} - {32 \over 9} L_1^{(0)} - 
{32 \over 9} L_2^{(0)} - {16 \over 9} L_3^{(0)} 
+ {16 \over 9} L_4^{(0)} \right. \right.
\nonumber \\ 
& & \left. \left. \phantom{xxxxxx} 
+ {64 \over 81} L_5^{(0)} - {160 \over 27} L_6^{(0)} 
- {256 \over 27} L_7^{(0)} - {128 \over 27} L_8^{(0)} 
+ {1 \over 486\pi^2} \log{M_\eta^2 \over \mu^2}\right] \right) \ ,
\nonumber 
\end{eqnarray}
then the counterterm amplitude, 
\begin{eqnarray}
& & F_\pi^4 M^{(6)2}_{\eta,{\rm CT}} = 
{2\over 9} M_\pi^6 \left( 6{\tilde B}_1 + 2 {\tilde B}_2 \right.
\nonumber \\ 
& & \left. + 16 {\tilde B}_3 - 12 {\tilde B}_4 + 9 {\tilde B}_5 - 9 
{\tilde B}_6 - 12 {\tilde B}_7 - 16 {\tilde B}_8 - 2 {\tilde B}_9 \right) 
\nonumber \\ 
& & + {8\over 9} M_\pi^4 M_K^2 \left( - 10 {\tilde B}_1 
- 3 {\tilde B}_2 \right.
\nonumber \\ 
& & \left. - 24 {\tilde B}_3 + 12 {\tilde B}_4 + 9 {\tilde B}_5 + 
+ 20 {\tilde B}_7 + 24 {\tilde B}_8 + 3 {\tilde B}_9 \right) 
\nonumber \\ 
& & + {8\over 9} M_\pi^2 M_K^4 \left( 20 {\tilde B}_1 
+ 36 {\tilde B}_3 - 6 {\tilde B}_4 - 27 {\tilde B}_5 + 
27 {\tilde B}_6 - 40 {\tilde B}_7 - 36 {\tilde B}_8 \right) 
\nonumber \\ 
& & + {32\over 9} M_K^6 \left( 
- 4 {\tilde B}_1 + 4 {\tilde B}_2 \right.
\nonumber \\ 
& & \left. - 4 {\tilde B}_3 + 6 {\tilde B}_4 + 9 {\tilde B}_5 + 9 {\tilde B}_6 
+ 8 {\tilde B}_7 + 4 {\tilde B}_8 - 4 {\tilde B}_9 \right) \ ,
\label{3b3}
\end{eqnarray}
and finally the YZ contribution, 
\begin{eqnarray}
\lefteqn{F_\pi^4 M^{(6)2}_{\eta,{\rm YZ}} =} 
\nonumber \\ 
& & - {M_\pi^4 \over 6} S_{\rm YZ} (M_\eta^2, M_\eta^2, M_\pi^2) 
- {(16 M_K^2 - 7 M_\pi^2)^2 \over 486} 
S_{\rm YZ} (M_\eta^2, M_\eta^2, M_\eta^2) 
\nonumber \\
& & - {1\over 8} R_{\rm YZ}(M_\eta^2; M_\pi^2; 
M_K^2; {2\over 3}(M_\pi^2 - M_K^2))  
\nonumber \\
& & - {1\over 8} R_{\rm YZ}(M_\eta^2; M_\eta^2; M_K^2;
{2\over 3}(3 M_K^2 - M_\pi^2)) \ \ .
\label{3b2}
\end{eqnarray}
\subsection{\bf Relation to ${\overline {MS}}$ Renormalization} 
Our formulae for the hypercharge polarization functions 
${\hat \Pi}_{8}^{(1,0)}$ displayed earlier 
in this appendix refer to the ${\overline\lambda}$ renormalization
used throughout this paper.  To obtain corresponding expressions in 
the ${\overline {MS}}$ renormalization of Ref.~\cite{betal97}, 
it suffices to follow the discussion in Sect.~VI-C for the 
isospin polarization functions. 

However some additional analysis is required in order to 
compare our ${\overline\lambda}$-subtracted decay constant 
and mass formulae to the ${\overline {MS}}$ scheme.  We 
shall omit detailed derivation and simply display the 
results, as the procedure mirrors that used to obtain 
Eq.~(\ref{msb11}).  What is needed is the following set of 
relations between the ${\overline\lambda}$-subtracted 
constants $\{ {\tilde B}_\ell \}$ and the associated 
${\overline {MS}}$ quantities 
${\tilde B}_\ell^{{\overline{\rm MS}}}$ for $\ell = 1, \dots , 9$,
\beq
{\tilde B}_\ell =  {\tilde B}_\ell^{{\overline{\rm MS}}} 
- \Delta {\tilde B}_\ell   \ \ ,
\eeq
where 
\beq
\Delta {\tilde B}_\ell = {C \over (4 \pi)^4} \left( 
-{175\over 576}, {19\over 96}, {691\over 5184}, {43\over 192}, 
{10\over 27}, - {10\over 81}, - {167\over 288}, 
{371\over 1296}, -{9\over 32} \right) \ \ .
\eeq
As expected from our previous discussion, to obtain 
the masses and decay constants in the 
${\overline {MS}}$ renormalization of Ref.~\cite{betal97} 
one needs only make the replacements ${\tilde B}_\ell  \to 
{\tilde B}_\ell^{{\overline{\rm MS}}}$ and omit all 
dependence on the constant $C$.
 
\vspace{1.0cm}

\begin{center}
{\bf Figure Captions}
\end{center}

\vspace{0.3cm}

\begin{flushleft}
Fig.~1 \hspace{0.2cm} Lowest-order graphs for the axialvector propagator.
 \\
\vspace{0.3cm}
Fig.~2 \hspace{0.2cm} One-loop graphs.  \\
\vspace{0.3cm}
Fig.~3 \hspace{0.2cm} Generic corrections to the axialvector propagator. \\
\vspace{0.3cm}
Fig.~4 \hspace{0.2cm} Two-loop 1PI non-sunset graphs. \\
\vspace{0.3cm}
Fig.~5 \hspace{0.2cm} The two-loop 1PI sunset graph.  \\ 
\vspace{0.3cm}
Fig.~6 \hspace{0.2cm} Two-loop 1PI vertex graphs.  \\ 
\vspace{0.3cm}
Fig.~7 \hspace{0.2cm} Two-loop 1PI self-energy graphs.  \\ 
\vspace{0.3cm}
\end{flushleft}
\vfill \eject

\begin{references}
\bibitem{ecker} This program is under study for 
chiral $SU(N)\times SU(N)$. G. Ecker, priv. comm.
\end{references}
\end{document}